\documentclass[aps,floatfix]{revtex4}

\usepackage[overload]{textcase}
\usepackage[dvips]{graphics}
\usepackage{amssymb}
\usepackage{amsmath}
\usepackage{graphicx,multirow}
\usepackage{axodraw}
\usepackage{pstricks}
\usepackage{color}
\makeatletter
\newcounter{subsubsubsection}[subsubsection]
\def\subsubsubsectionmark#1{}

\def\subsubsubsection{\@startsection
     {subsubsubsection}{4}{\z@} {-3.25ex plus -1
     ex minus -.2ex}{1.5ex plus .2ex}{\normalsize\it}}
\def\l@subsubsubsection{\@dottedtocline{4}{4.8em}
     {4.2em}}
\newcounter{subsubsubsubsection}[subsubsubsection]
\def\subsubsubsubsectionmark#1{}

\def\subsubsubsubsection{\@startsection
     {subsubsubsubsection}{5} {\z@} {-3.25ex plus -1
     ex minus -.2ex}{1.5ex plus .2ex}{\normalsize\bf}}
\def\l@subsubsubsubsection{\@dottedtocline{5}
     {5.8em}{5.2em}}

\def\etal{{\it et al.}}

\def\>{\rangle}
\def\<{\langle}

\def\B{{\cal B}}

\def\I{{\cal I}}

\def\L{{\cal L}}
\def\M{{\cal M}}
\def\O{{\cal O}}

\def\Y{{\cal Y}}
\def\alf{\alpha}
\def\betpp{\beta_{++}}
\def\betpm{\beta_{+-}}
\def\betmp{\beta_{-+}}
\def\betmm{\beta_{--}}
\def\gm{\gamma}
\def\pvec{\vec p}

\def\pl{p_\lambda}
\def\pr{p_\rho}
\def\plvec{\vec{p}_\lambda}
\def\prvec{\vec{p}_\rho}
\def\plsvec{\vec{p}_\Lambda}

\def\kvec{\vec k}
\def\lr{l_\rho}
\def\mr{m_\rho}
\def\lrp{l_{\rho'}}
\def\mrp{m_{\rho'}}

\def\Eb{E_b}
\def\Es{E_s}
\def\Lb{\Lambda_b}
\def\Lc{\Lambda_c}
\def\Ls{\Lambda}
\def\sh{\hat{s}}
\def\zh{\hat{z}}
\def\ym{y^{-}}
\def\yp{y^{+}}
\def\qmb{m_b}
\def\qms{m_s}
\def\mlh{\hat{m}_\ell}
\def\mbh{\hat{m}_b}
\def\mlb{m_{\Lambda_b}}
\def\mls{m_{\Lambda}}
\def\plb{p_{\Lambda_b}}
\def\pls{p_{\Lambda}}
\def\slb{s_{\Lambda_b}}
\def\sls{s_{\Lambda}}
\def\Re{\text{Re}}
\def\vvp{v\cdot v'}
\def\vq{\mlb-\mls\vvp}
\def\vpq{\mlb\vvp-\mls}
\def\detvva{\eta^{VV(\alf)}}
\def\detaaa{\eta^{AA(\alf)}}
\def\detvvpp{\eta^{VV(\betpp)}}
\def\detaapp{\eta^{AA(\betpp)}}
\def\detvvpm{\eta^{VV(\betpm)}}
\def\detaapm{\eta^{AA(\betpm)}}
\def\detvvmp{\eta^{VV(\betmp)}}
\def\detaamp{\eta^{AA(\betmp)}}
\def\detvvmm{\eta^{VV(\betmm)}}
\def\detaamm{\eta^{AA(\betmm)}}
\def\detvag{\eta^{VA(\gm)}}
\def\detavg{\eta^{AV(\gm)}}
\def\={&=&}
\def\<{\langle}
\def\>{\rangle}
\def\jp{{J^P}}
\def\slash#1{#1 \hskip -0.5em / }
\def\beq{\begin{equation}}
\def\eeq{\end{equation}}
\def\nn{\nonumber}
\def\beqy{\begin{eqnarray}}
\def\eeqy{\end{eqnarray}}
\def\beqynn{\begin{eqnarray*}}
\def\eeqynn{\end{eqnarray*}}

\begin{document}

\markboth{L. Mott, W. Roberts}
{Rare dileptonic decays of $\Lambda_b$ in a quark model}

%
%

\title{RARE DILEPTONIC DECAYS OF $\Lambda_b$ IN A QUARK MODEL}

\author{L. Mott}
\address{Department of Physics, Florida State University, Tallahassee, FL 32306}

\author{W. Roberts}
\address{Department of Physics, Florida State University, Tallahassee, FL 32306}

\begin{abstract}
Hadronic form factors for the rare weak transitions $\Lambda_{b}\rightarrow\Lambda^{(*)}$ are calculated using a nonrelativistic quark model. The form factors are extracted in two ways. An analytic extraction using single component wave functions (SCA) with the quark current being reduced to its nonrelativistic Pauli form is employed in the first method. In the second method, the form factors are extracted numerically using the full quark model wave function (MCN) with the full relativistic form of the quark current.  Although there are differences between the two sets of form factors, both sets satisfy the relationships expected from the heavy quark effective theory (HQET). Differential decay rates, branching ratios and forward-backward asymmetries (FBAs) are calculated for the dileptonic decays $\Lambda_{b}\rightarrow\Lambda^{(*)}\ell^{+}\ell^{-}$, for transitions to both ground state and excited daughter baryons. Inclusion of the long distance contributions from charmonium resonances significantly enhances the decay rates. In the MCN model the $\Lambda(1600)$ mode is the dominant mode in the $\mu$ channel when charmonium resonances are considered; the $\Lambda(1520)$ mode is also found to have a comparable branching ratio to that of the ground state in the $\mu$ channel.

\end{abstract}


\maketitle

\section{Introduction}

The weak decays of heavy hadrons have been an important source of information on some of the fundamental parameters of the Standard Model (SM). In particular, semileptonic decays of heavy hadrons have been important in the extraction of some Cabibbo-Kobayashi-Maskawa (CKM) matrix elements \cite{rpp}. However, the description of these processes requires a number of {\it a priori} unknown form factors that parametrize the uncalculable (to date) nonperturbative QCD dynamics. The precision to which these form factors can be calculated or modeled limits the accuracy with which the CKM matrix elements can be extracted. In this regard, the heavy quark effective theory (HQET) \cite{hqet}, quark models \cite{isgw,barik}, QCD sum rules \cite{qcdsr}, lattice QCD \cite{lattice}, etc. have been employed to improve the modeling of these form factors. 

In the same way, decays of heavy hadrons induced by flavor-changing neutral currents (FCNC), the so-called rare decays, have been a subject of significant interest in recent years. Processes like $b\to s\gamma$ and $b\to s\ell^+\ell^-$ are forbidden at tree level, and at one-loop level are suppressed by the Glashow-Iliopoulos-Maiani (GIM) mechanism. These decays therefore receive their main contribution from one-loop diagrams with a virtual top quark and a $W$ boson. Therefore, they provide valuable information about the CKM matrix elements $V_{ts}$ and $V_{tb}$.

In addition, because these decays occur at loop level, they are sensitive to new physics beyond the SM. In these rare decays, new physics can appear either through new contributions to the Wilson coefficients that enter into the effective Hamiltonian that describes these decays, or through new operators in the effective Hamiltonian that arise from sources beyond the SM.
For these reasons, these decays are promising candidates for looking for new physics beyond the SM. Thus, the values of the Wilson coefficients are crucial to the determination of any new physics. As with the case of semileptonic decays, the accuracy to which these parameters can be extracted is limited by our knowledge of the form factors that are used to parametrize the hadronic matrix elements. Many experimentally measurable quantities such as branching ratios, forward-backward asymmetries, lepton polarization asymmetries, etc. can be analyzed to more precisely determine SM parameters and to look for new physics beyond the SM. It is known that most observables will depend sensitively on the form factors; thus, the accuracy to which these form factors can be calculated is paramount to the determination of any new physics inferred from these decays.

There have been many theoretical investigations of rare $b\to s$ transitions in the the meson sector \cite{buras,buchalla,falk,greub,cho,robertsledroit,roberts,hewett,aliev1,aliev2,aliev3,aliev4,aliev5,aliev6,melikhov1,melikhov2,melikhov3,huangli,burdman,goto,huang1,huang2,yan,huang3,huang4,colangelo1,colangelo2,chua,ali,chang,itlan,beneke,kruger,safir,chen3,chen4,chen5,chen6,frank,arda,choudhury,bpsw,ghinculov,bashiry,defazio,ferrandes}. Melikhov {\it et al.} have used a relativistic constituent quark model to treat the dileptonic decays of the $B$ meson \cite{melikhov1}. In \cite{chen3}, perturbative QCD (pQCD) is used to estimate the form factors for $B\to K^{(*)}$. Light-cone sum rules (LCSRs) have been employed for these transitions as well \cite{aliev1,ali,arda,yan}. The form factors obtained from the various models have been used to calculate observables such as branching ratios (BRs), forward-backward asymmetries (FBAs) and lepton polarization asymmetries (LPAs). These observables have been calculated both within the SM and for various scenarios that arise beyond the standard model, such as supersymmetric (SUSY) models \cite{yan,chua,chen4}, models with universal extra dimensions (UEDs) \cite{ferrandes}, and other new physics (NP) scenarios \cite{chen6,arda}.

In the baryon sector, there have been a few techniques employed in the computation of the form factors for the rare $\Lb\to\Ls$ transitions \cite{aslam,wang,chen,chen1,chen2,he,zolf,mannel,huang,cheng}. In much of what has been put forth, HQET is used to reduce the number of independent form factors to two universal form factors valid for all currents, and a model is then employed to compute these two form factors. Aslam {\it et al.} \cite{aslam}, Wang {\it et al.} \cite{wang} and Huang {\it et al.} \cite{huang} have used LCSRs to obtain the form factors for $\Lb\to\Ls$. QCD sum rules (QCDSRs) have also been employed in obtaining these form factors \cite{chen,chen1,chen2,cheng,zolf}. A number of authors \cite{chen,chen2,mannel} have also used pole model parametrizations (PM) for the form factors. In \cite{he} the form factors for these transitions have been estimated using pQCD. The MIT bag model (BM) has been used in \cite{cheng}. To the best of our knowledge, no transitions to excited state $\Ls$'s have been explored.

Experimentally, the exclusive radiative transition $B\to K^*(892)\gamma$ was first observed by CLEO \cite{cleo1}. The branching ratio for this mode has been measured recently by both Belle \cite{belle1} and BaBar \cite{babar1} with an average branching ratio ${\cal B}(B^{0}\to K^{0*}(892)\gamma)=(4.33\pm 1.9)\times 10^{-5}$. Several other decay modes such as $B\to K_1\gamma,K_2^*(1430)\gamma$, etc. \cite{babar2,belle2,belle3,babar3} have been observed as well. The mode $B_s^0\to\phi\gamma$ has been observed by Belle; its branching ratio was measured to be $\B(B_s^0\to\phi\gamma)=(57_{-19}^{+22})\times10^{-6}$ \cite{belle4}. Branching ratios for the inclusive process $B\to X_s\gamma$ have been measured by CLEO \cite{cleo4}, BaBar \cite{babar4}, and Belle \cite{belle5}. From these measurements, the average branching ratio is \cite{hfag1}
\beq
{\cal B}(B\to X_{s}\gamma)=(3.55\pm 0.24\pm 0.09)\times 10^{-4}.\nn
\eeq
Branching ratios for the dileptonic decays have been measured by Belle \cite{belle6}, BaBar \cite{babar5}, and CDF \cite{cdf1}. From the BaBar and Belle data, $\B(B\to K\ell^+\ell^-)=(0.45\pm0.04)\times10^{-6}$ and  $\B(B\to K^*(892)\ell^+\ell^-)=(1.08\pm0.11)\times10^{-6}$. Branching ratios for the inclusive process $B\to X_{s}\ell^{+}\ell^{-}$ have also been measured by both Belle \cite{belle7} and BaBar \cite{babar6}; the average branching ratio is \cite{hfag2}
\beq
{\cal B}(B\to X_{s}\ell^{+}\ell^{-})=(3.66_{-0.77}^{+0.76})\times10^{-6}.\nn
\eeq
Note that this means that decays to the $K$ and $K^*$ account for less than 50\% of the rare dileptonic decays of the $B$ meson. The Belle Collaboration \cite{belle8} has also measured the forward-backward asymmetry in $B\to K^{*}\ell^{+}\ell^{-}$ and extracted ratios of Wilson coefficients from those data.

The experimental situation for $b\to s$ transitions in the baryon sector is less rich than in the meson sector. The CDF collaboration recently reported the first observation of the baryonic FCNC process $\Lb\to\Ls\mu^+\mu^-$ \cite{cdf2}. The branching ratio for this mode was measured to be $\B(\Lb\to\Ls\mu^+\mu^-)=(1.73\pm0.42\pm0.55)\times10^{-6}$. The LHCb Collaboration \cite{lhcb} estimate that with 2 fb$^{-1}$ of data taken in one year, there should be 750 $\Lambda_{b}\to\Lambda\gamma$ events and nearly six times as many events in excited hyperons. For the dileptonic decay mode, 800 events are expected for decays to the ground state $\Lambda$; no potential yields were reported for excited states.

In this paper, we examine the rare weak dileptonic decays of $\Lb$ baryons to ground state and excited $\Ls$ baryons. Some of the motivation for this study has been outlined above. It will also be useful to examine the sensitivity of some of the lepton asymmetries to the form factors. It has been shown for rare meson decays that there are observables, such as lepton polarization asymmetries, that are largely independent of the form factors in certain limits \cite{buras,robertsledroit,roberts,aliev1,aliev2,aliev3,aliev4}. Such quantities thus offer largely model-independent ways to examine the physics content of some of the Wilson coefficients \cite{roberts}. 

To this end, we use two approximations to compute the form factors for $\Lb\to\Ls^{(*)}$ transitions. First, we use the approximation employed in \cite{pervin,pervin1}. In that work, analytic form factors for $\Lb\to\Lc^{(*)}$, $\Lc\to\Ls^{(*)}$, $\Lb\to N^{(*)}$  and $\Lc\to N^{(*)}$ were calculated using single component wave functions obtained from a variational diagonalization of a quark model Hamiltonian. In that calculation, quark operators were reduced to their nonrelativistic Pauli form. The decay rates obtained using the form factors extracted in that calculation were in reasonable agreement with experimental results for the semileptonic decays $\Lb\to\Lc\ell\overline{\nu}_\ell$ and $\Lc\to\Ls\ell\overline{\nu}_\ell$. 

The second approximation we use for computing the form factors is an extension of this method; we keep the full relativistic form of the quark spinors and use the full quark model wave function in a numerical extraction of the form factors. Because this method uses fewer approximations, the form factors obtained in this way should give more reliable results than the form factors obtained from the first method. We present the results of both methods to demonstrate the sensitivity of the observables to the form factors. The goal here is to find observables that may be less dependent on the form factors.

The rest of this paper is organized as follows: in Section \ref{sec:merfba}, we discuss the hadronic matrix elements, decay rates, and forward-backward asymmetries. Section \ref{sec:hqet} gives a brief overview of HQET and presents HQET predictions for the relationships among the form factors for the transitions we investigate. In Section \ref{sec:model}, we describe the quark model used to obtain the form factors, including some description of the Hamiltonian. The two methods we employ to obtain the form factors are also briefly discussed. Numerical results such as form factors, differential decay rates, branching ratios, and forward-backward asymmetries are presented in Section \ref{sec:results}. We present our conclusions and outlook in Section \ref{sec:concl}. Some details of the calculation are shown in the Appendices.

\section{Matrix Elements, Decay Rates, and Forward-Backward Asymmetries\label{sec:merfba}}

\subsection{Matrix Elements\label{sec:bme}}

The amplitude for the dileptonic decay of the $\Lb$ baryon can be written
\begin{equation}
i{\M}(\Lb\to\Ls\ell^+\ell^-)=\frac{G_{F}}{\sqrt{2}}\frac{\alpha_{em}}{2\pi}V_{tb}V^{*}_{ts}(H_{1}^{\mu}L_{\mu}^{(V)}+H_{2}^{\mu}L_{\mu}^{(A)}),
\end{equation}
where $H_{1}^{\mu}$ and $H_{2}^{\mu}$ contain the hadronic matrix elements
\begin{eqnarray}
H_{1}^{\mu}&=&-\frac{2 m_{b}}{q^{2}}C_{7}(m_{b})T_R^{\mu}+C_{9}(m_{b})J_{L}^{\mu},
\label{eq:h1mu} \\
H_{2}^{\mu}&=&C_{10}(m_{b})J_{L}^{\mu}.
\label{eq:h2mu}
\end{eqnarray}
The $C_i$ are the Wilson coefficients, 
\begin{equation}
T_R^\mu=\<\Ls(\pls,\sls)\mid\overline{s}i\sigma^{\mu\nu}q_{\nu}(1+\gamma_{5})b\mid\Lb(\plb,\slb)\>,
\label{eq:trmu}
\end{equation}
and $J_L$ is the matrix element of the standard $V-A$ current
\begin{equation}
J_{L}^{\mu}=\<\Ls(\pls,\sls)\mid \overline{s}\gamma^{\mu}(1-\gamma_{5})b\mid\Lb(\plb,\slb)
\rangle.
\label{eq:jlmu}
\end{equation}
$L_{\mu}^{(V)}$ and $L_{\mu}^{(A)}$ are the vector and axial vector leptonic currents, respectively, written as
\begin{eqnarray}
L_{\mu}^{(V)}&=&\overline{u}_{\ell}(p_{-},s^-)\gamma_{\mu}v_{\ell}(p_{+},s^+), \\
L_{\mu}^{(A)}&=&\overline{u}_{\ell}(p_{-},s^-)\gamma_{\mu}\gamma_{5}v_{\ell}(p_{+},s^+
),
\end{eqnarray}
where $s^\pm$ are the spin projections for the lepton and the antilepton.

In our analysis of the dileptonic decays, we will include the long distance contributions coming from the charmonium resonances $J/\psi$, $\psi'$, $\ldots$ etc. To include these resonant contributions, we replace the Wilson coefficient $C_9$ in Eq. \ref{eq:h1mu} with the effective coefficient
\begin{equation}
C_{9}^{eff}=C_9+Y_{SD}(z,s')+Y_{LD}(s'),
\end{equation}
where $z=m_c/m_b$ and $s'=q^2/m_b^2$. $Y_{SD}$ contains the short distance (SD) contributions from the four-quark operators far from the charmonium resonance regions and $Y_{LD}$ are the long distance (LD) contributions from the four-quark operators near the resonances. The SD term can be calculated reliably in the perturbative theory, but the same cannot be done for the LD contributions. The LD contributions are usually parametrized using a Breit-Wigner formalism by making use of vector meson dominance (VMD) and the factorization approximation (FA). The explicit expressions for $Y_{SD}$ and $Y_{LD}$ are \cite{buras,aslam,wang,chen,chen1,chen2}
\begin{eqnarray}
Y_{SD}(z,s')&=&(3C_1+C_2+3C_3+C_4+3C_5+C_6)h(z,s')-\nonumber \\&&\frac{1}{2}(4C_3+4C_4+3C_5+C_6)h(1,s')-\frac{1}{2}(C_3+3C_4)h(0,s')+\nonumber \\&&\frac{2}{9}(3C_3+C_4+3C_5+C_6), \\
Y_{LD}(s')&=&-\frac{3\pi}{\alpha_{em}^2}(3C_1+C_2+3C_3+C_4+3C_5+C_6)\nn\\ && \times\sum_{j=J/\psi,\psi',\ldots}k_j\frac{m_j\Gamma(j\rightarrow\ell^+\ell^-)}{q^2-m_j^2+im_j\Gamma_j},\nn\\
\end{eqnarray}
where $m_j$, $\Gamma_j$, and $\Gamma(j\rightarrow\ell^+\ell^-)$ are the masses, total widths, and partial widths of the resonances, respectively,
\begin{eqnarray}
h(z,s')&=&-\frac{8}{9}\ln z+\frac{8}{27}+\frac{16z^2}{9s'}-\frac{2}{9}\left(2+\frac{4z^2}{s'}\right)\left| 1-\frac{4z^2}{s'}\right|^{1/2}\nonumber \\&&\times\left\{\Theta\left(1-\frac{4z^2}{s'}\right) \left[\ln\left(\frac{1+\sqrt{1-\frac{4z^2}{s'}}}{1-\sqrt{1-\frac{4z^2}{s'}}}\right)-i\pi\right] +2\Theta\left(\frac{4z^2}{s'}-1\right)\right.\nonumber \\&&\times\left.\arctan\left(\frac{1}{\sqrt{\frac{4z^2}{s'}-1}}\right)
\right\}, \nonumber \\
h(0,s')&=&\frac{8}{27}-\frac{4}{9}\ln s'+\frac{4}{9}i\pi,
\end{eqnarray}
and $k_j$ are phenomenological parameters introduced to compensate for VMD and FA; they are chosen so as to reproduce the correct branching ratio of $\B(B\to K^{(*)}V\to K^{(*)}\ell^+\ell^-)=\B(B\to K^{(*)}V)\B(V\to\ell^+\ell^-)$, for $V=J/\psi,\psi'$. Since none of the analogous branching ratios of the $\Lambda_b$ have yet been measured, we apply the phenomenological factors obtained from decays of the $B$ mesons to the decays of the $\Lambda_b$. For the lowest resonances $J/\psi$ and $\psi'$, we use $k=1.65$ and $k=2.36$, respectively; for the higher resonances, we use the average of $J/\psi$ and $\psi'$ (see \cite{ali,zolf}). 

The matrix elements in Eqs. \ref{eq:trmu}, and \ref{eq:jlmu} contain the currents $\overline{s}\gamma^\mu b$, $\overline{s}\gamma^\mu\gamma_5 b$, $\overline{s}i\sigma^{\mu\nu} b$ and $\overline{s}i\sigma^{\mu\nu}\gamma_5 b$. In this work, we examine decays to daughter baryons with $\jp=1/2^+,\,\,1/2^-,\,\,3/2^-,3/2^+,\,\,5/2^+$. We make this choice because in the quark model we use to calculate the matrix elements, these states have the most significant overlaps with the initial ground state within the spectator quark approximation.

A baryon with angular momentum and parity $J^P$ may be represented by a generalized Rarita-Schwinger field (or spinor-tensor) $u_{\mu_1\ldots\mu_n}(p_{\Lambda})$, with $n=J-1/2$ indices. These spinor-tensors are symmetric in all of the indices, and satisfy the conditions
\begin{eqnarray}
\slash{p}_{\Lambda}u_{\mu_1\ldots\mu_n}(p_{\Lambda})&=&m_{\Lambda}u_{\mu_1\ldots\mu_n}(p_{\Lambda}),\,\,\,\,\, \gamma^{\mu_1}u_{\mu_1\ldots\mu_n}(p_{\Lambda})=0, \nonumber \\
p_{\Lambda}^{\mu_1}u_{\mu_1\ldots\mu_n}(p_{\Lambda})&=&0,\,\,\,\,\, g^{\mu\nu}u_{\mu\nu\ldots\mu_n}(p_{\Lambda})=0.
\end{eqnarray}
For parity considerations, it is necessary to divide the spinor-tensors into two classes. Those with natural parity are called tensor, while those with unnatural parity are labeled as pseudo-tensor. Here, a state of total angular momentum $J$ is said to have natural parity if $P=(-1)^{J-1/2}$, unnatural parity otherwise. 

For transitions between the ground state and any state with $\jp=1/2^+$, the matrix elements of these currents are
\begin{eqnarray}
\langle \Lambda\mid \overline{s}\gamma^{\mu}b\mid
\Lambda_{b}\rangle&=&\overline{u}(p_{\Lambda},s_{\Lambda})\bigg[F_{1}(q^{2}
)\gamma^{\mu}+F_{2}(q^{2})v^\mu +F_{3}(q^{2})v'^\mu\bigg]u(p_{\Lambda_{b}},s_{\Lambda_{b}}),
\label{eq:f12p} \\
\langle \Lambda\mid \overline{s}\gamma^{\mu}\gamma_{5}b\mid
\Lambda_{b}\rangle&=&\overline{u}(p_{\Lambda},s_{\Lambda})\bigg[G_{1}(q^{2}
)\gamma^{\mu}+G_{2}(q^{2})v^\mu +G_{3}(q^{2})v'^\mu\bigg]\gamma_{5}u(p_{\Lambda_{b}},s_{\Lambda_{b}}),\nn\\
\label{eq:g12p} \\
\langle \Lambda\mid \overline{s}i\sigma^{\mu\nu}b\mid
\Lambda_{b}\rangle&=&\overline{u}(p_{\Lambda},s_{\Lambda}){\cal T}^{\mu\nu}u(p_{\Lambda_{b}},s_{\Lambda_{b}}), 
\label{eq:h12p}
\end{eqnarray}
where we have used $v=\plb/\mlb$, $v'=\pls/\mls$, and
\beqy
{\cal T}^{\mu\nu}&=&H_{1}(q^{2}
)i\sigma^{\mu\nu} +H_{2}(q^{2})(v^\mu\gamma^\nu-v^\nu\gamma^\mu)+H_{3}(q^{2})(v'^\mu\gamma^\nu-v'^\nu\gamma^\mu)
+ \nonumber \\ && H_{4}(q^{2})(v^\mu v'^\nu-v^\nu v'^\mu).
\eeqy
The $F_i$, $G_i$, and $H_i$ are the form factors of interest, and these are functions of the square of the four-momentum transfer $q^2=(\plb-\pls)^2$ between the initial and final baryons. Since
\beq
\sigma^{\mu\nu}\gamma_{5}=\frac{i}{2}\varepsilon^{\mu\nu\alpha\beta}\sigma_{\alpha\beta},
\label{eq:tid}
\eeq
the matrix elements involving the current $\overline{s}i\sigma^{\mu\nu}\gamma_{5}b$ can be related to those involving $\overline{s}i\sigma^{\mu\nu}b$. 

Similarly, the matrix elements for decays to daughter baryons with $J^P=3/2^{-}$ are given by
\begin{eqnarray}
\langle \Lambda\mid \overline{s}\gamma^{\mu}b\mid
\Lambda_{b}\rangle&=&\overline{u}_\alpha (p_{\Lambda},s_{\Lambda})\bigg[v^\alpha\bigg(F_{1}\gamma^{\mu}+F_{2}v^\mu +F_{3}v'^\mu\bigg)+\nn\\&&F_4 g^{\alpha\mu}\bigg]u(p_{\Lambda_{b}},s_{\Lambda_{b}}),\nn\\
\label{eq:f32m} \\
\langle \Lambda\mid \overline{s}\gamma^{\mu}\gamma_{5}b\mid
\Lambda_{b}\rangle&=&\overline{u}_\alpha (p_{\Lambda},s_{\Lambda})\bigg[v^\alpha\bigg(G_{1}\gamma^{\mu}+G_{2}v^\mu +G_{3}v'^\mu\bigg)+\nn\\&&G_4 g^{\alpha\mu}\bigg]\gamma_{5}u(p_{\Lambda_{b}},s_{\Lambda_{b}}),\nn\\
\label{eq:g32m} \\
\langle \Lambda\mid \overline{s}i\sigma^{\mu\nu}b\mid
\Lambda_{b}\rangle&=&\overline{u}_\alpha (p_{\Lambda},s_{\Lambda}){\cal T}^{\alpha\mu\nu}u(p_{\Lambda_{b}},s_{\Lambda_{b}}),
\label{eq:h32m}
\end{eqnarray}
where
\begin{eqnarray}
{\cal T}^{\alpha\mu\nu}&=&v^\alpha\bigg[H_{1}i\sigma^{\mu\nu} +H_{2}(v^\mu\gamma^\nu-v^\nu\gamma^\mu) +H_{3}(v'^\mu\gamma^\nu-v'^\nu\gamma^\mu) 
+\nn\\&&H_{4}(v^\mu v'^\nu-v^\nu v'^\mu)\bigg]+H_5 (g^{\alpha\mu}\gamma^\nu-g^{\alpha\nu}\gamma^\mu)+ H_6 (g^{\alpha\mu}v^\nu-g^{\alpha\nu}v^\mu).
\end{eqnarray}
For decays to $J^P=5/2^{+}$, the matrix elements are
\begin{eqnarray}
\langle \Lambda\mid \overline{s}\gamma^{\mu}b\mid
\Lambda_{b}\rangle&=&\overline{u}_{\alpha\beta} (p_{\Lambda},s_{\Lambda})v^\alpha\bigg[v^\beta\bigg(F_{1}\gamma^{\mu}+F_{2}v^\mu +F_{3}v'^\mu\bigg) +\nn\\&&F_4 g^{\beta\mu}\bigg]u(p_{\Lambda_{b}},s_{\Lambda_{b}}),\nn\\
\label{eq:f52p} \\
\langle \Lambda\mid \overline{s}\gamma^{\mu}\gamma_{5}b\mid
\Lambda_{b}\rangle&=&\overline{u}_{\alpha\beta} (p_{\Lambda},s_{\Lambda})v^\alpha\bigg[v^\beta\bigg(G_{1}\gamma^{\mu}+G_{2}v^\mu +G_{3}v'^\mu\bigg)+\nn\\&&G_4 g^{\beta\mu}\bigg]\gamma_{5}u(p_{\Lambda_{b}},s_{\Lambda_{b}}),\nn\\
\label{eq:g52p} \\
\langle \Lambda\mid \overline{s}i\sigma^{\mu\nu}b\mid
\Lambda_{b}\rangle&=&\overline{u}_{\alpha\beta} (p_{\Lambda},s_{\Lambda}){\cal T}^{\alpha\beta\mu\nu}u(p_{\Lambda_{b}},s_{\Lambda_{b}}),
\label{eq:h52p}
\end{eqnarray}
where
\begin{eqnarray}
{\cal T}^{\alpha\beta\mu\nu}&=&v^\alpha\bigg\{v^\beta\bigg[H_{1}i\sigma^{\mu\nu} +H_{2}\left(v^\mu\gamma^\nu-v^\nu\gamma^\mu\right) +H_{3}(v'^\mu\gamma^\nu-v'^\nu\gamma^\mu)
+\nn\\&&H_{4}(v^\mu v'^\nu-v^\nu v'^\mu)\bigg] +H_5 (g^{\beta\mu}\gamma^\nu-g^{\beta\nu}\gamma^\mu)+ H_6 (g^{\beta\mu}v^\nu-g^{\beta\nu}v^\mu)\bigg\}.
\end{eqnarray}

Thus far, only the matrix elements involving spinors with natural parity, i.e. spinors with parity $(-1)^{J-1/2}$, have been presented. The equations involving spinors with unnatural parity can be found by inserting $\gamma_5$ to the left of the parent baryon spinor in the equations for natural parity.

The matrix elements for the tensor currents can be written in a more convenient form. If Eq. \ref{eq:h12p} is contracted on both sides with the four-momentum transfer $q_\nu$, use of the equations of motion leads to
\begin{equation}
\langle \Lambda\mid \overline{s}i\sigma^{\mu\nu}q_{\nu}b\mid
\Lambda_{b}\rangle=\overline{u}(p_{\Lambda},s_{\Lambda})\bigg[F^{T}_{1}(q^{2}
)\gamma^{\mu}+F^{T}_{2}(q^{2})v^\mu +F^{T}_{3}(q^{2})v'^\mu\bigg]u(p_{\Lambda_{b}},s_{\Lambda_{b}}),
\label{eq:ft12p} \\
\end{equation}
where, 
\begin{eqnarray}
F^{T}_1&=&-\left(\mlb+\mls\right)H_1-(\vq)H_2-(\vpq)H_3, \nn \\
F^{T}_2&=&\mlb H_1+\left(\mlb-\mls\right) H_2+(\vpq) H_4, \nn \\
F^{T}_3&=&\mls H_1+\left(\mlb-\mls\right) H_3-(\vq) H_4,
\end{eqnarray}
valid for states with $\jp=1/2^+$.
For the axial tensor, a similar procedure leads to
\begin{equation}
\langle \Lambda\mid \overline{s}i\sigma^{\mu\nu}\gm_5 q_{\nu}b\mid
\Lambda_{b}\rangle=\overline{u}(\pls,s_{\Lambda})\bigg[G^{T}_{1}(q^{2}
)\gm^{\mu}+G^{T}_{2}(q^{2})v^\mu +G^{T}_{3}(q^{2})v'^\mu\bigg]\gm_5u(\plb,s_{\Lambda_{b}}),
\label{eq:gt12p} \\
\end{equation}
with
\begin{eqnarray}
G^{T}_1&=&\left(\mlb-\mls\right)H_1-\mls(1-\vvp)H_2-\mlb(1-\vvp)H_3, \nn \\
G^{T}_2&=&\mlb H_1-\mls H_2-\mlb H_3, \nn \\
G^{T}_3&=&\mls H_1+\mls H_2+\mlb H_3.
\end{eqnarray}
Similarly, for transitions to states with $\jp=1/2^-$, the matrix elements for the tensor and axial tensor currents become
\beqy
\<\Ls\mid\overline{s}i\sigma^{\mu\nu}q_\nu b\mid\Lb\>\=\overline{u}(\pls,\sls)\bigg[F^{T}_{1}(q^{2}
)\gamma^{\mu}+F^{T}_{2}(q^{2})v^\mu +F^{T}_{3}(q^{2})v'^\mu\bigg]\gm_5 u(\plb,\slb),\nn\\
\label{eq:ft12m} \\
\<\Ls\mid\overline{s}i\sigma^{\mu\nu}\gm_5 q_\nu b\mid\Lb\>\=\overline{u}(\pls,\sls)\bigg[G^{T}_1(q^2)\gm^\mu +G^{T}_2(q^2)v^\mu+G^{T}_3(q^2)v'^\mu\bigg]u(\plb,\slb),\nn\\ 
\label{eq:gt12m}
\eeqy
respectively, where
\beqy
F^{T}_1\=\left(\mlb-\mls\right)H_1-(\vq)H_2-(\vpq)H_3,\nn\\
F^{T}_2\=\mlb H_1-\left(\mlb+\mls\right)H_2+(\vpq)H_4,\nn\\
F^{T}_3\=\mls H_1-\left(\mlb+\mls\right)H_3-(\vq)H_4,
\eeqy
and
\beqy
G^{T}_1\=-\left(\mlb+\mls\right)H_1+\mls(1+\vvp)H_2+\mlb(1+\vvp)H_3,\nn\\
G^{T}_2\=\mlb H_1-\mls H_2-\mlb H_3,\nn\\
G^{T}_3\=\mls H_1-\mls H_2-\mlb H_3.
\eeqy
For transitions to states with $\jp=3/2^-$, we obtain
\beqy
\<\Lambda(\pls)\mid\overline{s}i\sigma^{\mu\nu}q_\nu b\mid\Lambda_b(\plb)\> &=& \overline{u}_\alpha(\pls,\sls)\bigg[v^\alpha\bigg(F^{T}_1\gm^\mu+F^{T}_2v^\mu+F^{T}_3v'^\mu\bigg) +\nn\\&&F^{T}_4g^{\alpha\mu}\bigg]u(\plb,\slb),
\label{eq:ft32m} \\
\<\Lambda(\pls)\mid\overline{s}i\sigma^{\mu\nu}\gm_5q_\nu b\mid\Lambda_b(\plb)\> &=& \overline{u}_\alpha(\pls,\sls)\bigg[v^\alpha\bigg(G^{T}_1\gm^\mu+G^{T}_2v^\mu+G^{T}_3v'^\mu\bigg)+\nn\\&&G^{T}_4g^{\alpha\mu}\bigg]\gm_5u(\plb,\slb),
\label{eq:gt32m}
\eeqy
while for transitions to states with $\jp=5/2^+$, we have
\beqy
\<\Lambda(\pls)\mid\overline{s}i\sigma^{\mu\nu}q_\nu b\mid\Lambda_b(\plb)\> &=& \overline{u}_{\alpha\beta}(\pls,\sls)v^\alpha\bigg[v^\beta\bigg(F^{T}_1\gm^\mu+F^{T}_2v^\mu+F^{T}_3v'^\mu\bigg) +\nn\\&&F^{T}_4g^{\beta\mu}\bigg]u(\plb,\slb),
\label{eq:ft52p} \\
\<\Lambda(\pls)\mid\overline{s}i\sigma^{\mu\nu}\gm_5q_\nu b\mid\Lambda_b(\plb)\> &=& \overline{u}_{\alpha\beta}(\pls,\sls)v^\alpha\bigg[v^\beta\bigg(G^{T}_1\gm^\mu+G^{T}_2v^\mu+G^{T}_3v'^\mu\bigg) +\nn\\&&G^{T}_4g^{\beta\mu}\bigg]\gm_5u(\plb,\slb),
\label{eq:gt52p}
\eeqy
where $F^{T}_i$ and $G^{T}_i$ are given by
\beqy
F^{T}_1&=&-\left(\mlb+\mls\right) H_1-(\vq) H_2-(\vpq) H_3-\mlb H_5, \nn \\
F^{T}_2&=&\mlb H_1+\left(\mlb-\mls\right) H_2+(\vpq) H_4-\mlb H_6, \nn \\
F^{T}_3&=&\mls H_1+\left(\mlb-\mls\right) H_3-(\vq) H_4, \nn \\
F^{T}_4&=&\left(\mlb-\mls\right) H_5+(\vq) H_6, \nn \\
G^{T}_1&=&\left(\mlb-\mls\right) H_1-\mls(1-\vvp)H_2-\mlb(1-\vvp)H_3+\mlb H_5+\mls H_6, \nn \\
G^{T}_2&=&\mlb H_1-\mls H_2-\mlb H_3, \nn \\
G^{T}_3&=&\mls H_1+\mls H_2+\mlb H_3-\mls H_6, \nn \\
G^{T}_4&=&\left(\mlb+\mls\right) H_5+\mls(1+\vvp) H_6.
\eeqy
For $\jp=3/2^+$, a state with unnatural parity, we have
\beqy
\<\Lambda(\pls)\mid\overline{s}i\sigma^{\mu\nu}q_\nu b\mid\Lambda_b(\plb)\> &=& \overline{u}_\alpha(\pls,\sls)\bigg[v^\alpha\bigg(F^{T}_1\gm^\mu+F^{T}_2v^\mu+F^{T}_3v'^\mu\bigg) +\nn\\&&F^{T}_4g^{\alpha\mu}\bigg]\gm_5u(\plb,\slb),
\label{eq:ft32p} \\
\<\Lambda(\pls)\mid\overline{s}i\sigma^{\mu\nu}\gm_5q_\nu b\mid\Lambda_b(\plb)\> &=& \overline{u}_\alpha(\pls,\sls)\bigg[v^\alpha\bigg(G^{T}_1\gm^\mu+G^{T}_2v^\mu+G^{T}_3v'^\mu\bigg) +\nn\\&&G^{T}_4g^{\alpha\mu}\bigg]u(\plb,\slb),
\label{eq:gt32p}
\eeqy
with
\beqy
F^{T}_1&=&\left(\mlb-\mls\right) H_1-(\vq) H_2-(\vpq) H_3-\mlb H_5, \nn \\
F^{T}_2&=&\mlb H_1-\left(\mlb+\mls\right) H_2+(\vpq) H_4-\mlb H_6, \nn \\
F^{T}_3&=&\mls H_1-\left(\mlb+\mls\right) H_3-(\vq) H_4, \nn \\
F^{T}_4&=&-\left(\mlb+\mls\right) H_5+(\vq) H_6, \nn \\
G^{T}_1&=&-\left(\mlb+\mls\right) H_1+\mls(1+\vvp)H_2+\mlb(1+\vvp)H_3+\mlb H_5+\mls H_6, \nn \\
G^{T}_2&=&\mlb H_1-\mls H_2-\mlb H_3, \nn \\
G^{T}_3&=&\mls H_1-\mls H_2-\mlb H_3-\mls H_6, \nn \\
G^{T}_4&=&-\left(\mlb-\mls\right) H_5-\mls(1-\vvp) H_6.
\eeqy

We can now use these redefined tensor form factors to write Eqs. \ref{eq:h1mu} and \ref{eq:h2mu} in a simplified form.


For transitions to states with $J=1/2$, these two equations become
\beqy
H_1^\mu&=&\overline{u}(\pls,\sls)\bigg[\gm^\mu\bigg(A_1+B_1\gm_5\bigg)+v^\mu\bigg(A_2+B_2\gm_5\bigg)+v'^\mu\bigg(A_3+B_3\gm_5\bigg)\bigg]u(\plb,\slb),\nn\\ \\
H_2^\mu&=&\overline{u}(\pls,\sls)\bigg[\gm^\mu\bigg(D_1+E_1\gm_5\bigg)+v^\mu\bigg(D_2+E_2\gm_5\bigg)+v'^\mu\bigg(D_3+E_3\gm_5\bigg)\bigg]u(\plb,\slb).\nn\\
\eeqy
For transitions to states with $J=3/2$, 
\beqy
H_1^\mu&=&\overline{u}_\alpha(\pls,\sls)\bigg[v^\alpha\bigg(\gm^\mu\bigg(A_1+B_1\gm_5\bigg)+v^\mu\bigg(A_2+B_2\gm_5\bigg)+v'^\mu\bigg(A_3+B_3\gm_5\bigg)\bigg)\nn\\&& +g^{\alpha\mu}\bigg(A_4+B_4\gm_5\bigg)\bigg]u(\plb,\slb), \\
H_2^\mu&=&\overline{u}_\alpha(\pls,\sls)\bigg[v^\alpha\bigg(\gm^\mu\bigg(D_1+E_1\gm_5\bigg)+v^\mu\bigg(D_2+E_2\gm_5\bigg)+v'^\mu\bigg(D_3+E_3\gm_5\bigg)\bigg)\nn\\&& +g^{\alpha\mu}\bigg(D_4+E_4\gm_5\bigg)\bigg]u(\plb,\slb),
\eeqy
while for those with $J=5/2$,
\beqy
H_1^\mu&=&\overline{u}_{\alpha\beta}(\pls,\sls)v^\alpha\bigg[v^\beta\bigg(\gm^\mu\bigg(A_1+B_1\gm_5\bigg)+v^\mu\bigg(A_2+B_2\gm_5\bigg)+v'^\mu\bigg(A_3+B_3\gm_5\bigg)\bigg)\nn\\&& +g^{\beta\mu}\bigg(A_4+B_4\gm_5\bigg)\bigg]u(\plb,\slb), \\
H_2^\mu&=&\overline{u}_{\alpha\beta}(\pls,\sls)v^\alpha\bigg[v^\beta\bigg(\gm^\mu\bigg(D_1+E_1\gm_5\bigg)+v^\mu\bigg(D_2+E_2\gm_5\bigg)+v'^\mu\bigg(D_3+E_3\gm_5\bigg)\bigg)\nn\\&& +g^{\beta\mu}\bigg(D_4+E_4\gm_5\bigg)\bigg]u(\plb,\slb).
\eeqy
For states with natural parity spinors,
\beqy
A_i&=&-\frac{2m_b}{q^2}C_7F^{T}_i+C_9F_i,\nn\\
B_i&=&-\frac{2m_b}{q^2}C_7G^{T}_i-C_9G_i,\nn\\
D_i&=&C_{10}F_i, \,\,\,\,\ E_i=-C_{10}G_i,
\label{eq:natpar_dilep}
\eeqy
and for states with unnatural parity,
\beqy
A_i&=&-\frac{2m_b}{q^2}C_7G^{T}_i-C_9G_i,\nn\\
B_i&=&-\frac{2m_b}{q^2}C_7F^{T}_i+C_9F_i,\nn\\
D_i&=&-C_{10}G_i, \,\,\,\,\ E_i=C_{10}F_i.
\label{eq:unnatpar_dilep}
\eeqy

\subsection{Decays Rates and Forward-Backward Asymmetries\label{sec:drfba}}

The decay rate is given by
\begin{equation}
d\Gamma=\frac{1}{2m_{\Lambda_b}}\left(\prod_{f}{\frac{d^{3}p_{f}}{(2\pi)^{3}}\frac{1}{2E_{f}}}\right)(2\pi)^{4}\delta^{(4)}\left(p_{\Lambda_b}-\sum_{f}{p_{f}}\right)|\overline{\cal M}|^2.
\end{equation}
For the case of unpolarized baryons, $|\overline{\cal M}|^2$ is the squared amplitude averaged over the initial polarization and summed over the final polarizations. 
\begin{figure}
\centerline{\includegraphics[width=3.0in]{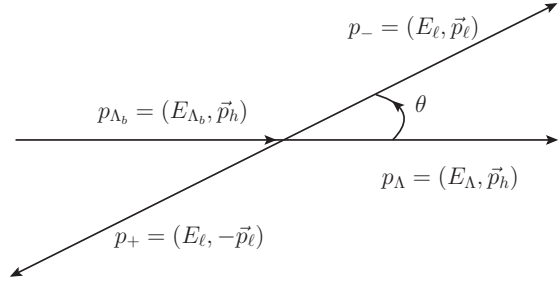}}
\caption{Kinematics of the dilepton rest frame.}\label{fig:dilep}
\end{figure}

For dileptonic decays, the squared average amplitude is
\beq
|\overline{\cal M}|^2= \frac{G_F^2\alpha_{em}^2}{2^4\pi^2}\left|V_{tb}V_{ts}^{*}\right|^2\left(H_a^{\mu\nu}L_{a\mu\nu}+H_b^{\mu\nu}L_{b\mu\nu}+H_c^{\mu\nu}L_{c\mu\nu}+H_d^{\mu\nu}L_{d\mu\nu}\right).
\eeq
For unpolarized leptons, the leptonic tensors are
\begin{eqnarray}
L_{a\mu\nu}&=&\sum_{spin} L_\mu^{(V)\dag} L_\nu^{(V)}=4\left[p_{+\mu}p_{-\nu}+p_{+\nu}p_{-\mu}-(p_{-}\cdot p_{+}+m_\ell^2)g_{\mu\nu}\right], \\
L_{b\mu\nu}&=&\sum_{spin} L_\mu^{(A)\dag} L_\nu^{(A)}=4\left[p_{+\mu}p_{-\nu}+p_{+\nu}p_{-\mu}-(p_{-}\cdot p_{+}-m_\ell^2)g_{\mu\nu}\right], \\
L_{c\mu\nu}&=&\sum_{spin} L_\mu^{(V)\dag} L_\nu^{(A)}=4i\varepsilon_{\mu\nu\alpha\beta} p_{-}^{\alpha} p_{+}^{\beta}, \\
L_{d\mu\nu}&=&\sum_{spin} L_\mu^{(A)\dag} L_\nu^{(V)}=4i\varepsilon_{\mu\nu\alpha\beta} p_{-}^{\alpha} p_{+}^{\beta}.
\end{eqnarray}
The hadronic tensors $H_f^{\mu\nu}$ are 
\begin{eqnarray}
H_a^{\mu\nu}&=&\sum_{pol} H_1^{\mu\dag} H_1^{\nu},\,\,\,\,\,
H_b^{\mu\nu}=\sum_{pol} H_2^{\mu\dag} H_2^{\nu}, \\
H_c^{\mu\nu}&=&\sum_{pol} H_1^{\mu\dag} H_2^{\nu},\,\,\,\,\,
H_d^{\mu\nu}=\sum_{pol} H_2^{\mu\dag} H_1^{\nu}.
\end{eqnarray}
The most general Lorentz structure for each of these hadronic tensors is 
\beq
H^{\mu\nu}_f=-\alpha^f g^{\mu\nu}+\beta_{++}^f Q^\mu Q^\nu+\beta_{+-}^f Q^\mu q^\nu+\beta_{-+}^f q^\mu Q^\nu+\beta_{--}^f q^\mu q^\nu+i\gamma^f\varepsilon^{\mu\nu\alpha\beta}Q_\alpha q_\beta,
\label{eq:hmunu}
\eeq
where $Q\equiv\plb+\pls$ and $q=\plb-\pls$ is the 4-momentum transfer.

We carry out our calculations in the dilepton rest frame (see Fig. \ref{fig:dilep}). In this frame
\beqy
E_{\Lb}&=&\frac{\mlb}{2\sqrt{\sh}}(1-r+\sh),\,\,\,\, E_{\Ls}=\frac{\mlb}{2\sqrt{\sh}}(1-r-\sh),\nn\\
p_h&=&\frac{\mlb}{2}\sqrt{\frac{\phi(\sh)}{\sh}},\,\,\,\,E_\ell=\frac{\mlb}{2}\sqrt{\sh},\,\,\,\,p_\ell=\frac{\mlb}{2}\sqrt{\sh\psi(\sh)}.\nn
\eeqy
$p_h$ is the 3-momentum of either baryon in this frame. In addition, we have defined
\beqy
\sh&\equiv&q^2/\mlb^2, \,\,\,\, r\equiv m_\Lambda^2/\mlb^2, \,\,\,\, \hat m_\ell\equiv m_\ell/m_{\Lambda_b}, \nonumber\\
\phi(\sh)&=&(1-r)^2-2(1+r)\sh+\sh^2,\,\,\,\,\psi(\sh)=1-4 \hat m_\ell^2/\sh.\nn
\eeqy
Performing the contractions, the differential decay rate becomes
\begin{equation}
\frac{d^2\Gamma}{d\sh d\zh}= \frac{m_{\Lambda_b}G_F^2\alpha_{em}^2}{2^{13}\pi^5}\left|V_{tb}V_{ts}^{*}\right|^2 \sqrt{\phi(\sh)\psi(\sh)} {\cal F}_0 (\sh,\zh),
\label{eq:d2gam0}
\end{equation}
where $\zh=\cos\theta$. In these decays, $4\hat m_\ell^2\leq\sh\leq (1-\sqrt{r})^2$ and $-1\leq\zh\leq 1$. The normalized rate ${\cal F}_0 (\sh,\zh)$ has the form
\begin{equation}
{\cal F}_0 (\sh,\zh)={\cal I}_0 (\sh)+\zh {\cal I}_1 (\sh)+\zh^2 {\cal I}_2 (\sh),
\label{eq:r0}
\end{equation}
where
\begin{equation}
{\cal I}_0(\sh)=\alpha^a A_\alpha+\beta_{++}^a A_{++}+\alpha^b B_\alpha+\beta_{++}^b B_{++}+\beta_{+-}^b B_{+-}+\beta_{-+}^b B_{-+}+\beta_{--}^b B_{--},
\label{eq:i0}
\end{equation}
and
\begin{eqnarray}
A_\alpha&=&4m_{\Lambda_b}^2 (2\hat m_\ell^2+\sh), \,\,\,\,\
A_{++}=2m_{\Lambda_b}^4 \phi(\sh), \nonumber \\
B_\alpha&=&4m_{\Lambda_b}^2 (\sh-6\hat m_\ell), \,\,\,\,\
B_{++}=2m_{\Lambda_b}^4 \left(\phi(\sh)+4\hat m_\ell^2 \left(2(1+r)-\sh\right)\right), \nonumber \\
B_{+-}&=&B_{-+}=8m_{\Lambda_b}^4 \hat m_\ell^2 (1-r), \,\,\,\,\
B_{--}=8m_{\Lambda_b}^4 \hat m_\ell^2 \sh.
\end{eqnarray}
The terms proportional to $\zh$ and $\zh^2$ are
\begin{eqnarray}
{\cal I}_1(\sh)&=&4\mlb^4\sh\sqrt{\phi(\sh)\psi(\sh)}(\gm^c+\gm^d),
\label{eq:i1} \\
{\cal I}_2(\sh)&=&-2\mlb^4\phi(\sh)\psi(\sh)(\beta_{++}^a +\beta_{++}^b),
\label{eq:i2}
\end{eqnarray}
respectively. Integrating out the $\zh$ dependence from Eq. \ref{eq:d2gam0}, the decay rate becomes
\begin{equation}
\frac{d\Gamma}{d\sh}=\frac{m_{\Lambda_b}G_F^2\alpha_{em}^2}{2^{12}\pi^5}\left|V_{tb}V_{ts}^{*}\right|^2 \sqrt{\phi(\sh)\psi(\sh)}{\cal R}_0(s),
\label{eq:dgam}
\end{equation}
where
\beq
{\cal R}_0(\sh)={\cal I}_0(\sh)+\frac{1}{3}{\cal I}_2(\sh).
\label{eq:r0s}
\eeq
The functions ${\cal I}_0$ and ${\cal I}_2$ are given in Eqs. \ref{eq:i0} and \ref{eq:i2}, respectively. The explicit forms of the coefficients $\alpha$, $\beta_{\pm\pm}$, and $\gamma$ are given in \ref{sec:hadten}.

The forward-backward asymmetry (FBA) is defined as
\begin{equation}
{\cal A}_{FB}(\sh)=\frac{1}{d\Gamma/d\sh}\left[\int_{0}^{1} d\zh\frac{d^2\Gamma}{d\sh d\zh}-\int_{-1}^{0} d\zh\frac{d^2\Gamma}{d\sh d\zh}\right].
\label{eq:afb1}
\end{equation}
Using Eqs. \ref{eq:d2gam0} and \ref{eq:r0}, we obtain
\begin{equation}
{\cal A}_{FB}(\sh)=\frac{{\cal I}_1(\sh)}{2\left[{\cal I}_0(\sh)+\frac{1}{3}{\cal I}_2(\sh)\right]},
\label{eq:afb2}
\end{equation}
where ${\cal I}_1$ is given in Eq. \ref{eq:i1}. Since this observable is written as a ratio, it might be expected to be less dependent on the form factors than the differential decay rate. HQET considerations may make it even less model-dependent \cite{chen1}, so that it may provide a (somewhat) model-independent way of extracting the Wilson coefficients. Since ${\cal A}_{FB}$ also depends on the chirality of the hadronic and leptonic currents, this observable is also sensitive to any new physics effects beyond the Standard Model.

\section{HQET\label{sec:hqet}}

Heavy quark effective theory (HQET) is an effective tool for examining the phenomenology of hadrons containing a single heavy quark. It has been applied to hadronic matrix elements in many processes at higher and higher order in the $1/m_Q$ expansion, with $m_Q$ being the mass of the heavy quark. In this section we present the leading-order relationships among the HQET form factors and those presented in Section \ref{sec:bme} for the heavy to light transitions $\Lambda_b\rightarrow\Lambda^{(*)}$. The properties of the spinor-tensors that are used to represent the daughter baryons have been discussed near the start of section \ref{sec:bme}.

To leading order in HQET, transitions between a heavy baryon and a light one are described in terms of only two form factors. For any current operator $\Gamma$, the hadronic matrix elements involving tensor states can be written as \cite{mrr}
\begin{equation}
 \langle \Lambda^{(*)}(p_{\Lambda^{(*)}})\mid \overline{s}\Gamma b\mid
\Lambda_{b}(v)\rangle=\overline{u}_{\mu_1\ldots\mu_n}(p_{\Lambda^{(*)}})M^{\mu_1\ldots\mu_n}\Gamma
u(v) \label{eq:hqetme}
\end{equation}
where $v$ is the velocity of the $\Lb$ baryon and $M^{\mu_1\ldots\mu_n}$ is the most general tensor consistent with HQET that can be constructed from the available kinematic variables. We may not use any factors of $\gamma^{\mu_i}$, $p_{\Lambda^{(*)}}^{\mu_i}$, or $g^{\mu_i\mu_j}$ in constructing $M^{\mu_1\ldots\mu_n}$; therefore, it must take the form
\begin{equation}
M^{\mu_1\ldots\mu_n}=v^{\mu_1}\ldots v^{\mu_n}\bigg[\xi^{(n)}_{1}(v\cdot p_{\Lambda^{(*)}})+\slash{v}\xi^{(n)}_{2}(v\cdot p_{\Lambda^{(*)}})\bigg],
\end{equation}
For the case of a pseudo-tensor daughter baryon, the matrix element has the same form as Eq. \ref{eq:hqetme} except that $M^{\mu_1\ldots\mu_n}$ is now a pseudo-tensor. A pseudo-tensor when sandwiched between spinors can be constructed by multiplying an ordinary tensor by $\gamma_5$; thus, for transitions involving a pseudo-tensor spinor,
\begin{equation}
M^{\mu_1\ldots\mu_n}=v^{\mu_1}\ldots v^{\mu_n}\bigg[\zeta^{(n)}_{1}(v\cdot p_{\Lambda^{(*)}})+\slash{v}\zeta^{(n)}_{2}(v\cdot p_{\Lambda^{(*)}})\bigg]\gamma_5.
\end{equation}
Using these forms of the matrix elements, the form factors defined in Section \ref{sec:bme} are found to satisfy the following  relations in the limit when the $b$ quark is infinitely heavy.

For any state with $J^P=1/2^{+}$,
\begin{eqnarray}
F_3&=&G_3=H_3=H_4=0, \,\,\,\, F_2=G_2=-H_2=2\xi^{(0)}_2, \nonumber \\ F_1&=&\xi^{(0)}_1-\xi^{(0)}_2, \,\,\,\, G_1=H_1=\xi^{(0)}_1+\xi^{(0)}_2,
\label{eq:hqet12p}
\end{eqnarray}
while for $J^P=1/2^{-}$, we have
\begin{eqnarray}
F_3&=&G_3=H_3=H_4=0, \,\,\,\, F_2=G_2=H_2=-2\zeta^{(0)}_2, \nonumber \\ F_1&=&-\left[\zeta^{(0)}_1+\zeta^{(0)}_2\right], \,\,\,\, G_1=-H_1=-\left[\zeta^{(0)}_1-\zeta^{(0)}_2\right].
\label{eq:hqet12m}
\end{eqnarray}
For $J^P=3/2^{-}$,
\begin{eqnarray}
F_3&=&G_3=H_3=F_4=G_4=H_4=H_5=H_6=0, \,\,\,\, F_2=G_2=-H_2=2\xi^{(1)}_2, \nonumber \\ F_1&=&\xi^{(1)}_1-\xi^{(1)}_2, \,\,\,\, G_1=H_1=\xi^{(1)}_1+\xi^{(1)}_2.
\label{eq:hqet32m}
\end{eqnarray}
For $J^P=3/2^{+}$,
\begin{eqnarray}
F_3&=&G_3=H_3=F_4=G_4=H_4=H_5=H_6=0, \,\,\,\, F_2=G_2=H_2=-2\zeta^{(1)}_2, \nonumber \\ F_1&=&-\left[\zeta^{(1)}_1+\zeta^{(1)}_2\right], \,\,\,\, G_1=-H_1=-\left[\zeta^{(1)}_1-\zeta^{(1)}_2\right].
\label{eq:hqet32p}
\end{eqnarray}
For $J^P=5/2^{+}$,
\begin{eqnarray}
F_3&=&G_3=H_3=F_4=G_4=H_4=H_5=H_6=0, \,\,\,\, F_2=G_2=-H_2=2\xi^{(2)}_2, \nonumber \\ F_1&=&\xi^{(2)}_1-\xi^{(2)}_2, \,\,\,\, G_1=H_1=\xi^{(2)}_1+\xi^{(2)}_2.
\label{eq:hqet52p}
\end{eqnarray}
Again, we stress that the above relationships are for heavy to light transitions only, and are valid in the limit of an infinitely heavy $b$ quark.

\section{The Model\label{sec:model}}

\subsection{Quark Model\label{sec:qmod}}

In the models considered here, a baryon state takes the form
\begin{eqnarray}
\mid B_{q}(\vec p_{B_{q}},s)\rangle&=&3^{-3/4}\sqrt{2E_{B_q}}\int d^{3}p_{\rho}d^{3}p_{\lambda}C^{A}\sum_{s_1,s_2,s_q}\chi_{s_1,s_2,s_q}^{J s}(\vec p_\rho,\vec p_\lambda)\nn\\&&\times\mid\left[q_1(\vec p_1,s_1)q_2(\vec p_2,s_2)q(\vec p,s_q)\right]_{B_{q}}\rangle,
\label{eq:bwf}
\end{eqnarray}
where $\vec p_\rho=\frac{1}{\sqrt{2}}(\vec p_1-\vec p_2)$, $\vec p_\lambda=\frac{1}{\sqrt{6}}(\vec p_1+\vec p_2-2\vec p)$ are the Jacobi momenta, $C^{A}$ is the antisymmetric color wave function, $J$ is the total angular momentum, $E_{B_q}$ is the energy of the baryon $B_q$, and the notation $\left[q_1 q_2 q\right]_{B_q}$ denotes the flavor wave function of the baryon $B_q$. For example, the flavor wave function for $\Ls_q$ has the form
\begin{equation}
\left[udq\right]_{\Lambda_q}=\frac{1}{\sqrt{2}}(ud-du)q,
\end{equation}
which is antisymmetric in the quarks $u$ and $d$. When coupled with the appropriate Clebsch-Gordan coefficients, the ket $\mid\left[\ldots\right]_{B_q}\>$ becomes the flavor-spin wave function for the baryon $B_q$. $\chi$ is the momentum space wave function, and is given by
\begin{equation}
\chi_{M_1 M_2 M_3}^{J M_J}(\vec p_\rho,\vec p_\lambda) =\sum_i \eta_i^{B_{q}} C_{\frac{1}{2}M_1\frac{1}{2}M_2}^{S^D_i M_D} C_{S^D_i M_D\frac{1}{2}M_3}^{S_i M_S} C_{L_i M_L S_i M_S}^{J M_J} \psi_{L_i M_L (n_\rho l_\rho n_\lambda l_\lambda)_i}(\vec p_\rho,\vec p_\lambda),
\label{eq:mswf1}
\end{equation}
where $S^D_i$ is the total spin of the pair of spectator quarks, $M_D$ is its projection, $S_i$ is the total spin of the quarks in the baryon, and $M_S$ is its projection.
In addition,
\begin{equation}
\psi_{L M_L n_\rho l_\rho n_\lambda l_\lambda}(\vec p_\rho,\vec p_\lambda)=\sum_{m_\rho m_\lambda} C_{l_\rho m_\rho l_\lambda m_\lambda}^{L M_L} \phi_{n_\rho l_\rho m_\rho}(\alpha_\rho;\vec p_\rho) \phi_{n_\lambda l_\lambda m_\lambda}(\alpha_\lambda;\vec p_\lambda).
\label{eq:mswf2}
\end{equation}
The $\eta_i^{B_q}$ are expansion coefficients that are determined by a diagonalization of a quark model Hamiltonian and the $C_{j_1 m_1 j_2 m_2}^{j m}$ are Clebsch-Gordan coefficients. The combination of momentum, spin and flavor wave functions is symmetric. The functions $\phi_{n l m}(\alpha;\vec p)$ are the basis functions. The basis used in this work is the harmonic oscillator basis,
\begin{equation}
\phi_{n l m}(\alpha;\vec p)=\exp\left(-p^2/2\alpha^2\right) {\cal L}_{n}^{l+1/2}\left(p^2/\alpha^2\right) {\cal Y}_{l m}(\vec p),
\label{eq:exbas}
\end{equation}
where $\alpha$ is the length parameter, which is determined by a minimization of the quark model Hamiltonian,
\beq
\L_n^{l+1/2}\left(p^2/\alpha^2\right)=N_{nl}(\alpha)L_n^{l+1/2}\left(p^2/\alpha^2\right),\,\,\,\, N_{nl}(\alpha)=\frac{(i)^l(-1)^n}{\alpha^{l+3/2}}\left[\frac{2\Gamma(n+1)}{\Gamma(n+l+3/2)}\right]^{1/2},
\eeq
$L_n^k$ are the associated Laguerre polynomials and ${\cal Y}_{l m}$ are the solid harmonics.

The states described by Eqs. \ref{eq:bwf}-\ref{eq:exbas} are obtained from a variational diagonalization of a quark model Hamiltonian. The Hamiltonian used in this model takes the form \cite{roper}
\begin{equation}
H=\sum_{i}{K_{i}}+\sum_{i<j}{\left(\hat V^{c}_{ij}+\hat V^{hyp}_{ij}\right)}+\hat V^{SO}+C_{qqq}.
\label{eq:qmham}
\end{equation}
$K_i$ is the nonrelativistic kinetic energy of the $i$th quark,
\begin{equation}
K_{i}=m_{i}+\frac{\vec p_{i}^{2}}{2m_{i}}.
\end{equation}
The spin-independent part of the potential has the linear-plus-Coulomb terms,
\begin{equation}
V^{c}_{ij}=\frac{1}{2}br_{ij}-\frac{2}{3}\frac{\alpha_{coul}}{r_{ij}},
\end{equation}
where $r_{ij}=|\vec r_i-\vec r_j|$. The spin-dependent parts describe the hyperfine and spin-orbit interactions. The hyperfine piece, which contains contact and tensor terms, is written as
\begin{eqnarray}
V^{hyp}_{ij}&=&\frac{2}{3}\frac{1}{m_{i}m_{j}}\bigg\{\frac{8\pi}{3}\alpha_{con}
\vec S_{i}\cdot\vec S_{j}\delta^{(3)}(\vec r_{ij})+\frac{\alpha_{tens}}{r_{ij}^{3}}\left[\frac{3(\vec
S_{i}\cdot\vec r_{ij})(\vec S_{j}\cdot\vec r_{ij})}{r_{ij}^{2}}-\vec
S_{i}\cdot\vec S_{j}\right]\bigg\},\nn\\
\end{eqnarray}
and the {\it ad-hoc} spin-orbit potential is chosen to be
\begin{equation}
V^{SO}=\frac{\alpha_{SO}}{\widetilde{m}_{B_q}^2(\rho^2+\lambda^2)}\vec L\cdot\vec S.
\end{equation}
Here $\widetilde{m}_{B_q}$ is the sum of the masses of the three quarks that make up the baryon $B_q$, $\vec\rho$ and $\vec\lambda$ are Jacobi coordinates conjugate to $\vec{p}_\rho$ and $\vec{p}_\lambda$, respectively, $\vec L$ is the total orbital angular momentum, and $\vec S$ is the total spin of the baryon. The parameters $C_{qqq}$, $b$, $\alpha_{coul}$, $\alpha_{con}$, $\alpha_{tens}$, $\alpha_{SO}$, and $m_i$ are obtained from a fit to the experimental spectrum of baryon states.

\subsection{Form Factors\label{sec:ff}}

In order to extract the form factors, we need to calculate matrix elements that take the form
\beqy
\langle \Ls(\plsvec,s^\prime)|\bar{s}\Gamma b|\Lb(0,s)\rangle&=&3^{-3/2}\sqrt{2E_{\Ls}}\sqrt{2\mlb}\int d^3\pr' d^3\pl' d^3\pr d^3\pl  C^{A*} C^A \nn\\ &&\times\sum_{s'_1 s'_2 s_{q'}}\sum_{s_1 s_2 s_q} \chi_{s'_1 s'_2 s_{q'}}^{J's'*}(\prvec',\plvec') \nn\\ &&\times
\< q'_1 q'_2 s\mid\overline{s}\Gamma b\mid q_1 q_2 b\> \chi_{s_1 s_2 s_q}^{Js}(\prvec,\plvec)
\label{eq:me}
\eeqy
where 
\beq
\< q'_1 q'_2 s\mid\overline{s} \Gamma b\mid q_1 q_2 b\> =\<q'_1 q'_2\mid q_1 q_2\> \< s\mid\overline{s}\Gamma b\mid b\>.\nn
\eeq
The matrix element $\<q'_1 q'_2\mid q_1 q_2\>$ gives $\delta$-functions in spin, momentum and flavor in the spectator approximation.

We calculate the matrix elements of Eq. \ref{eq:me} using two approximations. In the first approximation, we use the method outlined by Pervin {\it et al.} in \cite{pervin,pervin1}. The form factors are extracted by performing a nonrelativistic reduction of the quark spinors, keeping terms up to ${\cal O}(1/m_q)$. The hadronic matrix elements are then calculated analytically by using single component wave functions. The matrix elements calculated using the quark model wave functions are then used to extract the form factors defined in Section \ref{sec:bme}. Due to the choice of basis, the resulting form factors have the general form of a polynomial times a Gaussian. Once the form factors have been extracted, the polynomials are truncated to the zeroth order in the daughter baryon momentum $\pls$. In fact, although the analytic expressions for the form factors shown in \cite{pervin, pervin1} were shown for single-component wave functions, the results for the decay rates shown in those articles were calculated using all components of the wave functions.

In the second method we use the full multi-component wave functions found from the diagonalization of Eq. \ref{eq:qmham}. We also keep the full relativistic form of the quark spinors. Although giving a full analytic treatment of the hadronic matrix elements becomes challenging in this case, much of the calculation can still be done analytically and only a couple of the integrations need to be performed numerically. The advantage of this approach is that it avoids the truncation of the quark currents which may be justified for the $b$ quark but which is much less justifiable for the $s$ quark. We will compare the results from the two models with the expectations of HQET.

We work in the $\Lb$ rest frame where the initial quark momenta can be written in terms of the Jacobi momenta as
\beqy
\pvec_1=\frac{1}{\sqrt{2}}\prvec+\frac{1}{\sqrt{6}}\plvec,\,\,\,\, \pvec_2=-\frac{1}{\sqrt{2}}\prvec+\frac{1}{\sqrt{6}}\plvec,\,\,\,\, \pvec=-\sqrt{\frac{2}{3}}\plvec.
\eeqy
Implementing the spectator approximation and integrating over the final Jacobi momenta together lead to
\beqy
\langle \Lambda\mid\overline{s}\Gamma b\mid\Lambda_b\rangle&=&\sum_{h',h} a_{h'}^{\Lambda^*} a_{h}^{\Lambda_b} \delta_{s_{1}'s_{1}}\delta_{s_{2}'s_{2}} (-1)^{l_{\lambda'}+l_{\lambda}} \nn\\&&\times U_{n_\rho l_\rho m_\rho}^{n_{\rho'} l_{\rho'} m_{\rho'}}(\alpha_\rho,\alpha_{\rho'})W_{\Gamma;n_\lambda l_\lambda m_\lambda s_q}^{n_{\lambda'} l_{\lambda'} m_{\lambda'} s_{q'}}(\alpha_\lambda,\alpha_{\lambda'}),
\label{eq:qmme}
\eeqy
where the coefficients $a_{h(h')}$ are the products of the normalization of the baryon state (the $\sqrt{2E_{B_q}}$ of Eq. \ref{eq:bwf}), the expansion coefficients (the $\eta_i^{B_{q}}$ of Eq. \ref{eq:mswf1}), and the various Clebsch-Gordan coefficients that appear in the parent (daughter) baryon wave function, and the indices $h(h')$ contain all the relevant quantum numbers being summed over for the parent (daughter) baryon state. $U_{n_\rho l_\rho m_\rho}^{n_{\rho'} l_{\rho'} m_{\rho'}}$ is the spectator overlap,
\begin{equation}
U_{n_\rho l_\rho m_\rho}^{n_{\rho'} l_{\rho'} m_{\rho'}}(\alpha_\rho,\alpha_{\rho'})=\int d^{3} p_\rho \phi_{n_{\rho'} l_{\rho'} m_{\rho'}}^{*}(\alpha_{\rho'};\vec p_{\rho}) \phi_{n_\rho l_\rho m_\rho}(\alpha_\rho;\vec p_\rho).
\label{eq:uint}
\end{equation}
This integral can be done analytically and is given in \ref{sec:shme}. $W_{\Gamma;n_\lambda l_\lambda m_\lambda s_q}^{n_{\lambda'} l_{\lambda'} m_{\lambda'} s_{q'}}$ is the interaction overlap,
\begin{eqnarray}
W_{\Gamma;n_\lambda l_\lambda m_\lambda s_q}^{n_{\lambda'} l_{\lambda'} m_{\lambda'} s_{q'}}(\alpha_\lambda,\alpha_{\lambda'})&=&\int d^{3} p \phi_{n_{\lambda'} l_{\lambda'} m_{\lambda'}}^{*}(\beta';\frac{2m_q}{\widetilde m_{\Lambda}}\vec p_\Lambda+\vec p)\nn\\&&\times \langle s(\vec p_\Lambda+\vec p,s_{q'})\mid\overline{s}\Gamma b\mid b(\vec p,s_q)\rangle \phi_{n_\lambda l_\lambda m_\lambda}(\beta;\vec p),
\label{eq:wint}
\end{eqnarray}
where $\beta^{(\prime)}=\sqrt{2/3}\alpha_\lambda^{(\prime)}$ is the reduced length parameter for the parent (daughter) baryon, $\widetilde m_\Lambda=m_s+2m_q$, $m_s$ is the mass of the strange quark and $m_q$ is the mass of each light quark. Using the basis functions in Eq. \ref{eq:exbas}, we obtain
\begin{eqnarray*}
W_{\Gamma;n_\lambda l_\lambda m_\lambda s_q}^{n_{\lambda'} l_{\lambda'} m_{\lambda'} s_{q'}}(\alpha_\lambda,\alpha_{\lambda'})&=&\int d^{3} p \exp\left(-\frac{p'^2}{2\beta'^2}-\frac{p^2}{2\beta^2}\right) {\cal L}_{n_{\lambda'}}^{l_{\lambda'}+\frac{1}{2}*}\left(\frac{p'^2}{\beta'^2}\right){\cal Y}_{l_{\lambda'} m_{\lambda'}}^{*}(\vec p') \\ &&  \times \langle s(\vec p_\Lambda+\vec p,s_{q'})\mid\overline{s}\Gamma b\mid b(\vec p,s_q)\rangle{\cal L}_{n_{\lambda}}^{l_{\lambda}+\frac{1}{2}}\left(\frac{p^2}{\beta^2}\right) {\cal Y}_{l_{\lambda} m_{\lambda}}(\vec p) ,
\end{eqnarray*}
where $\vec p'=(2m_q/\widetilde m_{\Lambda})\vec p_\Lambda+\vec p$. The angular dependence in the exponential is eliminated by making the substitutions
\begin{equation}
\vec p=\vec k+c\vec p_\Lambda,\,\,\,\, \vec p'=\vec k+c'\vec p_\Lambda,
\label{eq:pkswitch}
\end{equation}
where
\begin{equation}
c=-\frac{m_q \alpha_{\lambda}^2}{\widetilde m_{\Lambda} \alpha_{\lambda\lambda'}^2},\,\,\,\, c'=\frac{m_q \alpha_{\lambda'}^2}{\widetilde m_{\Lambda} \alpha_{\lambda\lambda'}^2},
\end{equation}
and $\alpha_{\lambda\lambda'}=\sqrt{(\alpha_{\lambda}^{2}+\alpha_{\lambda'}^{2})/2}$. This leads to
\begin{eqnarray}
W_{\Gamma;n_\lambda l_\lambda m_\lambda s_q}^{n_{\lambda'} l_{\lambda'} m_{\lambda'} s_{q'}}(\alpha_\lambda,\alpha_{\lambda'})&=&\exp\left(-\frac{3m_q^2}{2\widetilde m_\Lambda^2}\frac{p_\Lambda^2}{\alpha_{\lambda\lambda'}^2}\right) \int d^{3} k e^{-\kappa_\lambda^2 k^2} {\cal L}_{n_{\lambda'}}^{l_{\lambda'}+\frac{1}{2}*}\left(\frac{p'^2}{\beta'^2}\right){\cal Y}_{l_{\lambda'} m_{\lambda'}}^{*}(\vec p') \nonumber \\ &&  \times  \langle s(\vec p_\Lambda+\vec p,s_{q'})\mid\overline{s}\Gamma b\mid b(\vec p,s_q)\rangle{\cal L}_{n_{\lambda}}^{l_{\lambda}+\frac{1}{2}}\left(\frac{p^2}{\beta^2}\right){\cal Y}_{l_{\lambda} m_{\lambda}}(\vec p),
\label{eq:wint2}
\end{eqnarray}
where $\kappa_\lambda=\alpha_{\lambda\lambda'}/\alpha_{\lambda'} \alpha_\lambda$. The major difference between the two approaches to extracting the form factors is the way in which Eq. \ref{eq:wint2} is handled.

\subsubsection{SCA model}

As an example of how the matrix elements are computed in the SCA approximation, we examine the ground state-to-ground state transition. Using single component wave functions, the spectator overlap is
\begin{equation}
U_{0 0 0}^{0 0 0}(\alpha_\rho,\alpha_{\rho'})=\left(\frac{\alpha_\rho \alpha_{\rho'}}{\alpha_{\rho\rho'}^2}\right)^{3/2},
\end{equation}
where $\alpha_{\rho\rho'}=\sqrt{(\alpha_{\rho}^{2}+\alpha_{\rho'}^{2})/2}$. The interaction overlap, Eq. \ref{eq:wint2}, becomes
\beqy
W_{\Gamma;0 0 0 s_q}^{0 0 0 s_{q'}}(\alpha_\lambda,\alpha_{\lambda'})&=&\frac{1}{(\pi\beta'\beta)^{3/2}} \exp\left(-\frac{3m_q^2}{2\widetilde m_\Lambda^2}\frac{p_\Lambda^2}{\alpha_{\lambda\lambda'}^2}\right)\nn\\&&\times\int d^3 k e^{-\kappa_\lambda^2 k^2} \langle s(\vec p_\Lambda+\vec p,s_{q'})\mid\overline{s}\Gamma b\mid b(\vec p,s_q)\rangle.
\label{eq:wint3}
\eeqy
In this approximation, a nonrelativistic expansion of the spinors in the quark current $\<s(\plsvec+\pvec,s_{q'})\mid\overline{s}\Gamma b\mid b(\pvec,s_q)\>$ is carried out. For the resulting matrix element, only terms up to $\O(\frac{1}{\qmb\qms})$ are kept. For example, the current $\<s(\plsvec+\pvec,-1/2)\mid\overline{s}\gamma_{+}\gamma_5 b\mid b(\pvec,+1/2)\>$ is reduced to the form
\beq
\<s(\plsvec+\pvec,-1/2)\mid\overline{s}\gamma_{+}\gamma_5 b\mid b(\pvec,+1/2)\>=-\frac{1}{\qmb\qms}\left[\sqrt{\frac{2\pi}{3}}(\pls)_{+}\Y_{11}(\pvec)+\sqrt{\frac{4\pi}{15}}\Y_{22}(\pvec)\right], \label{eq:av1_sca}
\eeq
where 
\beq
\gamma_+=-\frac{1}{\sqrt{2}}\left(\gamma_1+i\gamma_2\right),\,\,\,\, (\pls)_+=-\frac{1}{\sqrt{2}}\left((\pls)_1+i(\pls)_2\right).
\eeq
The $\Y_{lm}(\pvec)$ can be written in terms of $\Y_{lm}(\kvec)$ by use of an addition theorem. A quark current will thus have the general form
\begin{equation}
\langle s(\vec p_\Lambda+\vec p,s_{q'})\mid\overline{s}\Gamma b\mid b(\vec p,s_q)\rangle=\sum_{\ell=0}^{2} \sum_{m_\ell=-\ell}^{\ell} \left(a_{\Gamma;\ell m_\ell}^{s_{q'}s_q}+b_{\Gamma;\ell m_{\ell}}^{s_{q'}s_q}p^2\right){\cal Y}_{\ell m_{\ell}}(\vec k),
\label{eq:qc1}
\end{equation}
where $a_{\Gamma;\ell m_\ell}$ and $b_{\Gamma;\ell m_{\ell}}$ depend on $m_b$, $m_s$, and $\vec p_\Lambda$; these coefficients are truncated at $\O(\frac{1}{\qmb\qms})$. From Eq. \ref{eq:pkswitch}, $p^2=c^2p_\Lambda^2+2c\vec p_\Lambda\cdot\vec k+k^2$. The dot product is proportional to $\sum(-1)^m {\cal Y}_{1 -m}(c\vec p_\Lambda){\cal Y}_{1 m}(\vec k)$. Thus, the integrals in Eq. \ref{eq:wint3} have the basic form of an integral over a product of spherical harmonics times an integral of a Gaussian times a polynomial, both of which can be performed analytically.

The form factors for the vector and axial vector currents have the same forms as those published in \cite{pervin}. For the tensor currents, the form factors we obtain are given in \ref{sec:aff}.

\subsubsection{MCN model\label{sec:mcnff}}

In the MCN model, we keep both the full quark model wave function and the full relativistic form of the quark spinors. We therefore need a more general treatment of the interaction overlap, which is written as (see \ref{sec:shme})
\begin{eqnarray}
W_{\Gamma;n_\lambda l_\lambda m_\lambda s_q}^{n_{\lambda'} l_{\lambda'} m_{\lambda'} s_{q'}}(\alpha_\lambda,\alpha_{\lambda'})&=&\exp\left(-\frac{3m_q^2}{2\widetilde m_\Lambda^2}\frac{p_\Lambda^2}{\alpha_{\lambda\lambda'}^2}\right) \nn\\&&\times \sum_{\lambda'\mu_{\lambda'}}\sum_{\lambda\mu_\lambda} B_{l_{\lambda'} m_{\lambda'}}^{\lambda' \mu_{\lambda'} *}(c'\vec p_{\Lambda}) B_{l_{\lambda} m_{\lambda}}^{\lambda \mu_{\lambda}}(c\vec p_{\Lambda}) {\cal J}_{\Gamma;n_\lambda l_\lambda s_q:\lambda \mu_\lambda}^{n_{\lambda'} l_{\lambda'} s_{q'}:\lambda' \mu_{\lambda'}}(\kappa_\lambda^2),\nn\\
\end{eqnarray}
where the coefficients $B_{l m}^{\lambda \mu_\lambda}$ are given in Eq. \ref{eq:bcoef} and
\begin{eqnarray}
{\cal J}_{\Gamma;n_\lambda l_\lambda s_q:\lambda \mu_\lambda}^{n_{\lambda'} l_{\lambda'} s_{q'}:\lambda' \mu_{\lambda'}}(\kappa_\lambda^2)&=&\int d^3 k e^{-\kappa_\lambda^2 k^2} {\cal L}_{n_{\lambda'}}^{l_{\lambda'}+\frac{1}{2}*}\left(\frac{p'^2}{\beta'^2}\right) {\cal Y}_{\lambda' \mu_{\lambda'}}^{*}(\vec k) \nn\\&&\times \langle s(\vec p_\Lambda+\vec p,s_{q'})\mid\overline{s}\Gamma b\mid b(\vec p,s_q)\rangle  {\cal L}_{n_{\lambda}}^{l_{\lambda}+\frac{1}{2}}\left(\frac{p^2}{\beta^2}\right) {\cal Y}_{\lambda \mu_{\lambda}}(\vec k).\nn\\
\label{eq:jint}
\end{eqnarray}

The full relativistic form of Eq. \ref{eq:av1_sca} is
\beq
\<s(\plsvec+\pvec,-1/2)\mid\overline{s}\gamma_{+}\gamma_5 b\mid b(\pvec,+1/2)\>=-\frac{\sqrt{\frac{8\pi}{3}}(\pls)_{+}\Y_{11}(\pvec)+\sqrt{\frac{16\pi}{15}}\Y_{22}(\pvec)}{\sqrt{\Eb}\sqrt{\Es}\sqrt{\Eb+\qmb}\sqrt{\Es+\qms}},\label{eq:gm5p1}
\eeq
where $\Eb(\Es)$ is the relativistic energy of the $b(s)$ quark. Thus, the current $\langle s(\vec p_\Lambda+\vec p,s_{q'})\mid \overline{s}\Gamma b\mid b(\vec p,s_q)\rangle$ can still be written as a linear combination of solid harmonics with maximum $\ell=2$. However, the coefficients of this expansion no longer take as simple a form as in Eq. \ref{eq:qc1}. The most general form of the quark current is
\begin{equation}
\langle s(\vec p_\Lambda+\vec p,s_{q'})\mid \overline{s}\Gamma b\mid b(\vec p,s_q)\rangle=\sum_{\ell=0}^{2} \sum_{m_\ell=-\ell}^{\ell} \xi_{\Gamma;\ell m_\ell}^{s_{q'} s_q}(\vec k,\vec p_\Lambda) {\cal Y}_{\ell m_\ell}(\vec k).
\end{equation}
The coefficients $\xi_\Gamma$ contain terms of the type $k^N E_b^{\pm 1/2}(E_b+m_b)^{\pm 1/2} E_s^{\pm 1/2}(E_s+m_s)^{\pm 1/2}$, where $N$ is a nonnegative integer. Both $E_b$ and $E_s$ are functions of $\vec p_\Lambda\cdot\vec k$, which complicates a full analytic treatment of the matrix elements. The normalized Laguerre polynomials in Eq. \ref{eq:jint} are also functions of $\vec p_\Lambda\cdot\vec k$. The  coefficients $\xi$ and the normalized Laguerre polynomials can be expanded in terms of spherical harmonics so that the momentum and angular integrals can be performed separately, with the angular term being done analytically and the momentum integrals being treated numerically. The details of this semianalytic treatment of the matrix elements are given in \ref{sec:shme}.

\section{Results\label{sec:results}}

\subsection{Form Factors}

The parameters for the quark model wave functions used in this work are taken from \cite{roper} and are given in Tables \ref{hampar} and \ref{qmtab}. For ease of use, all of the form factors calculated using the MCN model are parametrized to have the form
\beq
F(\sh)=(a_0+a_2\pls^2+a_4\pls^4)\exp\left(-\frac{3m_q^2}{2\tilde m_\Lambda^2}\frac{p_\Lambda^2}{\alpha_{\lambda\lambda'}^2}\right),\,\,\,\, \pls=\frac{\mlb}{2}\sqrt{\phi(\sh)},
\label{eq:ffparam}
\eeq
where $\pls$ is the daughter baryon momentum in the $\Lb$ rest frame. The parameters $a_0$, $a_2$ and $a_4$ for the vector and axial vector form factors are given in Table \ref{vaff}. The corresponding parameters for the tensor form factors are given in Table \ref{tff}. We have also tried a parametrization that included a term in $\pls^6$, but the coefficient of that term was significantly smaller than the coefficient of the $\pls^4$ term for all form factors.

\begin{table}
\caption{Hamiltonian parameters obtained from a fit to a selection of known baryons.}
{\begin{tabular}{|c|c|c|c|c|c|c|c|c|c|}
\hline\small $m_q$ &\small $m_s$ &\small $m_c$ &\small $m_b$ &\small $b$ &\small $\alf_{coul}$ &\small $\alf_{con}$ &\small $\alf_{SO}$ &\small $\alf_{tens}$ &\small $C_{qqq}$ \\
\small (${\text{GeV}}$) &\small (${\text{GeV}}$) &\small (${\text{GeV}}$) &\small (${\text{GeV}}$) &\small (${\text{GeV}^2}$) & & &\small (${\text{GeV}}$) & &\small ($\text{GeV}$) \\ \hline
\small $0.2848$ &\small $0.5553$ &\small $1.8182$ &\small $5.2019$ &\small $0.1540$ &\small $\approx0.0$ &\small $1.0844$ &\small $0.9321$ &\small $-0.2230$ &\small $-1.4204$ \\ \hline
\end{tabular}\label{hampar}}
\end{table}

\begin{table}
\caption{Baryon masses and wave function size parameters, $\alf_\rho$ and $\alf_\lambda$, for states with different $\jp$. All values are in GeV.}
{\begin{tabular}{|c|c|c|c|c|}
\hline\small State, $J^{P}$ &\small Experiment &\small Model &\small $\alf_\lambda$  &\small $\alf_\rho$ \\ \hline
\small$\Lambda_b(5620)\,1/2^{+}$ &\small $5.62$ &\small $5.61$ &\small $0.443$ &\small $0.385$  \\ \hline
\small$\Lambda(1115)\,1/2^{+}$ &\small $1.12$ &\small $1.10$ &\small $0.387$ &\small $0.372$  \\ \hline
\small$\Lambda(1600)\,1/2^{+}$ &\small $1.60$ &\small $1.71$ &\small $0.387$ &\small $0.372$  \\ \hline
\small$\Lambda(1405)\,1/2^{-}$ & \small$1.41$ &\small $1.48$ &\small $0.333$ &\small $0.320$  \\ \hline
\small$\Lambda(1520)\,3/2^{-}$ &\small $1.52$ &\small $1.53$ &\small $0.333$ &\small $0.308$  \\ \hline
\small$\Lambda(1890)\,3/2^{+}$ &\small $1.89$ &\small $1.81$ &\small $0.325$ &\small $0.303$  \\ \hline
\small$\Lambda(1820)\,5/2^{+}$ &\small $1.82$ &\small $1.81$ &\small $0.325$ &\small $0.303$  \\ \hline
\end{tabular}\label{qmtab}}
\end{table}

\begin{table}
\caption{Coefficients in the parametrization of the vector and axial-vector form factors obtained in the MCN approach.}
{\begin{tabular}{|c|c|c|c|c|c|c|c|c|c|}
\hline                      & $a_n(\text{GeV}^{-n})$ & $F_1$ & $F_2$ & $F_3$ & $F_4$ & $G_1$ & $G_2$ & $G_3$ & $G_4$ \\ \hline
                            &  $a_0$ & $1.21$ & $-0.202$ & $-0.0615$ & $-$ & $0.927$ & $-0.236$ & $0.0756$ & $-$ \\
$\Lambda_b\to\Lambda(1115)$ &  $a_2$ & $0.319$ & $-0.219$ & $0.00102$ & $-$ & $0.104$ & $-0.233$ & $0.0195$ & $-$ \\
                            &  $a_4$ & $-0.0177$ & $0.0103$ & $-0.00139$ & $-$ & $-0.00553$ & $0.0110$ & $-0.00115$ & $-$ \\ \hline
                            &  $a_0$ & $0.467$ & $-0.381$ & $0.0501$ & $-$ & $0.114$ & $-0.394$ & $-0.0433$ & $-$ \\
$\Lambda_b\to\Lambda(1600)$ &  $a_2$ & $0.615$ & $-0.281$ & $-0.0295$ & $-$ & $0.300$ & $-0.307$ & $0.0478$ & $-$ \\
                            &  $a_4$ & $0.0568$ & $-0.0399$ & $-0.00163$ & $-$ & $0.0206$ & $-0.0445$ & $0.00566$ & $-$ \\ \hline
                            &  $a_0$ & $0.246$ & $-0.984$ & $0.118$ & $-$ & $1.15$ & $-0.874$ & $0.00871$ & $-$ \\
$\Lambda_b\to\Lambda(1405)$ &  $a_2$ & $0.238$ & $-0.0257$ & $0.0237$ & $-$ & $0.260$ & $-0.0264$ & $-0.0196$ & $-$ \\
                            &  $a_4$ & $0.00976$ & $0.0173$ & $-0.000692$ & $-$ & $-0.00303$ & $0.0159$ & $-0.000977$ & $-$ \\ \hline
                            &  $a_0$ & $-1.66$ & $0.544$ & $0.126$ & $-0.0330$ & $-0.964$ & $0.625$ & $-0.183$ & $0.0530$ \\
$\Lambda_b\to\Lambda(1520)$ &  $a_2$ & $-0.295$ & $0.194$ & $0.00799$ & $-0.00977$ & $-0.100$ & $0.219$ & $-0.0380$ & $0.0161$ \\
                            &  $a_4$ & $0.00924$ & $-0.00420$ & $-0.000365$ & $0.00211$ & $0.00264$ & $-0.00508$ & $0.00351$ & $-0.00221$ \\ \hline
                            &  $a_0$ & $-0.460$ & $1.33$ & $-0.232$ & $0.0485$ & $-1.71$ & $1.14$ & $0.0193$ & $-0.0153$ \\
$\Lambda_b\to\Lambda(1890)$ &  $a_2$ & $-0.271$ & $0.00439$ & $-0.0315$ & $0.0140$ & $-0.284$ & $0.00990$ & $0.0374$ & $-0.00770$ \\
                            &  $a_4$ & $-0.0116$ & $-0.0149$ & $0.000345$ & $-0.00218$ & $-0.00146$ & $-0.0134$ & $-0.000343$ & $-0.000236$ \\ \hline
                            &  $a_0$ & $2.48$ & $-0.952$ & $-0.202$ & $0.0810$ & $1.25$ & $-1.12$ & $0.355$ & $-0.143$ \\
$\Lambda_b\to\Lambda(1820)$ &  $a_2$ & $0.362$ & $-0.238$ & $-0.0119$ & $0.00573$ & $0.122$ & $-0.272$ & $0.0446$ & $-0.0197$ \\
                            &  $a_4$ & $-0.00639$ & $0.00224$ & $0.000303$ & $-0.000169$ & $-0.00134$ & $0.00303$ & $-0.00103$ & $0.000440$ \\ \hline
\end{tabular}\label{vaff}}
\end{table}

\begin{table}
\caption{Coefficients in the parametrization of the tensor form factors obtained in the MCN approach.}
{\begin{tabular}{|c|c|c|c|c|c|c|c|}
\hline                      & $a_n(\text{GeV}^{-n})$ & $H_1$ & $H_2$ & $H_3$ & $H_4$ & $H_5$ & $H_6$ \\ \hline
                            &  $a_0$ & $0.936$ & $0.227$ & $-0.0757$ & $-0.0174$ & $-$ & $-$ \\
$\Lambda_b\to\Lambda(1115)$ &  $a_2$ & $0.0722$ & $0.265$ & $-0.0195$ & $-0.00986$ & $-$ & $-$ \\
                            &  $a_4$ & $-0.00643$ & $-0.0101$ & $0.00116$ & $-0.000524$ & $-$ & $-$ \\ \hline
                            &  $a_0$ & $0.121$ & $0.389$ & $0.0421$ & $0.00676$ & $-$ & $-$ \\
$\Lambda_b\to\Lambda(1600)$ &  $a_2$ & $0.313$ & $0.295$ & $-0.0479$ & $-0.0242$ & $-$ & $-$ \\
                            &  $a_4$ & $0.0101$ & $0.0550$ & $-0.00565$ & $-0.00404$ & $-$ & $-$ \\ \hline
                            &  $a_0$ & $-1.13$ & $-0.872$ & $0.00645$ & $-0.112$ & $-$ & $-$ \\
$\Lambda_b\to\Lambda(1405)$ &  $a_2$ & $-0.256$ & $-0.0241$ & $-0.0197$ & $-0.00215$ & $-$ & $-$ \\
                            &  $a_4$ & $0.00288$ & $0.0158$ & $-0.000965$ & $0.00151$ & $-$ & $-$ \\ \hline
                            &  $a_0$ & $-1.08$ & $-0.507$ & $0.187$ & $0.0772$ & $-0.0517$ & $0.0206$ \\
$\Lambda_b\to\Lambda(1520)$ &  $a_2$ & $-0.0732$ & $-0.246$ & $0.0295$ & $0.0267$ & $-0.0173$ & $0.00679$ \\
                            &  $a_4$ & $0.00464$ & $0.00309$ & $-0.00107$ & $-0.00217$ & $0.00259$ & $-0.000220$ \\ \hline
                            &  $a_0$ & $1.68$ & $1.13$ & $0.0214$ & $0.198$ & $-0.0147$ & $0.0331$ \\
$\Lambda_b\to\Lambda(1890)$ &  $a_2$ & $0.280$ & $0.00710$ & $0.0380$ & $-0.00103$ & $-0.00818$ & $0.00674$  \\
                            &  $a_4$ & $0.00154$ & $-0.0134$ & $-0.000450$ & $-0.00155$ & $-0.000234$ & $-0.00239$ \\ \hline
                            &  $a_0$ & $1.55$ & $0.830$ & $-0.355$ & $-0.160$ & $0.143$ & $-0.0581$ \\
$\Lambda_b\to\Lambda(1820)$ &  $a_2$ & $0.0959$ & $0.298$ & $-0.0446$ & $-0.0327$ & $0.0198$ & $-0.0205$ \\
                            &  $a_4$ & $-0.00427$ & $-0.0000926$ & $0.00103$ & $0.000739$ & $-0.000441$ & $0.00221$ \\ \hline
\end{tabular}\label{tff}}
\end{table}

\begin{figure}
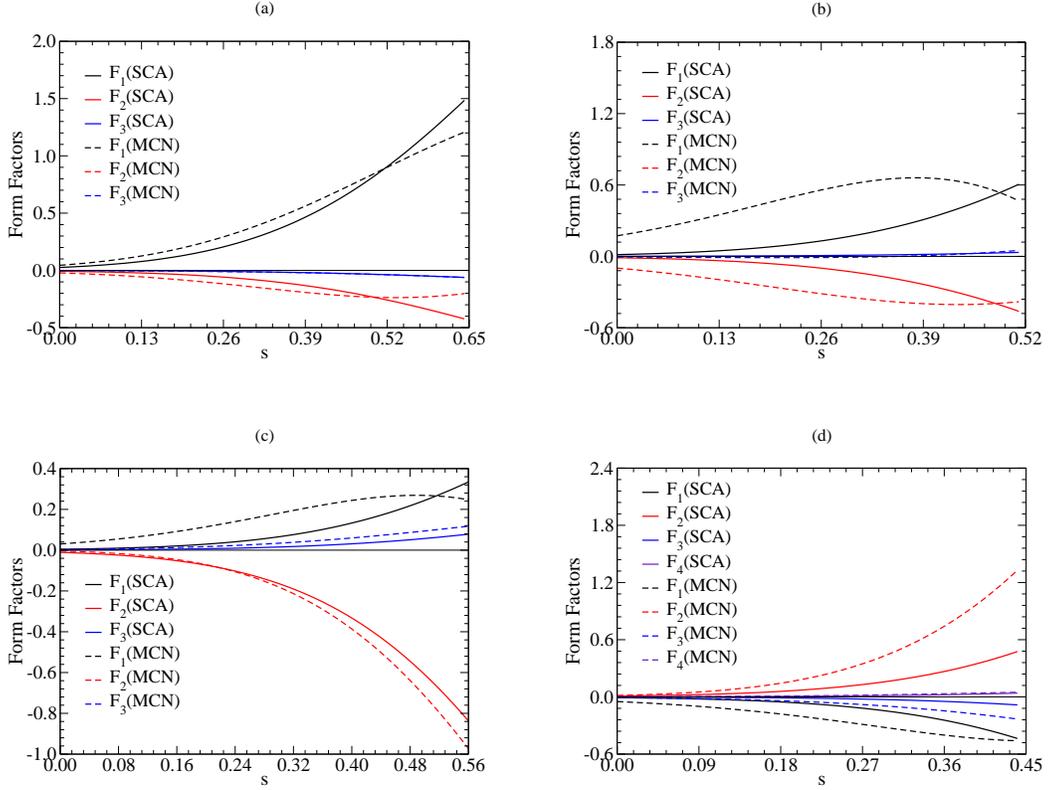

\centerline{\includegraphics[width=2.5in]{vff12p_sca_mcn.eps}\hskip30pt \includegraphics[width=2.5in]{vff12p1r_sca_mcn.eps}}\vskip25pt
\centerline{\includegraphics[width=2.5in]{vff12m_sca_mcn.eps}\hskip30pt \includegraphics[width=2.5in]{vff32p_sca_mcn.eps}}
\caption{Comparison of vector form factors calculated using the SCA and MCN models. These form factors are for decays to (a) $\Lambda(1115),\,\, J^P=1/2^{+}$, (b) $\Lambda(1600),\,\, J^P=1/2_1^{+}$, (c) $\Lambda(1405),\,\, J^P=1/2^{-}$, and (d) $\Lambda(1890),\,\, J^P=3/2^{+}$.}\label{fig:vff}
\end{figure}

Figure \ref{fig:vff} shows a comparison of the vector form factors obtained using the two models for transitions to the ground state (Fig. \ref{fig:vff}(a)), the first radial excitation (Fig. \ref{fig:vff}(b)), the $\Lambda(1405)$ (Fig. \ref{fig:vff}(c)) and the $\Lambda(1820)$ (Fig. \ref{fig:vff}(d)). In the SCA model, all of the kinematic dependence of the form factors takes the form of a Gaussian in the momentum of the daughter baryon, calculated in the rest frame of the parent. Thus, all of the SCA form factors have magnitudes that monotonically increase as $\sh$ is increased. In contrast with this, form factors obtained in the MCN have additional dependence on the momentum of the daughter baryon, as parametrized in Eq. \ref{eq:ffparam}. Depending on the relative sizes of $a_0$ and $a_2$ in this expression, the shapes of the MCN form factors can be quite different from those obtained in the SCA model. The most striking differences occur for the transitions to the radially excited state, shown in Fig. \ref{fig:vff}(b).

For all transitions examined, $F_3$ (and $F_4$, where appropriate) is very small in both models, consistent with the expectations of HQET. For states with natural parity, $F_1$ is the largest form factor, while for states with unnatural parity, $F_2$ is largest. Among the axial form factors, $G_1$ and $G_2$ are the largest, independent of the transition considered, and $G_3$ and $G_4$ (where appropriate) are again much smaller. This can be seen by examining the coefficients given in Table \ref{vaff}. The SCA form factors all follow a similar trend. Among the tensor form factors, $H_1$ and $H_2$ are the largest, with the other form factors all being significantly smaller, in both models, for all transitions considered.

\subsubsection{Comparison with HQET}

It is of considerable interest to compare the two sets of form factors we obtain with the expectations of HQET. In the following paragraphs, we carry out this comparison.

\subsubsubsection{$\it\jp=1/2^+$}

\begin{figure}
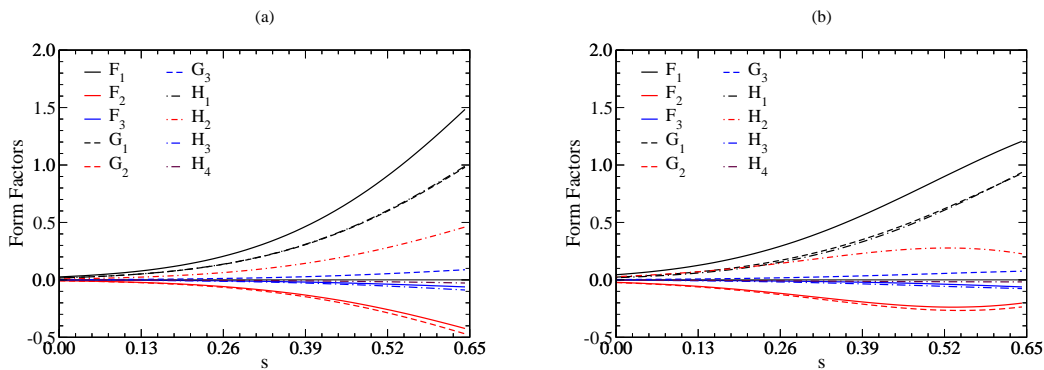

\centerline{\includegraphics[width=2.5in]{ff_12p_sca.eps}\hskip30pt \includegraphics[width=2.5in]{ff_12p_mcn.eps}}
\caption{Form Factors for $\Lambda_{b}\rightarrow\Lambda(1115),\,\, J^P=1/2^{+}$ as a function of $\hat s=q^{2}/m_{\Lambda_{b}}^{2}$. Graphs (a) and (b) show the form factors obtained using SCA and MCN models, respectively.}\label{fig:ff12p}
\end{figure}

For decays to daughter baryons with $\jp=1/2^+$, the leading order HQET predictions are shown in Eq. \ref{eq:hqet12p}.
Figures \ref{fig:ff12p}(a) and (b) show the form factors for decays to the ground state obtained using both the SCA and MCN models, respectively. From the figures, we see that both models satisfy the leading order predictions of HQET. In both models, $F_2\approx G_2$ through most of the kinematic range. We also note that $G_1$ and $H_1$ are virtually indistinguishable in both models.

\begin{figure}
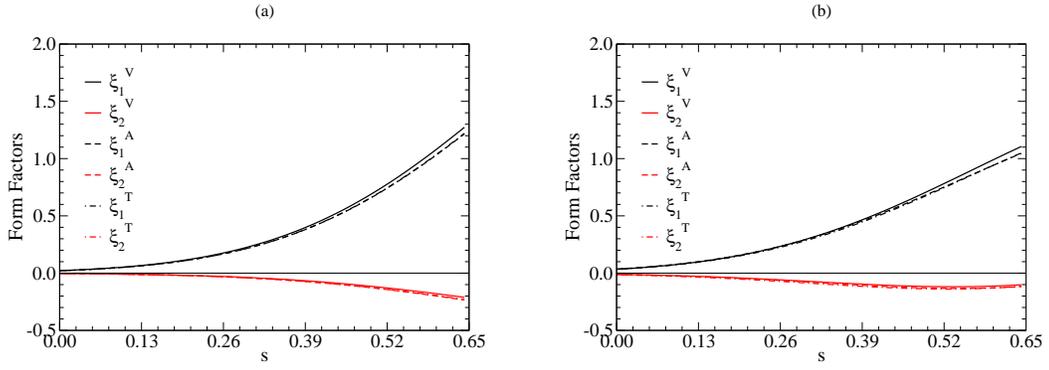

\centerline{\includegraphics[width=2.5in]{hqet_ff_12p_sca.eps}\hskip30pt
\includegraphics[width=2.5in]{hqet_ff_12p_mcn.eps}}
\caption{HQET Form Factors for $\Lambda_{b}\rightarrow\Lambda(1115),\,\, J^P=1/2^{+}$ as a function of $\hat s=q^{2}/m_{\Lambda_{b}}^{2}$. The graphs show the HQET form factors calculated using the vector, axial vector, and tensor form factors using (a) SCA and (b) MCN models.}\label{fig:hqet12p}
\end{figure}

The expressions for the leading order HQET predictions can be inverted to give
\begin{eqnarray}
\xi_1^{(0)}&=&F_1+F_2/2=G_1-G_2/2 =H_1+H_2/2,\nonumber\\
\xi_2^{(0)}&=&F_2/2=G_2/2=-H_2/2. \nonumber 
\end{eqnarray}
It is useful to extract the $\xi_i$ independently from the vector, axial vector and tensor form factors, and compare the three different forms obtained in this way. Figure \ref{fig:hqet12p} shows  the three extractions for these form factors from both the SCA and MCN models for transitions to the ground state. As can be seen, the form factors obtained using the axial vector and tensor form factors are virtually identical in both models. The set obtained from the vector form factors is also in very good agreement with the other sets as well. This shows clearly that the form factors obtained from both the SCA and MCN models satisfy the leading order expectations of HQET. However, the curves for the $\xi^{(0)}$ from the three extractions should not be expected to be identical, since the expressions for the $F_i$, $G_i$ and $H_i$ in terms of universal HQET `Isgur-Wise' type functions will receive corrections due to the finite mass of the $b$ quark. This holds for all of the angular momentum states we consider.

One HQET expectation is that the form factors for $\Lb\to\Ls$ should be the same as those for $\Lambda_c\to\Lambda$, up to terms of order $1/m_q$, where $m_q$ is the mass of the heavy quark. At the `nonrecoil point', or at maximum $q^2$, the CLEO collaboration has extracted the ratio $\xi_2^{(0)}/\xi_1^{(0)}=-0.25\pm0.14\pm0.08$ \cite{cleo3}. From our analysis, we find that
\beqy
\xi_2^{V(0)}/\xi_1^{V(0)}&=&-0.166(-0.092), \,\,\,\, \xi_2^{A(0)}/\xi_1^{A(0)}=-0.193(-0.113), \nn\\ \xi_2^{T(0)}/\xi_1^{T(0)}&=&-0.188(-0.108),\nn
\eeqy
within the SCA (MCN) model, at the nonrecoil point. For both the SCA and MCN models, our values for this ratio are consistent with the value reported by CLEO. We note here that $\left|\xi_2^{(0)}/\xi_1^{(0)}\right|$ is smaller in the MCN model than in the SCA model for all three (vector, axial vector, tensor) form factor scenarios.

\begin{figure}
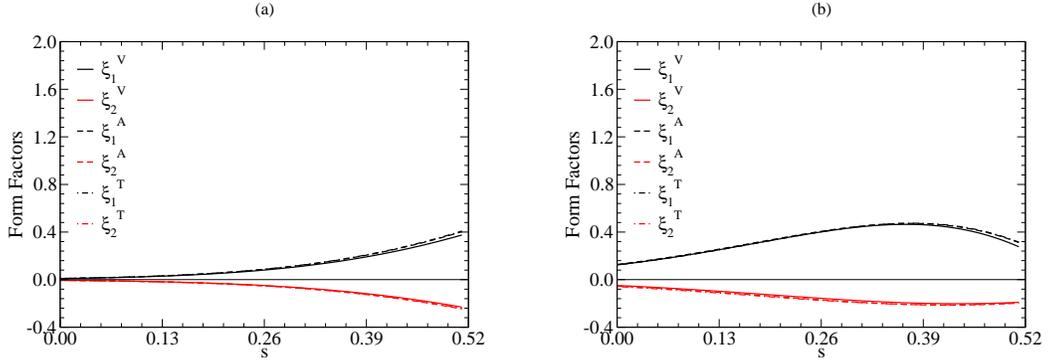

\centerline{\includegraphics[width=2.5in]{hqet_ff_12p1r_sca.eps}\hskip30pt
\includegraphics[width=2.5in]{hqet_ff_12p1r_mcn.eps}}
\caption{Same as Fig. \ref{fig:hqet12p}, but for $\Lambda_{b}\rightarrow\Lambda(1600),\,\, J^P=1/2^{+}$.}\label{fig:hqet12p1r}
\end{figure}

The HQET form factors for transitions to the first radial excitation obtained from the two models we use are shown in Figure \ref{fig:hqet12p1r}. For each model, the three possible extractions of the $\xi_i^{(0')}$ agree with each other reasonably well over the entire kinematically allowed range. Not surprisingly, the $\xi_i^{(0')}$ extracted from the two models are very different. 

The shape of the function $\xi_1^{(0')}$ in the MCN model can be understood by taking the limit in which the strange quark is also heavy. In that limit, the transitions to states with $J^P=1/2^+$ are described by a single form factor. Furthermore, for transitions to the ground state, that form factor, the Isgur-Wise function, is normalized to unity at the nonrecoil point. Since the wave function of the radially excited state must be orthogonal to that of the ground state, the corresponding Isgur-Wise function for transitions to the radially excited state must vanish at the nonrecoil point, up to corrections  proportional to the inverse of the heavy quark masses. In the SCA model, the truncated form factors still exhibit the expected behavior of being ${\cal O}(1/m_q)$ near the non-recoil point, but the more precise form factors of the MCN model show this behavior more clearly.

The ratios $\xi_2^{(0')}/\xi_1^{(0')}$ at zero recoil for this state for the three extractions within the SCA (MCN) model are
\beqy
\xi_2^{V(0')}/\xi_1^{V(0')}&=&-0.615(-0.689), \,\,\,\, \xi_2^{A(0')}/\xi_1^{A(0')}=-0.611(-0.633), \nn\\ \xi_2^{T(0')}/\xi_1^{T(0')}&=&-0.590(-0.617).\nn
\eeqy
For the three form factor scenarios, $\left|\xi_2^{(0')}/\xi_1^{(0')}\right|$ is smaller in the SCA model than in the MCN model.

\subsubsubsection{$\it\jp=1/2^-$}

\begin{figure}[t]
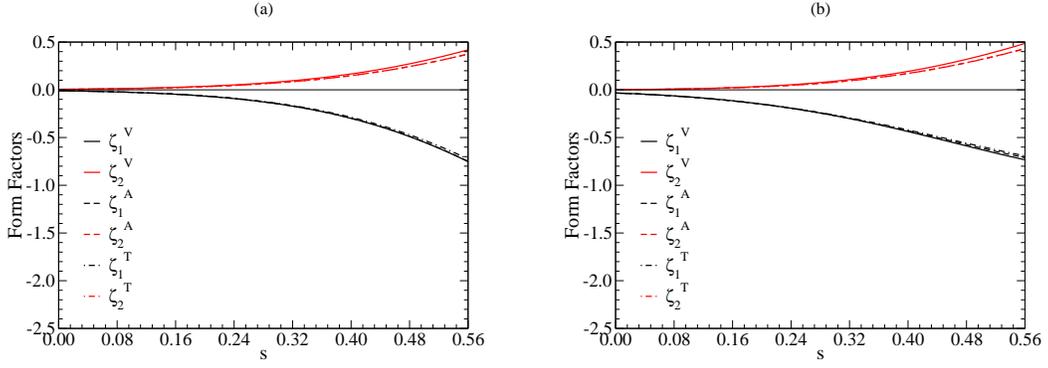

\centerline{\includegraphics[width=2.5in]{hqet_ff_12m_sca.eps}\hskip30pt
\includegraphics[width=2.5in]{hqet_ff_12m_mcn.eps}}
\caption{Same as Fig. \ref{fig:hqet12p}, but for $\Lambda_{b}\rightarrow\Lambda(1405),\,\, J^P=1/2^{-}$.}\label{fig:hqet12m}
\end{figure}

For decays to states with $\jp=1/2^-$, the leading-order HQET predictions are shown in Eq. \ref{eq:hqet12m}.
These relations can be inverted to give
\begin{eqnarray}
\zeta_1^{(0)}&=&-(F_1-F_2/2)=-(G_1+G_2/2)=H_1-H_2/2, \nonumber\\
\zeta_2^{(0)}&=&-F_2/2=-G_2/2=-H_2/2. \nonumber
\end{eqnarray}
Figure \ref{fig:hqet12m} shows the three versions of the extracted $\zeta_i^{(0)}$ for transitions to this state, for each model. There is good agreement among the three sets of form factors in both models. We note here that although $\zeta_1^{V(0)}$ and $\zeta_1^{A(0)}$ are indistinguishable over the entire kinematic range in the SCA model, they begin to deviate slightly from each other as they approach zero recoil in the MCN model. In both models, $\zeta_2^{A(0)}$ and $\zeta_2^{T(0)}$ are virtually identical. The values for the ratio $\zeta_2/\zeta_1$ at zero recoil for transitions to this state are 
\beqy
\zeta_2^{V(0)}/\zeta_1^{V(0)}&=&-0.555(-0.667), \,\,\,\, \zeta_2^{A(0)}/\zeta_1^{A(0)}=-0.499(-0.610), \nn\\ \zeta_2^{T(0)}/\zeta_1^{T(0)}&=&-0.517(-0.626),\nn
\eeqy
within the SCA (MCN) model. From the above, we see that $\left|\zeta_2^{(0)}/\zeta_1^{(0)}\right|$ is larger in the MCN model for all three scenarios.

\subsubsubsection{$\it\jp=3/2^-$}

\begin{figure}
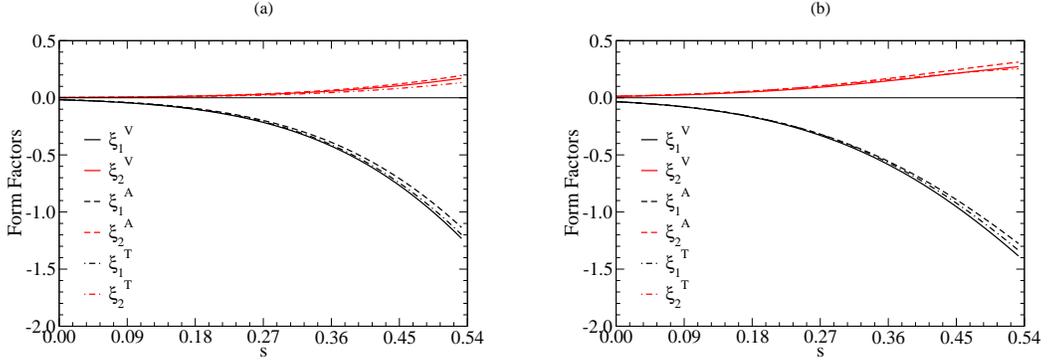

\centerline{\includegraphics[width=2.5in]{hqet_ff_32m_sca.eps}\hskip30pt
\includegraphics[width=2.5in]{hqet_ff_32m_mcn.eps}}
\caption{Same as Fig. \ref{fig:hqet12p}, but for $\Lambda_{b}\rightarrow\Lambda(1520),\,\, J^P=3/2^{-}$.}\label{fig:hqet32m}
\end{figure}

The leading-order HQET predictions for decays to states with $\jp=3/2^-$ are shown Eq. \ref{eq:hqet32m}.
By inverting these expressions, we find that
\begin{eqnarray}
\xi_1^{(1)}&=&F_1+F_2/2=G_1-G_2/2=H_1+H_2/2, \nonumber\\
\xi_2^{(1)}&=&F_2/2=G_2/2=-H_2/2. \nonumber
\end{eqnarray}
The three sets of HQET form factors for transitions to this state for each model are shown in Figure \ref{fig:hqet32m}.  The three extractions agree with each other throughout most of the kinematic range, in both models. For $\xi_2^{(1)}/\xi_1^{(1)}$ at the nonrecoil point, we find
\beqy
\xi_2^{V(1)}/\xi_1^{V(1)}&=&-0.139(-0.196), \,\,\,\, \xi_2^{A(1)}/\xi_1^{A(1)}=-0.172(-0.245), \nn\\ \xi_2^{T(1)}/\xi_1^{T(1)}&=&-0.111(-0.190).\nn
\eeqy
We see that the magnitude of this ratio is larger in the MCN model for all three form factor scenarios.

\subsubsubsection{$\it\jp=3/2^+$}

\begin{figure}
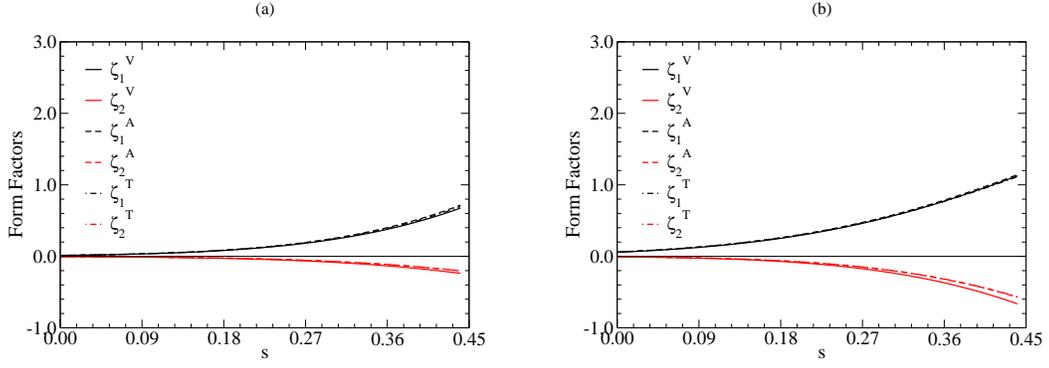

\centerline{\includegraphics[width=2.5in]{hqet_ff_32p_sca.eps}\hskip30pt
\includegraphics[width=2.5in]{hqet_ff_32p_mcn.eps}}
\caption{Same as Fig. \ref{fig:hqet12p}, but for $\Lambda_{b}\rightarrow\Lambda(1890),\,\, J^P=3/2^{+}$.}\label{fig:hqet32p}
\end{figure}

For decays to states with $\jp=3/2^+$, the leading-order HQET predictions are given in Eq. \ref{eq:hqet32p}.
Inverting these expressions leads to
\begin{eqnarray}
\zeta_1^{(1)}&=&-(F_1-F_2/2)=-(G_1+G_2/2)=H_1-H_2/2, \nonumber\\
\zeta_2^{(1)}&=&-F_2/2=-G_2/2=-H_2/2.\nn
\end{eqnarray}
Figure \ref{fig:hqet32p} shows the three extractions for the $\zeta_i^{(1)}$ for transitions to this state obtained in both models. There is good agreement among the three sets of form factors for both models. In the SCA model, the three sets of form factors are nearly the same over much of the kinematic range. However, in the MCN model we find that the $\zeta_1^{(1)}$ for all three scenarios are almost indistinguishable over the entire kinematic range, but $\zeta_2^{V(1)}$ deviates from the other two as they approach the nonrecoil point. The ratio $\zeta_2^{(1)}/\zeta_1^{(1)}$ at zero recoil for the three form factor scenarios are
\beqy
\zeta_2^{V(1)}/\zeta_1^{V(1)}&=&-0.353(-0.591), \,\,\,\, \zeta_2^{A(1)}/\zeta_1^{A(1)}=-0.285(-0.501), \nn\\ \zeta_2^{T(1)}/\zeta_1^{T(1)}&=&-0.293(-0.509),\nn
\eeqy
within the SCA (MCN) model. The magnitude of this ratio is larger in the MCN model for all three scenarios.

\subsubsubsection{$\it\jp=5/2^+$}

\begin{figure}
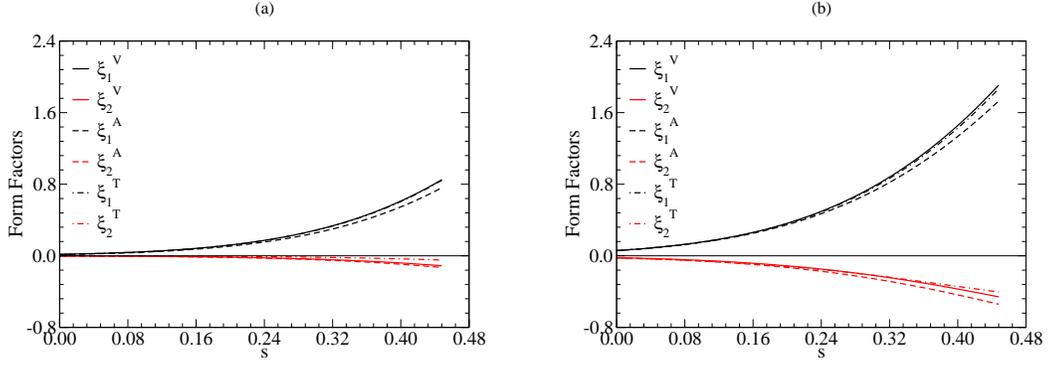

\centerline{\includegraphics[width=2.5in]{hqet_ff_52p_sca.eps}\hskip30pt
\includegraphics[width=2.5in]{hqet_ff_52p_mcn.eps}}
\caption{Same as Fig. \ref{fig:hqet12p}, but for $\Lambda_{b}\rightarrow\Lambda(1820),\,\, J^P=5/2^{+}$.}\label{fig:hqet52p}
\end{figure}

For decays to daughter baryons with $\jp=5/2^+$, the leading-order HQET predictions are shown in Eq. \ref{eq:hqet52p}.
These expressions can be inverted to give
\begin{eqnarray}
\xi_1^{(2)}&=&F_1+F_2/2=G_1-G_2/2=H_1+H_2/2, \nonumber\\
\xi_2^{(2)}&=&F_2/2=G_2/2=-H_2/2.\nn
\end{eqnarray}
The three sets of HQET form factors for transitions to this state are shown in Figure \ref{fig:hqet52p}. In both models, the three sets of form factors are nearly identical at small $\sh$. In the SCA model, $\xi_1^{V(2)}$ and $\xi_1^{T(2)}$ remain indistinguishable over the entire kinematic range while $\xi_1^{A(2)}$ deviates as the nonrecoil point is approached. It can also be seen that $\xi_2^{V(2)}$ and $\xi_2^{A(2)}$ remain fairly close over the kinematic range with $\xi_2^{T(2)}$ deviating from the others as they approach zero recoil. In the MCN model, the sets of form factors obtained from the vector and tensor form factors remain nearly the same throughout the entire kinematic range with those obtained from the axial vector form factors deviating as they approach zero recoil. The ratio $\xi_2^{(2)}/\xi_1^{(2)}$ at the nonrecoil point for the three scenarios are
\beqy
\xi_2^{V(2)}/\xi_1^{V(2)}&=&-0.130(-0.238), \,\,\,\, \xi_2^{A(2)}/\xi_1^{A(2)}=-0.172(-0.309), \nn\\ \xi_2^{T(2)}/\xi_1^{T(2)}&=&-0.056(-0.211),\nn
\eeqy
for the SCA (MCN) model. The magnitude of this ratio is larger in the MCN model for all three scenarios.

\subsection{Dileptonic Decays}

In this section, we present differential decays rates, branching ratios (BRs), and FBAs for dileptonic decays using the two sets of form factors that we have extracted. We present results obtained using Wilson coefficients that have been calculated in the standard model (SM), with both sets of form factors. We also examine one scenario that arises beyond the SM, namely a supersymmetric (SUSY) extension to the SM, but there we use the MCN form factors exclusively. In our numerical calculations, the SM values of the Wilson coefficients are taken from \cite{huang} and the SUSY values are taken from \cite{aslam}. These values are presented in Table \ref{wilco}. In \cite{huang}, the Wilson coefficients are evaluated using a naive dimensional regularization scheme at the scale $\mu=5.0\text{ GeV}$. The top  quark mass is taken to be $m_t=174\text{ GeV}$ and the cut-off $\Lambda_{\overline{MS}}=225\text{ MeV}$, where $\overline{MS}$ denotes the modified minimal subtraction scheme. The SUSY model used here is referred to as SUSYI in Ref. \cite{aslam}. SUSYI corresponds to the regions of parameter space where supersymmetry can destructively contribute and can change the sign of $C_7$, but contributions from neutral Higgs bosons are neglected.

Since the rates and FBAs for the $e$ and $\mu$ channels are essentially the same, in what follows, we present the results for the $\mu$ and $\tau$ channels only.

\begin{table}
\caption{SM and SUSY values for the Wilson coefficients. In the SUSY model we use, only $C_7$, $C_9$, and $C_{10}$ get modified from their SM values.}
{\begin{tabular}{|c|c|c|c|c|c|c|c|c|c|}
\hline &\small $C_1$ &\small $C_2$ &\small $C_3$ &\small $C_4$ &\small $C_5$ &\small $C_6$ &\small $C_7$ &\small $C_9$ &\small $C_{10}$ \\ \hline
\small SM &\small $-0.243$ &\small $1.105$ &\small $0.011$ &\small $-0.025$ &\small $0.007$ &\small $-0.031$ &\small $-0.312$ &\small $4.193$ &\small $-4.578$ \\ \hline
\small SUSY & & & & & & &\small $0.376$ &\small $4.767$ &\small $-3.735$ \\ \hline
\end{tabular}\label{wilco}}
\end{table}

\subsubsection{Decay Rates}

\begin{table}
\caption{Branching ratios for $\Lambda_b\rightarrow \Lambda^{(*)}\mu^{+}\mu^{-}$ in units of $10^{-6}$. The numbers in the column labeled SM1 are obtained using the SCA form factors with standard model Wilson Coefficients. The numbers in the column labeled SM2 are also obtained using SM Wilson coefficients, but the MCN form factors. The numbers in the column labeled SUSY are obtained using the MCN form factors with Wilson coefficients from a supersymmetric scenario. The column labeled LD refers to the long distance contributions of the charmonium resonances, with `a' indicating that these contributions have been neglected, and `b' indicating that they have been included. In this table, it is assumed that the $\Lambda(1600)$ is the first radial excitation. The lifetime of the $\Lambda_b$ is taken from the Particle Data Listings \cite{pdg}.}
{\begin{tabular}{|c|c|c|c|c|c|c|c|c|}
\hline\small State, $J^{P}$ &LD&SM1 &\small SM2 &\small SUSY &\small Aslam {\it et al.} \cite{aslam}  &\small Wang {\it et al.} \cite{wang}  &\small Chen {\it et al.} \cite{chen} & Experiment \cite{cdf2} \\
&& & & && &\small QCDSR $\,\,$\small PM& \\ \hline
\small$\Lambda(1115)\,1/2^{+}$ &a &\small $0.60$ &\small $0.70$ &\small $1.0$ &\small $5.9$ &\small $6.1$ &\small $2.1$ $\,\,\,\,\,\,\,\,\,\,\,$ $1.2$ & $1.73\pm0.42\pm0.55$. \\
&b &\small $21$ &\small $32$ &\small $32$ &\small $39$ &\small $46$ &\small $53$ $\,\,\,\,\,\,\,\,\,\,\,\,\,\,\,$ $36$ & $-$\\ \hline
\small$\Lambda(1600)\,1/2^{+}$ & a &\small $0.027$ &\small $0.32$ &\small $0.53$ & $-$ & $-$ & $-$& $-$ \\
& b &\small $2.6$ &\small $35$ &\small $35$ & $-$ & $-$ & $-$& $-$ \\ \hline
\small$\Lambda(1405)\,1/2^{-}$ & a &\small $0.094$ &\small $0.21$ &\small $0.32$ & $-$ & $-$ & $-$& $-$ \\
& b &\small $5.9$ &\small $19$ &\small $19$ & $-$ & $-$ & $-$ & $-$\\ \hline
\small$\Lambda(1520)\,3/2^{-}$ & a &\small $0.13$ &\small $0.21$ &\small $0.34$ & $-$ & $-$ & $-$ & $-$\\
& b &\small $14$ &\small $24$ &\small $24$ & $-$ & $-$ & $-$& $-$ \\ \hline
\small$\Lambda(1890)\,3/2^{+}$ & a &\small $0.018$ &\small $0.097$ &\small $0.17$ & $-$ & $-$ & $-$ & $-$\\
& b &\small $1.3$ &\small $5.8$ &\small $5.9$ & $-$ & $-$ & $-$& $-$ \\ \hline
\small$\Lambda(1820)\,5/2^{+}$ & a &\small $0.013$ &\small $0.082$ &\small $0.15$ & $-$ & $-$ & $-$ & $-$\\
& b &\small $0.84$ &\small $4.6$ &\small $4.7$ & $-$ & $-$ & $-$ & $-$\\ \hline
\end{tabular}\label{br1}}
\end{table}

\begin{table}
\caption{Branching ratios for $\Lambda_b\rightarrow \Lambda^{(*)}\tau^{+}\tau^{-}$ in units of $10^{-6}$. The columns are labeled as in Table \ref{br1}.}
{\begin{tabular}{|c|c|c|c|c|c|c|c|}
\hline\small State, $J^{P}$ &LD&\small SM1 &\small SM2 &\small SUSY &\small Aslam {\it et al.} \cite{aslam}  &\small Wang {\it et al.} \cite{wang}  &\small Chen {\it et al.} \cite{chen}  \\
&& & & && &\small QCDSR $\,\,$\small PM \\ \hline
\small$\Lambda(1115)\,1/2^{+}$ & a  &\small $0.22$ &\small $0.22$ &\small $0.38$ &\small $2.1$ &\small $2.4$ &\small $0.18$ $\,\,\,\,\,\,\,\,\,\,\,$ $0.26$ \\
&b &\small $0.59$ &\small $0.68$ &\small $0.86$ &\small $4.0$ &\small $4.3$ &\small $11$ $\,\,\,\,\,\,\,\,\,\,\,\,\,\,\,$ $9.0$ \\ \hline
\small$\Lambda(1600)\,1/2^{+}$ & a  &\small $<0.01$ &\small $<0.01$ &\small $0.016$ & $-$ & $-$ & $-$ \\
&b &\small $0.033$ &\small $0.12$ &\small $0.13$ & $-$ & $-$ & $-$ \\ \hline
\small$\Lambda(1405)\,1/2^{-}$ & a  &\small $0.023$ &\small $0.030$ &\small $0.055$ & $-$ & $-$ & $-$ \\
&b &\small $0.12$ &\small $0.22$ &\small $0.25$ & $-$ & $-$ & $-$ \\ \hline
\small$\Lambda(1520)\,3/2^{-}$ & a  &\small $0.013$ &\small $0.016$ &\small $0.031$ & $-$ & $-$ & $-$ \\
&b &\small $0.14$ &\small $0.18$ &\small $0.20$ & $-$ & $-$ & $-$ \\ \hline
\small$\Lambda(1890)\,3/2^{+}$ & a  &\small $<0.01$ &\small $<0.01$ &\small $<0.01$ & $-$ & $-$ & $-$ \\
&b &\small $<0.01$ &\small $<0.01$ &\small $<0.01$ & $-$ & $-$ & $-$ \\ \hline
\small$\Lambda(1820)\,5/2^{+}$ & a  &\small $<0.01$ &\small $<0.01$ &\small $<0.01$ & $-$ & $-$ & $-$ \\
&b &\small $<0.01$ &\small $<0.01$ &\small $<0.01$ & $-$ & $-$ & $-$ \\ \hline
\end{tabular}\label{br2}}
\end{table}

The branching ratios predicted for the $\mu$ and $\tau$ channels are presented in Tables \ref{br1} and \ref{br2}, respectively. Each table displays the results for the SM calculations using two models for the form factors, as well as one SUSY scenario with the MCN form factors. In addition, results obtained omitting and including the long distance (LD) contributions are presented. For ease of discussion, we will refer to the results obtained in the SM as SM1 for the SCA form factors, and SM2 for the MCN form factors. SM1a and SM2a will refer to results with LD contributions omitted, while SM1b and SM2b will refer to results with LD contributions included. Finally, SUSYa and SUSYb will refer to the results obtained using Wilson coefficients from the supersymmetric extension to the standard model discussed above, with the SUSYa (SUSYb) results obtained when LD contributions are omitted (included).  We also compare our model predictions with those made by other authors using SM Wilson coefficients, within the framework of light cone sum rules (LCSR) \cite{aslam,wang}, QCD sum rules (QCDSR) \cite{chen}, and a multipole model (PM) \cite{chen}, as well as with the recent experimental results from the CDF Collaboration \cite{cdf2}. Graphs of the differential decay rates $d\Gamma/d\sh$ are shown in Figs. \ref{fig:rate12p}-\ref{fig:rate52p}.

For decays to the ground state in the $\mu$ channel, the branching ratios we obtain in the SM scenario are smaller than those obtained by other authors \cite{aslam,wang,chen}. In the $\tau$ channel, our results in models SM1a and SM2a are smaller than the LCSR predictions for the ground state, but are comparable to the results obtained using QCDSR and PM. However, the predictions of the SM1b and SM2b models are much smaller than those of other models in this channel. The predictions of SM2 are larger than SM1 for decays to all $\Ls^{(*)}$, for both channels. Note, too, that SM1 predicts that decays to the ground state dominate the rare decays of the $\Lambda_b$, but SM2b indicates that decays to the first radial excitation are the dominant mode, with the decay rate to the $\Lambda(1520)$ being similar in magnitude to the decay rate to the ground state. In addition, the decay rate to the $\Lambda(1405)$ is only slightly smaller than the decay rate to the ground state. These results imply that searches for rare decays of the $\Lambda_b$, such as those planned by the LHCb collaboration, should include excited final states, as the decay rates to these states can be sizable. We note that this is consistent with the current experimental status in the rare decays of the $B$ meson, where decays to the $K$ and $K^*$ account for less than half of the inclusive dileptonic decay rate. In both the $\mu$ and $\tau$ channel, the SUSY BRs are larger than the SM results without LD contributions.

\subsubsection*{$\it\jp=1/2^+$}

\begin{figure}[t]
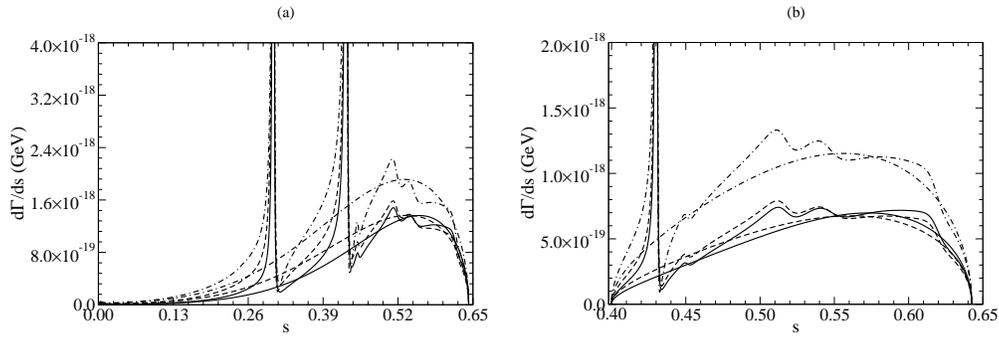

\centerline{\includegraphics[width=2.5in]{mu_drate_12p.eps}\,\,\,\,\,\,
\includegraphics[width=2.5in]{tau_drate_12p.eps}}
\caption{$d\Gamma/d\hat s$ for (a) $\Lambda_{b}\rightarrow\Lambda(1115)\mu^{+}\mu^{-}$ and (b) $\Lambda_{b}\rightarrow\Lambda(1115)\tau^{+}\tau^{-}$ without and with long distance (LD) contributions from charmonium resonances. The solid curves represent rates obtained from SM1, the dashed curves are from SM2, and the dot-dashed curves are from SUSY.}\label{fig:rate12p}
\end{figure}

Figs. \ref{fig:rate12p}(a) and (b) show the differential decay rates for decays to the ground state ($\Lambda(1115)$) in the $\mu$ and $\tau$ channels, respectively. In these graphs, the solid curves are obtained from SM1, the dashed curves are from SM2, while the dot-dashed curves are from SUSY. The resonance contributions enhance the rates for both decay channels, but the enhancement is much larger in the dimuon case. The SM1a and SM2a branching ratios are somewhat smaller than the experimental measurement, but the SM1b and SM2b branching ratios are significantly larger. Experimentally, regions in $\sh$ around the first two charmonium resonances are vetoed. This is because the non-leptonic decay rates of the $\Lambda_b$ to $\Lambda$ along with a vector charmonium, followed by the dileptonic decays of the charmonium, are much larger than the rare decays of interest. Such decays can be described by tree-level diagrams, and thus have much larger decay rates than rare decays. To compare our numbers with experiment, we should follow a similar procedure of vetoing the regions around the two lightest charmonium resonances. When this is done, the branching ratios that result are essentially identical to those that we report in the `a' scenarios. Thus, SM1a and SM2a are the numbers that should be compared with experiment.

\begin{figure}
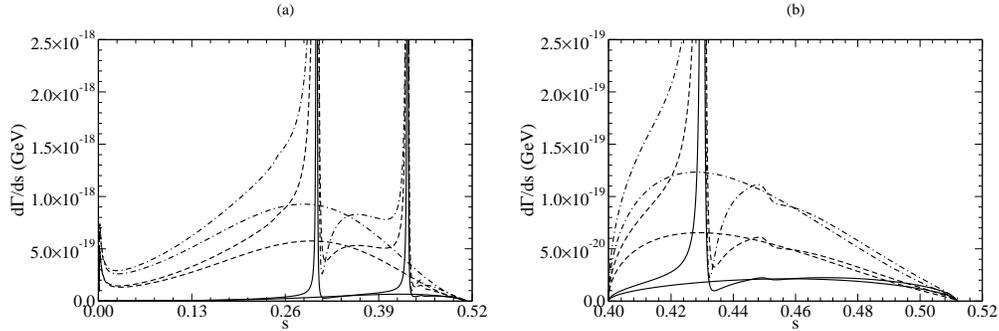

\centerline{\includegraphics[width=2.5in]{mu_drate_12p1r.eps}\,\,\,\,\,\, \includegraphics[width=2.5in]{tau_drate_12p1r.eps}}
\caption{Same as Fig. \ref{fig:rate12p} but for $\Lambda_{b}\rightarrow\Lambda(1600)$.}\label{fig:rate12p1r}
\end{figure}

Figure \ref{fig:rate12p1r} shows the differential decay rates for transitions to the first radial excitation, $\Lambda(1600)$. The SM1 predictions shown in Table \ref{br1} are much smaller than the SM2 predictions for both decay channels. Furthermore, SM2b predicts that decays to this state are the dominant rare decay mode of the $\Lambda_b$.  The truncations of the quark currents and the form factors in SM1 have lead to significant underestimates of the rates for decays to this state in both the $\mu$ and $\tau$ channels.

\subsubsection*{$\it\jp=1/2^-$}

\begin{figure}[t]
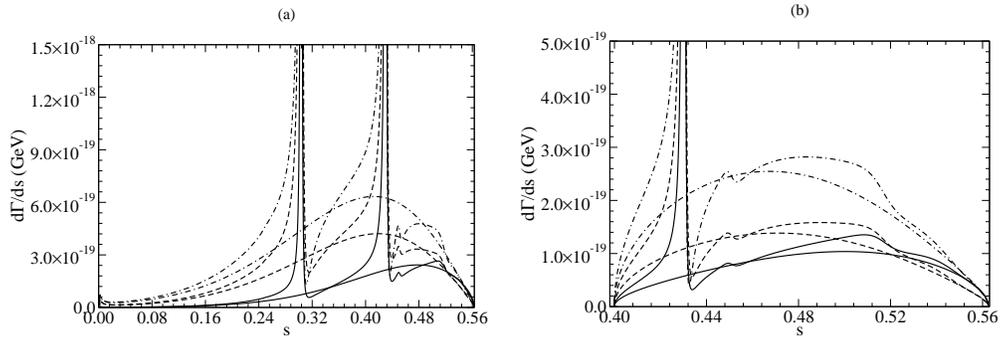

\centerline{\includegraphics[width=2.5in]{mu_drate_12m.eps}\,\,\,\,\,\, 
\includegraphics[width=2.5in]{tau_drate_12m.eps}}
\caption{Same as Fig. \ref{fig:rate12p} but for $\Lambda_{b}\rightarrow\Lambda(1405)$.}\label{fig:rate12m}
\end{figure}

Figs. \ref{fig:rate12m}(a) and (b) show the differential decay rates to the lowest-lying $1/2^-$ state, the $\Lambda(1405)$, assuming that it is a three-quark state (there is at least one other suggestion for the structure of this state in the literature \cite{l1405}). From the graphs, it can be seen that SM2 predicts a larger rate than SM1, and this is seen in Tables \ref{br1} and \ref{br2}. The SUSY rates are significantly larger in both decay channels over the entire kinematic range.

\subsubsection*{$\it\jp=3/2^-$}

\begin{figure}
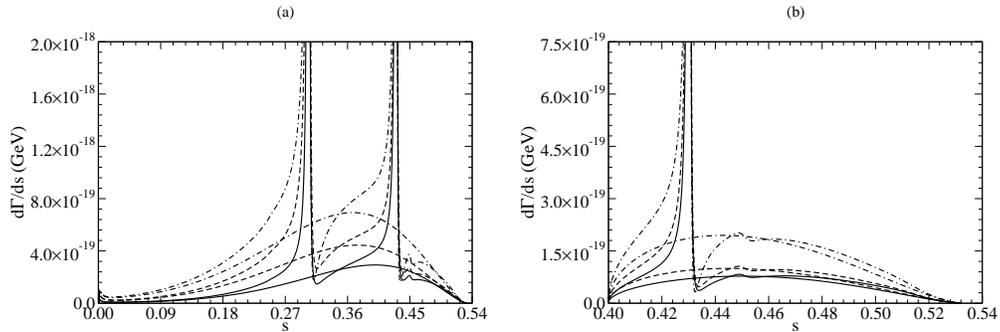

\centerline{\includegraphics[width=2.5in]{mu_drate_32m.eps}\,\,\,\,\,\,  \includegraphics[width=2.5in]{tau_drate_32m.eps}}
\caption{Same as Fig. \ref{fig:rate12p} but for $\Lambda_{b}\rightarrow\Lambda(1520)$.}\label{fig:rate32m}
\end{figure}

Figs. \ref{fig:rate32m}(a) and (b) show the differential decay rates for decays to $\Lambda(1520)$. The curves indicate that SM1 and SM2 make similar predictions for this channel, and this is borne out by the numbers in Tables \ref{br1} and \ref{br2}. The SUSY rates are larger in both the $\mu$ and $\tau$ channels. For this state, SM2b predicts that the decay rate into this state is very similar to the decay rate into the ground state.

\subsubsection*{$\it\jp=3/2^+$}

\begin{figure}[t]
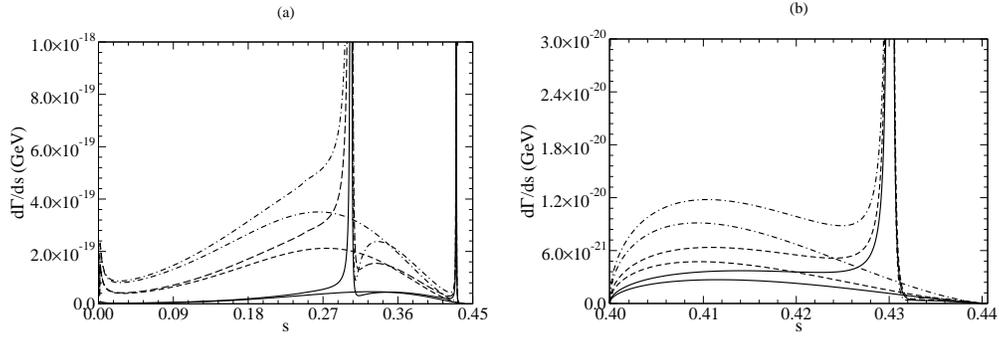

\centerline{\includegraphics[width=2.5in]{mu_drate_32p.eps}\,\,\,\,\,\,  \includegraphics[width=2.5in]{tau_drate_32p.eps}}
\caption{Same as Fig. \ref{fig:rate12p} but for $\Lambda_{b}\rightarrow\Lambda(1890)$.}\label{fig:rate32p}
\end{figure}

Figs. \ref{fig:rate32p}(a) and (b) show the differential decay rates for decays to $\Lambda(1890)$. In both the $\mu$ and $\tau$ channels, the predicted SM2 rates significantly larger than the SM1 rates. The SUSY rates are larger in both decay channels over the entire kinematic range. The SM2b rate for the $\mu$ channel into this mode is quite a bit smaller than the rate into the ground state. Nevertheless, this decay rate is not negligible.

\subsubsection*{$\it\jp=5/2^+$}

\begin{figure}
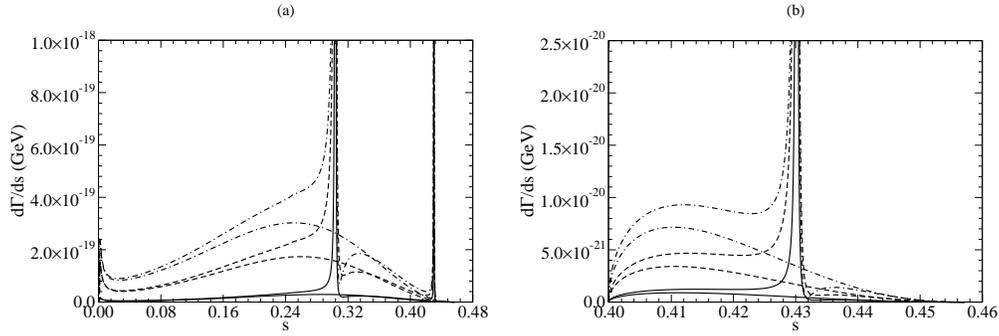

\vskip 0.5in
\centerline{\includegraphics[width=2.5in]{mu_drate_52p.eps}\,\,\,\,\,\,  \includegraphics[width=2.5in]{tau_drate_52p.eps}}
\caption{Same as Fig. \ref{fig:rate12p} but for $\Lambda_{b}\rightarrow\Lambda(1820)$.}\label{fig:rate52p}
\end{figure}

Figs. \ref{fig:rate52p}(a) and (b) show the differential decay rates for decays to $\Lambda(1820)$. As with many of the cases discussed, SM2 predicts larger rates for decays to this state than SM1, in the $\mu$ channel. In the $\tau$ channel, the predicted rates are negligible. The SUSY rates are larger in both decay channels over the entire kinematic range.

\subsubsection{Forward-Backward Asymmetries}

\begin{table}
\caption{Integrated forward-backward asymmetry for $\Lambda_b\rightarrow \Lambda^{(*)}\mu^{+}\mu^{-}$. The columns are labeled as in Table \ref{br1}.}
{\begin{tabular}{|c|c|c|c|c|c|c|}
\hline State, $J^{P}$ & LD& SM1 & SM2 & SUSY & Wang {\it et al.} \cite{wang}  & Chen {\it et al.} \cite{chen1}  \\ \hline
$\Lambda(1115)\,1/2^{+}$ & a & $-0.1255$ & $-0.1272$ & $-0.1800$ & $-0.0122$ & $-0.1338$ \\
& b & $-0.1128$ & $-0.1139$ & $-0.1582$ & $-0.0099$ & $-$ \\ \hline
$\Lambda(1600)\,1/2^{+}$ & a & $-0.0663$ & $-0.0889$ & $-0.1415$ & $-$ & $-$ \\
& b & $-0.0575$ & $-0.0793$ & $-0.1202$ & $-$ & $-$ \\ \hline
$\Lambda(1405)\,1/2^{-}$ & a & $0.0927$ & $0.1145$ & $0.1681$ & $-$ & $-$ \\
& b & $0.0822$ & $0.1021$ & $0.1455$ & $-$ & $-$ \\ \hline
$\Lambda(1520)\,3/2^{-}$ & a & $-0.0611$ &  $-0.0825$ & $-0.1371$ & $-$ & $-$ \\
& b & $-0.0566$ & $-0.0747$ & $-0.1165$ & $-$ & $-$ \\ \hline
$\Lambda(1890)\,3/2^{+}$ & a & $0.0410$ & $0.0586$ & $0.1116$ & $-$ & $-$  \\
& b & $0.0431$ & $0.0580$ & $0.0985$ & $-$ & $-$ \\ \hline
$\Lambda(1820)\,5/2^{+}$ & a & $-0.0074$ & $-0.0271$ & $-0.0946$ & $-$ & $-$  \\
& b & $-0.0133$ & $-0.0311$ & $-0.0846$ & $-$ & $-$ \\ \hline
\end{tabular}\label{afb1}}
\end{table}

\begin{table}
\caption{Integrated forward-backward asymmetry for $\Lambda_b\rightarrow \Lambda^{(*)}\tau^{+}\tau^{-}$. The columns are labeled as in Table \ref{br1}.}
{\begin{tabular}{|c|c|c|c|c|c|c|}
\hline State, $J^{P}$ &LD& SM1 & SM2 & SUSY & Wang {\it et al.} \cite{wang}  & Chen {\it et al.} \cite{chen1}  \\ \hline
$\Lambda(1115)\,1/2^{+}$ & a & $-0.0342$ & $-0.0330$ & $-0.0274$ & $-0.0067$ & $-0.0399$ \\
& b & $-0.0319$ & $-0.0307$ & $-0.0259$ & $-0.0062$ & $-$  \\ \hline
$\Lambda(1600)\,1/2^{+}$ & a & $-0.0119$ & $-0.0116$ & $-0.0098$ & $-$ & $-$ \\
& b & $-0.0106$ & $-0.0103$ & $-0.0088$ & $-$ & $-$  \\ \hline
$\Lambda(1405)\,1/2^{-}$ & a & $0.0221$ & $0.0243$ & $0.0197$ & $-$ & $-$  \\
& b & $0.0203$ & $0.0222$ & $0.0183$ & $-$ & $-$  \\ \hline
$\Lambda(1520)\,3/2^{-}$ & a & $-0.0072$ &  $-0.0098$ & $-0.0086$ & $-$ & $-$  \\
& b & $-0.0063$ & $-0.0086$ & $-0.0077$ & $-$ & $-$  \\ \hline
$\Lambda(1890)\,3/2^{+}$ & a & $0.0003$ & $0.0013$ & $0.0013$ & $-$ & $-$  \\
& b & $0.0002$ & $0.0010$ & $0.0010$ & $-$ & $-$  \\ \hline
$\Lambda(1820)\,5/2^{+}$ & a & $0.0021$ &  $0.0021$ & $0.0003$ & $-$ & $-$  \\
& b & $0.0020$ & $0.0021$ & $0.0003$ & $-$ & $-$  \\ \hline
\end{tabular}\label{afb2}}
\end{table}

\begin{table}
\caption{Zeroes of the forward-backward asymmetry for $\Lambda_b\rightarrow \Lambda^{(*)}\mu^{+}\mu^{-}$. The columns are labeled as in Table \ref{br1}.}
{\begin{tabular}{|c|c|c|c|c|c|}
\hline State, $J^{P}$ & LD& SM1 & SM2 & SUSY & Chen {\it et al.} \cite{chen2}  \\ \hline
$\Lambda(1115)\,1/2^{+}$ &a & $0.106$ & $0.098$ & $-$ & $0.109$ \\
&b & $0.088$ & $0.080$ & $-$ & $0.098$ \\ \hline
$\Lambda(1600)\,1/2^{+}$ &a & $0.143$ & $0.102$ & $-$ & $-$ \\
&b & $0.121$ & $0.085$ & $-$ & $-$ \\ \hline
$\Lambda(1405)\,1/2^{-}$ &a & $0.132$ & $0.107$ & $-$ & $-$ \\
&b & $0.111$ & $0.088$ & $-$ & $-$ \\ \hline
$\Lambda(1520)\,3/2^{-}$ &a & $0.117\,\,\,\,\,0.522$ &  $0.101\,\,\,\,\,0.525$ & $0.527$ & $-$ \\
&b & $0.097\,\,\,\,\,0.522$ & $0.085\,\,\,\,\,0.525$ & $0.527$ & $-$ \\ \hline
$\Lambda(1890)\,3/2^{+}$ &a & $0.116\,\,\,\,\,0.431$ & $0.103\,\,\,\,\,0.437$ & $0.439$ & $-$  \\
&b & $0.102\,\,\,\,\,0.430$ & $0.094\,\,\,\,\,0.430$ & $0.439$ & $-$ \\ \hline
$\Lambda(1820)\,5/2^{+}$ &a & $0.138\,\,\,\,\,0.415$ & $0.112\,\,\,\,\,0.424$ & $0.438$ & $-$  \\
&b & $0.121\,\,\,\,\,0.422$ & $0.099\,\,\,\,\,0.430$ & $0.441$ & $-$ \\ \hline
\end{tabular}\label{s01}}
\end{table}

\begin{table}
\caption{Zeroes of the forward-backward asymmetry for $\Lambda_b\rightarrow \Lambda^{(*)}\tau^{+}\tau^{-}$. The columns are labeled as in Table \ref{br1}.}
{\begin{tabular}{|c|c|c|c|c|}
\hline State, $J^{P}$ &LD& SM1 & SM2 & SUSY \\ \hline
$\Lambda(1520)\,3/2^{-}$ &a & $ 0.522$ &  $ 0.525$ & $0.527$ \\
&b & $0.522$ & $0.525$ & $0.527$ \\ \hline
$\Lambda(1890)\,3/2^{+}$ &a & $0.431$ & $0.437$ & $0.439$ \\
&b & $0.433$ & $0.437$ & $0.438$ \\ \hline
$\Lambda(1820)\,5/2^{+}$ &a & $0.415$ & $0.425$ & $0.439$ \\
&b & $0.412$ & $0.430$ & $0.440$ \\ \hline
\end{tabular}\label{s02}}
\end{table}

The differential forward-backward asymmetries (FBAs) ${\cal A}_{FB}(\sh)$ are shown in  Figs. \ref{fig:asym12p}-\ref{fig:asym52p}. The key to the curves is the same as the differential decay rates. In addition to the differential asymmetries, the zeroes in the asymmetries also contain information on the Wilson coefficients, and are therefore of interest. We will discuss these in some detail. Furthermore, it is useful to introduce the integrated forward-backward asymmetry $\<{\cal A}_{FB}\>$ in order to characterize the typical value of the FBA. This integrated FBA is defined as
\beq
\<{\cal A}_{FB}\>=\int_{4\mlh^2}^{(1-\sqrt{r})^2}{\cal A}_{FB}(\sh)d\sh.
\eeq

The integrated FBAs we obtain are shown in Tables \ref{afb1} and \ref{afb2}. The column labels have the same meaning as with the branching ratios shown in Tables \ref{br1} and \ref{br2}. We also compare our results with those of Wang {\it et. al.} \cite{wang} (LCSR) and Chen {\it et al.} \cite{chen1} (QCDSR). Tables \ref{s01} and \ref{s02} show the locations of the zeroes in the FBAs.

\subsubsection*{$\it\jp=1/2^+$}

\begin{figure}[t]
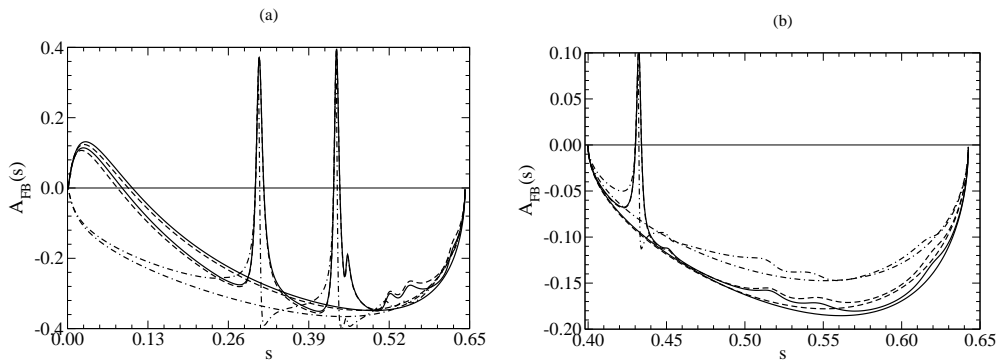

\centerline{\includegraphics[width=2.5in]{mu_fba_12p.eps}\,\,\,\,\,\,
\includegraphics[width=2.5in]{tau_fba_12p.eps}}
\caption{${\cal A}_{FB}(\hat s)$ for (a) $\Lambda_{b}\rightarrow\Lambda(1115)\mu^{+}\mu^{-}$ and (b) $\Lambda_{b}\rightarrow\Lambda(1115)\tau^{+}\tau^{-}$. The solid curves arise from the SM1 model, the dashed curves from SM2, and the dot-dashed curves from the SUSY scenario.}\label{fig:asym12p}
\end{figure}

\begin{figure}
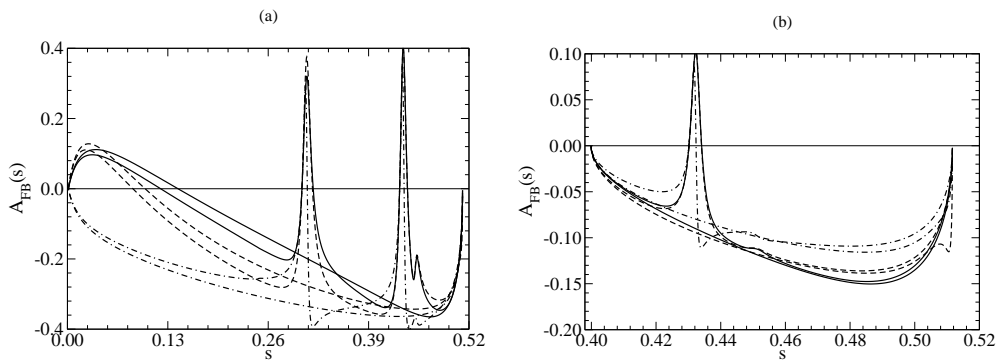

\centerline{\includegraphics[width=2.5in]{mu_fba_12p1r.eps}\,\,\,\,\,\, \includegraphics[width=2.5in]{tau_fba_12p1r.eps}}
\caption{Same as Fig. \ref{fig:asym12p} but for $\Lambda_{b}\rightarrow\Lambda(1600)$.}\label{fig:asym12p1r}
\end{figure}

Figs. \ref{fig:asym12p}(a) and (b) show the differential FBAs for decays to the ground state, in the $\mu$ and $\tau$ channels, respectively. The values of the integrated asymmetries, $\<{\cal A}_{FB}\>$, obtained in the SM1a and SM2a  models, agree well with the values reported by Chen {\it et al.} \cite{chen1} for both channels, but is roughly an order of magnitude larger than the value reported by Wang {\it et al.} \cite{wang}. Our values from the SM1b and SM2b models are also an order of magnitude larger than the corresponding value reported by Wang {\it et al.}  It is very interesting to note that for this final state, the integrated asymmetry appears to be largely independent of the model employed. When the results of other authors are considered, the results of Chen {\it et al.} confirm this conclusion, but the results of Wang {\it et al.} do not allow such a conclusion to be drawn. The curves of Figs. \ref{fig:asym12p}(a) and (b) also indicate that the FBA for decays to this state are largely independent of the form factors used.

The FBAs for decays to $\Lambda(1600)$ for the $\mu$ and $\tau$ channels are shown in Figs. \ref{fig:asym12p1r}(a) and (b), respectively. In the $\mu$ channel, the SM1 and SM2 models lead to significantly different FBAs over most of the kinematic range, and this is reflected in the values of the integrated FBAs. In addition, the locations of the zeroes are quite different in the two models. In the $\tau$ channel, the curves from the SM1 and SM2 models are closer than in the $\mu$ channel, and the integrated FBAs are essentially identical.

From Eq. \ref{eq:afb2}, the zeroes of the FBA occur when
\beq
{\cal A}_{FB}(\sh_0)=0\Rightarrow \I_1(\sh_0)=0.
\eeq
For decays to states with $J=1/2$, the expression for $\I_1$ is 
\beq
\I_1(\sh)=-16\mlb^4\sh\sqrt{\phi(\sh)\psi(\sh)}\left[\Re(A^{*}_1E_1)+\Re(B^{*}_1D_1)\right].
\eeq
Using Eq. \ref{eq:natpar_dilep}, the condition for the position of the zero(s) for decays to states with $\jp=1/2^+$ is
\beq
\Re(C_9^{*}C_{10})=\frac{2\mbh}{\sh_0}\Re(C_7^{*}C_{10})\bigg(\frac{F^T_1G_1-G^T_1F_1}{2\mlb F_1G_1}\bigg).
\label{eq:zero12}
\eeq
This relation holds for $\jp=1/2^-$ as well. From this relation, we see that $\sh_0$ depends on the two combinations of Wilson coefficients, $\Re\left( C_7^{*}C_{10}\right)$ and $\Re \left(C_9^{*}C_{10}\right)$, and a ratio of form factors. For this final state, there is a single possible zero when LD contributions are omitted: LD contributions introduce other zeroes, as can be clearly seen in Figs. \ref{fig:asym12p}(a) and \ref{fig:asym12p1r}(a). Using the form factors obtained in the two models we consider, the values of $\sh_0$ are shown in Table \ref{s01} for the $\mu$ channel. These values are in good agreement with those reported in \cite{chen2}.

For decays to states with $J=1/2$, apart from the endpoints and resonance regions, there are no zeroes in the FBAs in either channel for the SUSY model we use. The zeroes that occur in the SM scenarios can be traced to the opposite signs of $C_7$ and $C_9$. In the SUSY model that we employ here,  $C_7$ and $C_9$ have the same sign, and the condition for the zero in Eq. \ref{eq:zero12} can no longer be satisfied. In addition, in the $\tau$ channel, even in the SM1 and SM2 scenarios, no zeros are possible, apart from those induced by the resonance effects.

\subsubsection*{$\it\jp=1/2^-$}

\begin{figure}
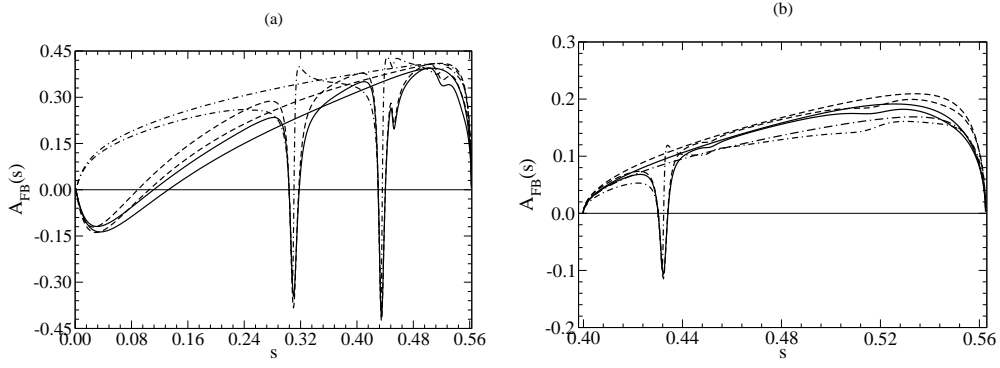

\centerline{\includegraphics[width=2.5in]{mu_fba_12m.eps}\,\,\,\,\,\, 
\includegraphics[width=2.5in]{tau_fba_12m.eps}}
\caption{Same as Fig. \ref{fig:asym12p} but for $\Lambda_{b}\rightarrow\Lambda(1405)$.}\label{fig:asym12m}
\end{figure}

The FBAs predicted for decays to the $\Lambda(1405)$ are shown in Fig. \ref{fig:asym12m}. In these figures, we see that the predictions from SM1 and SM2 are somewhat different, and the locations of the zero are also different (see Table \ref{s01}).  As with the decay to states with $J^P=1/2^+$, there are no zeros in the FBA in the SUSY scenario we consider here. There are also no zeros, apart from those induced by the resonance contributions, in the $\tau$ channel for any of the scenarios considered.

\subsubsection*{$\it\jp=3/2^-$}

\begin{figure}
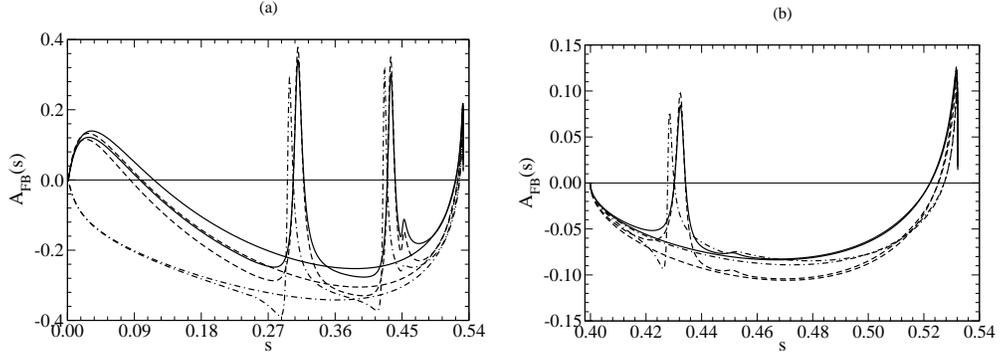

\centerline{\includegraphics[width=2.5in]{mu_fba_32m.eps}\,\,\,\,\,\,  \includegraphics[width=2.5in]{tau_fba_32m.eps}}
\caption{Same as Fig. \ref{fig:asym12p} but for $\Lambda_{b}\rightarrow\Lambda(1520)$.}\label{fig:asym32m}
\end{figure}

The predictions for the FBAs in the decays to the $\Lambda(1520)$ are shown in Fig. \ref{fig:asym32m}. The SM1 and SM2 models give slightly different asymmetries in the $\mu$ channel. Even without the LD contributions, the structure of the asymmetry arising from these decays is richer than in the decays to states with $J=1/2$.

For states with $J=3/2$
\beqy
\I_1(\sh)\=\frac{8}{3}\mlb^4\sh\sqrt{\phi(\sh)\psi(\sh)}\bigg[-\frac{\phi(\sh)}{r}\left[\Re(A^{*}_1E_1)+\Re(B^{*}_1D_1)\right]+\nn\\&&\ym(\sh)\left[\Re(A^{*}_1E_4)+\Re(B^{*}_4D_1)\right]+\yp(\sh)\left[\Re(A^{*}_4E_1)+\Re(B^{*}_1D_4)\right]+\nn\\&&2\left[\Re(A^{*}_4E_4)+\Re(B^{*}_4D_4)\right]\bigg],
\eeqy
where $\phi(\sh)=(1-r)^2-2(1+r)\sh+\sh^2$ and $y^\pm(\sh)=\left[(1\pm\sqrt{r})^2-\sh\right]/\sqrt{r}$. For $J^P=3/2^-$, the condition for the positions of the zeros is
\beq
\Re(C_9^{*}C_{10})=\frac{2\mbh}{\sh_0}\Re(C_7^{*}C_{10})\bigg(\frac{X_1}{2\mlb Y_1}\bigg),
\label{eq:zero32}
\eeq
where
\beqy
X_1\=-\frac{\phi(\sh_0)}{r}\left(F^T_1G_1-G^T_1F_1\right) +\ym(\sh_0)\left(F^T_1G_4-G^T_4F_1\right) +\nn\\&&\yp(\sh_0)\left(F^T_4G_1-G^T_1F_4\right) +2\left(F^T_4G_4-G^T_4F_4\right),\nn\\
Y_1\=-\frac{\phi(\sh_0)}{r}F_1G_1+\ym(\sh_0) F_1G_4+\yp(\sh_0) F_4G_1+2F_4G_4,\label{eq:zero32m}
\eeqy
As we can see in Fig. \ref{fig:asym32m}, apart from the resonance region, there are two zeroes for this mode in the $\mu$ channel in the SM. The zero at the larger value of $\sh$ is also present in the $\tau$ channel. This is quite different from the case with $J=1/2$ where there is only one zero in the $\mu$ channel and none for the $\tau$. The positions of the zeroes in the $\mu$ channel are shown in Table \ref{s01}. In the SUSY scenario that we explore, there is only one zero in the $\mu$ channel, and it sits at $\sh=0.527$ in both SUSY scenarios. In the $\tau$ channel, there is a single zero in the FBA, and its position is largely independent of the model used for the form factors.

\subsubsection*{$\it\jp=3/2^+$}

\begin{figure}
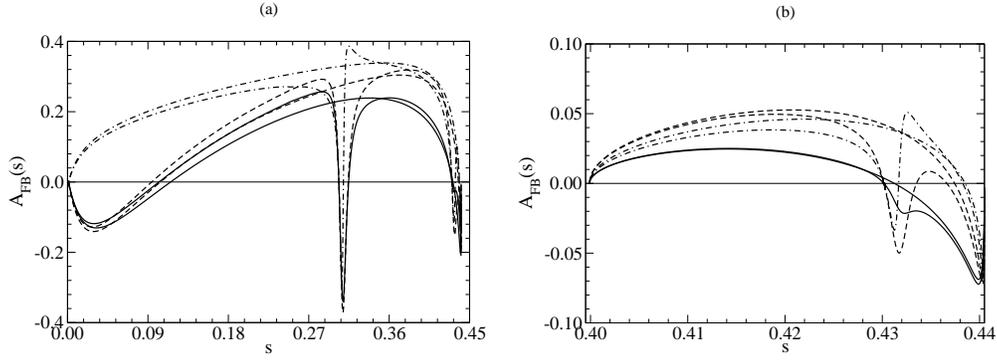

\centerline{\includegraphics[width=2.5in]{mu_fba_32p.eps}\,\,\,\,\,\,  \includegraphics[width=2.5in]{tau_fba_32p.eps}}
\caption{Same as Fig. \ref{fig:asym12p} but for $\Lambda_{b}\rightarrow\Lambda(1890)$.}\label{fig:asym32p}
\end{figure}

Fig. \ref{fig:asym32p}(a) shows the $\mu$ channel FBAs for the $\Lambda(1890)$ mode. There are some differences between the predictions of the SM1 and SM2 models, particularly at the larger values of $\sh$. These differences appear to be even larger in the $\tau$ channel (Fig. \ref{fig:asym32p}(b)), but the asymmetries predicted are quite small in both models.

For this decay mode, the condition for the zeroes in the asymmetry is the same as in Eq. \ref{eq:zero32}, but now
\beqy
X_1\=-\frac{\phi(\sh_0)}{r}\left(F^T_1G_1-G^T_1F_1\right) +\ym(\sh_0)\left(F^T_4G_1-G^T_1F_4\right) +\nn\\&&\yp(\sh_0)\left(F^T_1G_4-G^T_4F_1\right) +2\left(F^T_4G_4-G^T_4F_4\right),\nn\\
Y_1\=-\frac{\phi(\sh_0)}{r}F_1G_1+\ym(\sh_0) F_4G_1+\yp(\sh_0) F_1G_4+2F_4G_4,\label{eq:zero32p}
\eeqy
The positions of the zeroes in the FBA for decays to this state are also shown in Table \ref{s01} for the $\mu$ channel and Table \ref{s02} for the $\tau$ channel.

\subsubsection*{$\it\jp=5/2^+$}

\begin{figure}
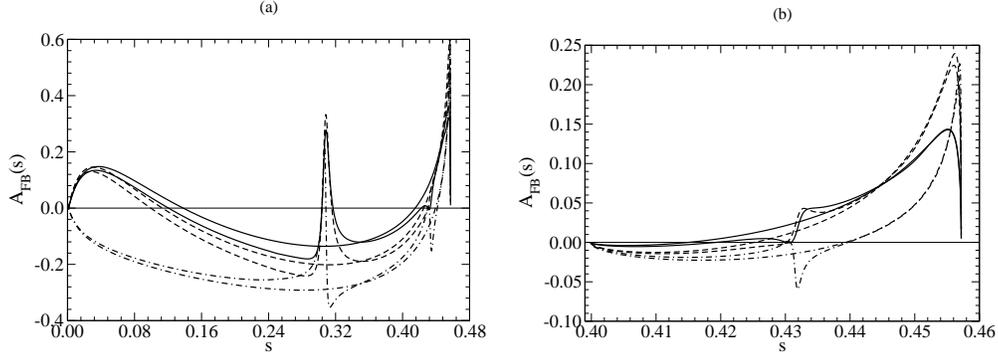

\centerline{\includegraphics[width=2.5in]{mu_fba_52p.eps}\,\,\,\,\,\,  \includegraphics[width=2.5in]{tau_fba_52p.eps}}
\caption{Same as Fig. \ref{fig:asym12p} but for $\Lambda_{b}\rightarrow\Lambda(1820)$.}\label{fig:asym52p}
\end{figure}

In the $\Lambda(1820)$ mode, SM1 and SM2 again predict slightly different FBAs in both the $\mu$ and $\tau$ channels (see Fig. \ref{fig:asym52p}). In the $\mu$ channel, the SM1 prediction for the position of the first zero is noticeably different from the prediction from SM2. However, the positions of the second zero are much closer in these two models.
 
For $J=5/2$, $\I_1$ takes the form
\beqy
\I_1(\sh)\=\frac{2}{5r}\mlb^4\sh\phi^{3/2}(\sh)\sqrt{\psi(\sh)}\bigg[-\frac{\phi(\sh)}{r}\left[\Re(A^{*}_1E_1)+\Re(B^{*}_1D_1)\right]-\nn\\&&2\left[\Re(A^{*}_1E_4)+\Re(B^{*}_4D_1)\right]+2\left[\Re(A^{*}_4E_1)+\Re(B^{*}_1D_4)\right]+\nn\\&&\Re(A^{*}_4E_4)+\Re(B^{*}_4D_4)\bigg].
\eeqy
The condition for the zero now becomes
\beq
\Re(C_9^{*}C_{10})=\frac{2\mbh}{\sh_0}\Re(C_7^{*}C_{10})\bigg(\frac{X_2}{2\mlb Y_2}\bigg),
\label{eq:zero52}
\eeq
where, for natural parity,
\beqy
X_2\=-\frac{\phi(\sh_0)}{r}\left(F^T_1G_1-G^T_1F_1\right) -2\left(F^T_1G_4-G^T_4F_1\right) +2\left(F^T_4G_1-G^T_1F_4\right) +\nn\\&&F^T_4G_4-G^T_4F_4,\nn\\
Y_2\=-\frac{\phi(\sh_0)}{r}F_1G_1-2 F_1G_4+2 F_4G_1+F_4G_4.\label{eq:zero52p}
\eeqy

We conclude this section with two general comments on the zeroes of the FBAs. The zeroes for the FBAs in decays to states with spin 1/2, along with the first zero in decays to the states with higher spin that we have considered, all lie relatively close to each other, despite the more complicated expression for the location of the zeroes for the states with higher spin. However, in Eqs. \ref{eq:zero32m}, \ref{eq:zero32p} and \ref{eq:zero52p}, terms in $F_4$, $G_4$, $F_4^T$ and $G_4^T$ are negligible (they are exactly zero in the limit of an infinitely heavy $b$ quark), so that the various expressions for the location of the zeroes all reduce to one that is identical to the case of spin 1/2, Eq. \ref{eq:zero12}. Furthermore
\beq
\frac{F^T_1G_1-G^T_1F_1}{2\mlb F_1G_1}\approx 1+{\cal O}\left(\frac{\xi_2}{\xi_1}\right)+{\cal O}\left(\frac{\Lambda_{\rm QCD}}{m_b}\right),
\eeq
for states with natural parity, or
\beq
\frac{F^T_1G_1-G^T_1F_1}{2\mlb F_1G_1}\approx 1+{\cal O}\left(\frac{\zeta_2}{\zeta_1}\right)+{\cal O}\left(\frac{\Lambda_{\rm QCD}}{m_b}\right),
\eeq
for states with unnatural parity, for the states we have examined. This means that the location of the zero, up to corrections ${\cal O}\left(\frac{\Lambda_{\rm QCD}}{m_b}\right)$ and ${\cal O}\left(\frac{\xi_2}{\xi_1}\right)$ or ${\cal O}\left(\frac{\zeta_2}{\zeta_1}\right)$, is approximately given by
\beq
\sh_0\approx -2\mbh\frac{\Re(C_7^{*}C_{10})}{\Re(C_9^{*}C_{10})},\label{eq:global0}
\eeq
independent of the angular momentum of the final state (at least, up to spin $5/2$), and of the form factors. Using the SM Wilson coefficients along with the physical mass of the $\Lb$ and the accepted mass of the $b$ quark, this gives a value of 0.121. This number, obtained in this simplifying limit, is in surprisingly good agreement with the values of the lower zeroes shown in Table \ref{s01}, for all states. Of course, there must be deviations from this simple limit, as the $b$ quarks is not infinitely heavy. However, it may be possible to systematically estimate the corrections to the (model independent?) value of 0.121.

The second comment on the zeroes follows immediately from the first: in the limit of an infinitely heavy $b$ quark, the location of the zeros is given by Eq. \ref{eq:global0} (up to the corrections mentioned), since $F_4$, $G_4$, $F_4^T$ and $G_4^T$ all vanish explicitly in this limit. This means that in this limit, there can only be one zero in the FBAs if the form factors are treated in the strict heavy-quark limit. This also means that the location of this second zero may be sensitively dependent on the form factors, the angular momentum of the state being considered and, of course, on the Wilson coefficients. Surprisingly, the results shown in Table \ref{s02} suggest that the most important dependence is the angular momentum of the daughter baryon.

\begin{figure}
\vskip 0.5in
\centerline{\includegraphics[width=3.5in]{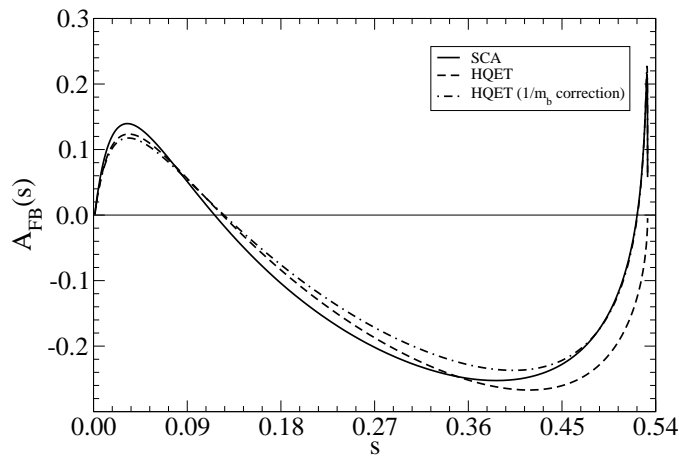}}
\caption{FBA for $\Lambda_{b}\rightarrow\Lambda(1520)$, obtained in model SM1a. The solid curve arises from the SCA form factors. The dashed curve is obtained by omitting any terms that are $\O(1/m_b)$ from the SCA form factors. The dot-dashed curve is obtained by allowing $F_4$ to be non-zero.}\label{fig:asym32mhqet}
\end{figure}

This is illustrated in Fig. \ref{fig:asym32mhqet}. In this graph, the solid curve arises from the SCA form factors. The dashed curve is obtained by omitting any terms that are $\O(1/m_b)$ from the SCA form factors. This means, for instance that $F_4$, $G_4$, $F_4^T$ and $G_4^T$ all vanish identically. It's clear that for these truncated form factors, there is only a single zero in the FBA. The dot-dashed curve arises when only $F_4$ is allowed to be non-zero, and the second zero reappears in the FBA.

\subsection{Sensitivity to Model Parameters}

One question that can be asked is how sensitive are the results presented above to the choice of parameters in the model. The easiest way to assess this is to vary one or more parameters from the values we have used, and note the effect on the calculated decay rates, for instance. We have carried out this exercise, changing the mass of the $b$ quark by 10\% from its value reported in Table \ref{hampar}. The result is that the decay rate to the ground state $\Lambda$ changes by less than 0.02\% of the reported value, in either model of the form factors. For decays to other states, the changes in the decay rate are similarly small. This suggests that the results that we have obtained are quite robust to changes in parameters, {\em within the context of this particular model}. It is extremely difficult for us to speculate on how the results would change if the model itself were to be changed. We should also emphasize that the parameters reported in Table \ref{hampar} are obtained in a fit to the baryon spectrum. A change in the mass of the $b$ quark of 10\% means that the model states are no longer good representations of the observed physical states.

\section{Conclusions and Outlook\label{sec:concl}}

In this work, we have examined the rare weak dileptonic decays of the $\Lb$ baryon using a constituent quark model. Analytic and numerical results for the form factors for the decays to ground and excited states with different quantum numbers  have been obtained and compared with the leading order predictions of HQET. We have shown that for $\Lb\to\Ls^{(*)}$, the leading order relations among form factors predicted by HQET are satisfied by the form factors for both the SCA and MCN models. For transitions considered, MCN model form factors are found to be larger than those in the SCA model.

We have examined the decay rates and FBAs for decays to a number of $\Lambda$ final states, in a number of scenarios. We have compared our results with the predictions of other authors, as well as the $\mu$-channel measurement reported by the CDF collaboration. Our predictions for decays to the ground state are smaller than the experimental value and the predictions of other authors in the $\mu$ channel. In the $\tau$ channel, our model results are in agreement with QCD sum rules and pole model predictions without LD contributions. However, when LD contributions are considered, our results are smaller than those of other authors. We comment on this further near the end of this section.

The SM2 models predict larger branching ratios for all daughter $\Ls$ decay modes than do the SM1 models. SM1 predicts that decays to the ground state are dominant in the $\mu$ channel. In SM2a, this mode is also dominant, but in SM2b, decays to $\Ls(1600)$ are dominant. SM2b also predicts that the decay rate to the $\Ls(1520)$ is comparable with the decay rate to the ground state, with the decay rate to the $\Ls(1405)$ being only slightly smaller. These results are consistent with the current status of the rare dileptonic decays of $B$ mesons, where decays to the two lowest lying kaons are insufficient to saturate the inclusive rate, and in fact account for less than 50\% of the inclusive rate. These results also suggest that it might be prudent for the LHCb collaboration and CDF to search for rare decays in modes other than the $\Ls(1115)$.

There is, however, a very important caveat. Decays in which the $q^2$ of the dileptons is near the mass-squared of the two lowest-lying vector charmonium resonances are inaccessible experimentally, as they are embedded in the much larger background coming from (tree-level) nonleptonic decays with vector charmonia in the final state. Our results obtained omitting the vector charmonia, the SM1a and SM2a models, are therefore closer to experimental reality, and these scenarios suggest that decays to the $\Lambda(1115)$ are indeed the dominant rare decay mode of the $\Lambda_b$. Nevertheless, the SM2a model that uses the more precise calculation of the form factors indicates that there will be a sizable fraction of rare decays into excited states of the $\Lambda$.

For some decay modes, such as decays to the $\Ls(1115)$, the FBAs are largely independent of the model choice for much of the kinematic range. For other modes, particularly the $\Ls(1600)$, there are significant differences between the predictions of the SM1 and SM2 models. The zero that occurs at lower values of $\sh$ in these FBAs turns out to be largely independent of the angular momentum of the daughter baryon, but with some dependence on the form factors. Nevertheless, this form factor dependence is surprisingly small. For decays to states with $J\geq3/2$, these FBAs in general have more than one zero (for the states we have considered, there is one additional zero). At leading order in HQET, the second zero does not exist. For decays to the ground state, it is known that the positions of the zeroes are modified in many scenarios that arise beyond the SM. This is also found to be true of the excited states, not surprisingly. Thus, it would be crucial to know the number of zeroes and their positions reliably (assuming that there will ever be sufficient statistics for the differential FBAs to be extracted with high precision). Our results indicate that the leading order predictions of HQET are misleading in this regard.

The work presented here can be extended in several directions. We can use these form factors to study any number of polarization observables that can arise in these decays for leptons and baryons. Like the FBAs, these asymmetries are functions of the ratio of form factors and may therefore be less sensitive to the details of the form factor models. Such observables could provide more model-independent ways of extracting some of the Wilson coefficients.

In comparing our rates with those of other authors, the main weakness of our model is highlighted, namely that the form factors predicted drop too rapidly as the momentum of the daughter baryon increases. This makes these form factors less reliable for nonleptonic decays such as $\Lb\to\Ls^{(*)}\pi$ or the radiative decays $\Lb\to\Ls^{(*)}\gamma$, in which $q^2$ is small or zero. This is an inherent problem with models of this kind that use the harmonic oscillator basis. We need to emphasize that this is NOT an inherent flaw of all quark models, but of the particular choice of basis. Furthermore, this weakness is exacerbated by the larger range of $q^2$ accessible in the rare decays of the $\Lambda_b$. It would be interesting to perform these calculations using wave functions computed from other bases, such as the Sturmian basis. The Sturmian basis leads to form factors whose kinematic dependence on $q^2$ appears in the form of multipoles. This has two significant advantages. First, it is common to expect form factors to have multipole dependence experimentally. Second, multipole forms would decrease significantly less rapidly as the momentum of the daughter baryon increases. These forms can therefore be expected to lead to larger branching fractions for many of the decay modes that we have considered. The semianalytic method developed here using the harmonic oscillator basis can be easily adapted for the Sturmian basis.

The sensitivity of the rare decays we have studied to new physics that arises beyond the Standard Model can be easily studied with the form factors that we have calculated herein. The operator product structure easily incorporates scenarios from beyond the Standard Model, such as supersymmetric models, models with fourth generation quarks, and models with universal extra dimensions. These types of investigations can help provide constraints on any possible new physics scenarios.

\section*{Acknowledgment} We gratefully acknowledge the support of the Department of Physics, the College of Arts and Sciences, and the Office of Research at Florida State University.
This research is supported by the U.S. Department of Energy under contracts 
DE-SC0002615.

\appendix

\section{Semianalytic Treatment of Hadronic Matrix Elements\label{sec:shme}}

The hadronic matrix elements (Eq. \ref{eq:qmme}) can be written in the form
\beqy
\langle \Lambda\mid\overline{s}\Gamma b\mid\Lambda_b\rangle&=&\sum_{h',h} a_{h'}^{\Lambda^*} a_{h}^{\Lambda_b} \delta_{s_{1}'s_{1}}\delta_{s_{2}'s_{2}} (-1)^{l_{\lambda'}+l_{\lambda}} \nn\\&&\times U_{n_\rho l_\rho m_\rho}^{n_{\rho'} l_{\rho'} m_{\rho'}}(\alpha_\rho,\alpha_{\rho'})W_{\Gamma;n_\lambda l_\lambda m_\lambda s_q}^{n_{\lambda'} l_{\lambda'} m_{\lambda'} s_{q'}}(\alpha_\lambda,\alpha_{\lambda'}),
\eeqy
where the coefficients $a_{h(h')}$ are the products of the normalization of the baryon state (the $\sqrt{2E_{B_q}}$ of Eq. \ref{eq:bwf}), the expansion coefficients (the $\eta_i^{B_{q}}$ of Eq. \ref{eq:mswf1}) and the various Clebsch-Gordan coefficients that appear in the parent (daughter) baryon wave function, and the indices $h(h')$ contain all the relevant quantum numbers being summed over for the parent (daughter) baryon state. $U_{n_\rho l_\rho m_\rho}^{n_{\rho'} l_{\rho'} m_{\rho'}}$ is the spectator overlap given by 
\begin{equation}
U_{n_\rho l_\rho m_\rho}^{n_{\rho'} l_{\rho'} m_{\rho'}}(\alpha_\rho,\alpha_{\rho'})=\int d^{3} p_\rho \phi_{n_{\rho'} l_{\rho'} m_{\rho'}}^{*}(\alpha_{\rho'};\vec p_{\rho}) \phi_{n_\rho l_\rho m_\rho}(\alpha_\rho;\vec p_\rho),
\end{equation}
and $W_{\Gamma;n_\lambda l_\lambda m_\lambda s_q}^{n_{\lambda'} l_{\lambda'} m_{\lambda'} s_{q'}}$ is the interaction overlap, which takes the form
\begin{eqnarray}
W_{\Gamma;n_\lambda l_\lambda m_\lambda s_q}^{n_{\lambda'} l_{\lambda'} m_{\lambda'} s_{q'}}(\alpha_\lambda,\alpha_{\lambda'})&=&\exp\left(-\frac{3m_q^2}{2\widetilde m_\Lambda^2}\frac{p_\Lambda^2}{\alpha_{\lambda\lambda'}^2}\right) \int d^{3} k e^{-\kappa_\lambda^2 k^2} {\cal L}_{n_{\lambda'}}^{l_{\lambda'}+\frac{1}{2}*}\left(\frac{p'^2}{\beta'^2}\right){\cal Y}_{l_{\lambda'} m_{\lambda'}}^{*}(\vec p') \nonumber \\ &&  \times \langle s(\vec p_\Lambda+\vec p,s_{q'})\mid\overline{s}\Gamma b\mid b(\vec p,s_q)\rangle  {\cal L}_{n_{\lambda}}^{l_{\lambda}+\frac{1}{2}}\left(\frac{p^2}{\beta^2}\right){\cal Y}_{l_{\lambda} m_{\lambda}}(\vec p).
\label{eq:wi1}
\end{eqnarray}
Here, $\kappa_\lambda=\alpha_{\lambda\lambda'}/\alpha_{\lambda'} \alpha_\lambda$ and $\alpha_{\lambda\lambda^\prime}=\sqrt{(\alpha_{\lambda}^{2}+\alpha_{\lambda^\prime}^{2})/2}$.

From the power series definition of the Laguerre polynomials, the integral over $\vec p_\rho$ can be done analytically to give
\begin{eqnarray}
U_{n_\rho l_\rho m_\rho}^{n_{\rho'} l_{\rho'} m_{\rho'}}(\alpha_\rho,\alpha_{\rho'})&=&\delta_{l_{\rho'} l_\rho} \delta_{m_{\rho'} m_\rho} N_{n_{\rho'} l_\rho}^{*}(\alpha_{\rho'}) N_{n_\rho l_\rho}(\alpha_\rho) \nn\\&&\times \sum_{r'=0}^{n_{\rho'}}\sum_{r=0}^{n_\rho}\frac{c_{n_{\rho'} r'}^{l_\rho+\frac{1}{2}}c_{n_\rho r}^{l_\rho+\frac{1}{2}}}{\alpha_{{\rho'}}^{2r'}\alpha_{\rho}^{2r}} {\cal I}\left(2(l_\rho+r'+r+1);\kappa_{\rho}^{2}\right),
\end{eqnarray}
where $N_{n l}(\alpha)$ are the normalization constants for the oscillator basis, $c_{n r}^{k}$ are the coefficients in the power series definition of the Laguerre polynomials,
\begin{equation}
c_{nr}^{k}=(-1)^r\frac{\Gamma(n+k+1)}{\Gamma(n-r+1)\Gamma(k+r+1)\Gamma(r+1)}.
\end{equation}
${\cal I}$ is the Gaussian integral,
\begin{equation}
{\cal I}(m;b)=\int_{0}^{\infty} x^m e^{-b x^2} dx=\frac{\Gamma((m+1)/2)}{2b^{(m+1)/2}},
\end{equation}
$\kappa_\rho=\alpha_{\rho\rho'}/\alpha_{\rho'} \alpha_\rho$, $\alpha_{\rho\rho'}=\sqrt{(\alpha_{\rho}^{2}+\alpha_{\rho'}^{2})/2}$, and the Kronecker deltas $\delta_{\lrp\lr},\,\, \delta_{\mrp\mr}$ are due to the orthogonality of the spherical harmonics.

We can write the solid harmonics in Eq. \ref{eq:wi1} in terms of solid harmonics over $\vec k$ by making use of the addition theorem,
\begin{equation}
{\cal Y}_{l m}(\vec a+\vec b)=\sum_{\lambda=0}^{l} \sum_{\mu_\lambda=-\lambda}^{\lambda} B_{l m}^{\lambda \mu_\lambda}(\vec a) {\cal Y}_{\lambda \mu_\lambda}(\vec b),
\label{eq:addth}
\end{equation}
where
\begin{eqnarray}
B_{l m}^{\lambda \mu_\lambda}(\vec a)&=&\frac{4\pi (2l+1)!!}{(2\lambda+1)!!(2l-2\lambda+1)!!} {\cal C}_{\lambda \mu_\lambda l-\lambda m-\mu_\lambda}^{l m} {\cal Y}_{l-\lambda m-\mu_\lambda}(\vec a),
\label{eq:bcoef} \\
{\cal C}_{l_1 m_1 l_2 m_2}^{l_3 m_3}&=&\langle Y_{l_3 m_3}\mid Y_{l_2 m_2}\mid Y_{l_1 m_1}\rangle,
\end{eqnarray}
and $Y_{l m}$ are the usual spherical harmonics. Eq. \ref{eq:wi1} therefore becomes
\begin{eqnarray}
W_{\Gamma;n_\lambda l_\lambda m_\lambda s_q}^{n_{\lambda'} l_{\lambda'} m_{\lambda'} s_{q'}}(\alpha_\lambda,\alpha_{\lambda'})&=&\exp\left(-\frac{3m_q^2}{2\widetilde m_\Lambda^2}\frac{p_\Lambda^2}{\alpha_{\lambda\lambda'}^2}\right) \nn\\&&\times \sum_{\lambda'\mu_{\lambda'}}\sum_{\lambda\mu_\lambda} B_{l_{\lambda'} m_{\lambda'}}^{\lambda' \mu_{\lambda'} *}(c'\vec p_{\Lambda}) B_{l_{\lambda} m_{\lambda}}^{\lambda \mu_{\lambda}}(c\vec p_{\Lambda}) {\cal J}_{\Gamma;n_\lambda l_\lambda s_q:\lambda \mu_\lambda}^{n_{\lambda'} l_{\lambda'} s_{q'}:\lambda' \mu_{\lambda'}}(\kappa_\lambda^2),\nn\\
\end{eqnarray}
where
\begin{eqnarray}
{\cal J}_{\Gamma;n_\lambda l_\lambda s_q:\lambda \mu_\lambda}^{n_{\lambda'} l_{\lambda'} s_{q'}:\lambda' \mu_{\lambda'}}(\kappa_\lambda^2)&=&\int d^3 k e^{-\kappa_\lambda^2 k^2} {\cal L}_{n_{\lambda'}}^{l_{\lambda'}+\frac{1}{2}*}\left(\frac{p'^2}{\beta'^2}\right) {\cal Y}_{\lambda' \mu_{\lambda'}}^{*}(\vec k) \nn\\&&\times \langle s(\vec p_\Lambda+\vec p,s_{q'})\mid\overline{s}\Gamma b\mid b(\vec p,s_q)\rangle  {\cal L}_{n_{\lambda}}^{l_{\lambda}+\frac{1}{2}}\left(\frac{p^2}{\beta^2}\right) {\cal Y}_{\lambda \mu_{\lambda}}(\vec k).\nn\\
\label{eq:jint2}
\end{eqnarray}

In its most general form, the quark current $\langle s(\vec p_\Lambda+\vec p,s_{q'})\mid \overline{s}\Gamma b\mid b(\vec p,s_q)\rangle$ can be written
\begin{equation}
\langle s(\vec p_\Lambda+\vec p,s_{q'})\mid \overline{s}\Gamma b\mid b(\vec p,s_q)\rangle=\sum_{\ell=0}^{2} \sum_{m_\ell=-\ell}^{\ell} \xi_{\Gamma;\ell m_\ell}^{s_{q'} s_q}(\vec k,\vec p_\Lambda) {\cal Y}_{\ell m_\ell}(\vec k).
\label{eq:gfqc}
\end{equation}
It was discussed in Section \ref{sec:mcnff} that the coefficients $\xi_\Gamma$ are complicated functions of $\vec p_{\Lambda}\cdot \vec k$. To see this, recall Eq. \ref{eq:gm5p1},
\beq
\<s(\plsvec+\pvec,-1/2)\mid\overline{s}\gamma_{+}\gamma_5 b\mid b(\pvec,+1/2)\>=-\frac{\sqrt{\frac{8\pi}{3}}(\pls)_{+}\Y_{11}(\pvec)+\sqrt{\frac{16\pi}{15}}\Y_{22}(\pvec)}{\sqrt{\Eb}\sqrt{\Es}\sqrt{\Eb+\qmb}\sqrt{\Es+\qms}}.\nn
\eeq
By use of Eq. \ref{eq:addth}, $\Y_{lm}(\pvec)$ can be written in terms of $\Y_{lm}(\kvec)$ as
\beqynn
\Y_{11}(\pvec)&=&\sqrt{3}c(\pls)_{+}\Y_{00}(\kvec)+\Y_{11}(\kvec),\\
\Y_{22}(\pvec)&=&\sqrt{\frac{15}{2}}c^2(\pls)_{+}^2\Y_{00}(\kvec)+\sqrt{10}c(\pls)_{+}\Y_{11}(\kvec)+\Y_{22}(\kvec).
\eeqynn
Now $\<s(\plsvec+\pvec,-1/2)\mid\overline{s}\gamma_{+}\gamma_5 b\mid b(\pvec,+1/2)\>$ becomes
\beqynn
\<s(\plsvec+\pvec,-1/2)\mid\overline{s}\gamma_{+}\gamma_5 b\mid b(\pvec,+1/2)\>&=&-f(k,\pls;x)\bigg[
\sqrt{8\pi}c(c+1)(\pls)_{+}^2\Y_{00}(\kvec)\\&&+\sqrt{\frac{8\pi}{3}}(2c+1)(\pls)_{+}\Y_{11}(\kvec)+\sqrt{\frac{16\pi}{15}}\Y_{22}(\kvec)\bigg],
\eeqynn
where,
\beqy
f(k,\pls;x)=\frac{1}{\sqrt{\Eb}\sqrt{\Es}\sqrt{\Eb+\qmb}\sqrt{\Es+\qms}},\nn
\eeqy
and
\beqynn
\Eb&=&\sqrt{p^2+\qmb^2}=(c^2\pls^2+k^2+\qmb^2+2c\pls k x)^{1/2},\\
\Es&=&\sqrt{(\pvec+\plsvec)^2+\qms^2}\\&=&[(c+1)^2\pls^2+k^2+\qms^2+2(c+1)\pls k x]^{1/2}.
\eeqynn
In the above, we have used $x=\hat p_{\Lambda}\cdot \hat k$, where $\hat p_{\Ls}=\plsvec/\pls$ and $\hat k=\kvec/k$. Matching $\<s(\plsvec+\pvec,-1/2)\mid\overline{s}\gamma_{+}\gamma_5 b\mid b(\pvec,+1/2)\>$ with the general form of the quark current (Eq. \ref{eq:gfqc}), we find that
\beqynn
\xi_{\gamma_{+}\gamma_5;00}^{-+}(\kvec,\plsvec)&=&-\sqrt{8\pi}c(c+1)(\pls)_{+}^2f(k,\pls;x),\\
\xi_{\gamma_{+}\gamma_5;11}^{-+}(\kvec,\plsvec)&=&-\sqrt{\frac{8\pi}{3}}(2c+1)(\pls)_{+}f(k,\pls;x),\\
\xi_{\gamma_{+}\gamma_5;22}^{-+}(\kvec,\plsvec)&=&-\sqrt{\frac{16\pi}{15}}f(k,\pls;x).
\eeqynn
The $\xi$ coefficients can each be expanded in terms of Legendre polynomials as
\beq
\xi_{\Gamma;\ell m_\ell}^{s_{q'} s_q}(\kvec,\plsvec)=\sum_{L=0}^{\infty}\zeta_{\Gamma;\ell m_\ell L}^{s_{q'}s_q}(k,\plsvec) P_L(x),\nn
\eeq
where
\beq
\zeta_{\Gamma;\ell m_\ell L}^{s_{q'}s_q}(k,\plsvec)=\left(\frac{2L+1}{2}\right)\int_{-1}^{+1}dx\xi_{\Gamma;\ell m_\ell}^{s_{q'} s_q}(\kvec,\plsvec) P_L(x).\nn
\eeq
Using
\beq
P_L(x)=\left(\frac{4\pi}{2L+1}\right)\sum_{M=-L}^{+L}Y_{LM}^*(\hat{p}_\Lambda)Y_{LM}(\hat{k})\nn
\eeq
we can write the $\xi$ coefficients in terms of $Y_{LM}(\hat{k})$,
\beq
\xi_{\Gamma;\ell m_\ell}^{s_{q'} s_q}(\vec k,\vec p_\Lambda)=\sum_{L=0}^{\infty}\sum_{M_L=-L}^{L} \zeta_{\Gamma;\ell m_\ell L M_L}^{s_{q'}s_q}(k,\vec p_\Lambda) Y_{L M_L}(\hat k),
\label{eq:xi}
\eeq
where
\beqy
\zeta_{\Gamma;\ell m_\ell L M_L}^{s_{q'}s_q}(k,\vec p_\Lambda)&=&\left(\frac{4\pi}{2L+1}\right)Y_{LM}^*(\hat{p}_\Lambda)\zeta_{lmL}(k,\plsvec)\nn \\
&=&2\pi Y_{L M_L}^{*}(\hat p_\Lambda) \int_{-1}^{+1} dx \xi_{\Gamma;\ell m_\ell}^{s_{q'} s_q}(\vec k,\vec p_\Lambda) P_{L}(x).
\label{eq:zeta}
\eeqy
The integral above is done numerically. The quark current can now be written as
\begin{equation}
\langle s(\vec p_\Lambda+\vec p,s_{q'})\mid \overline{s}\Gamma b\mid b(\vec p,s_q)\rangle=\sum_{\ell m_\ell}\sum_{L M_L} \zeta_{\Gamma;\ell m_\ell L M_L}^{s_{q'}s_q}(k,\vec p_\Lambda) Y_{L M_L}(\hat k) {\cal Y}_{\ell m_\ell}(\vec k).
\label{eq:qcex}
\end{equation}

We also note that the normalized Laguerre polynomials in Eq. \ref{eq:jint2} are functions of $\vec p_{\Lambda}\cdot \vec k$. These functions are expanded in terms of spherical harmonics as well, giving
\begin{equation}
{\cal L}_{n}^{l+\frac{1}{2}}\left(\frac{p^2}{\beta^2}\right)=N_{n l}(\beta)\sum_{L=0}^{n}\sum_{M=-L}^{L} \Lambda_{n l}^{L M}(k,c\vec p_\Lambda) Y_{L M}(\hat k),
\label{eq:nlex}
\end{equation}
where
\begin{equation}
\Lambda_{n l}^{L M}(k,c\vec p_\Lambda)=2\pi Y_{L M_L}^{*}(\hat p_\Lambda) \int_{-1}^{+1} dx L_{n}^{l+\frac{1}{2}}\left(\frac{p^2}{\beta^2}\right) P_{L}(x).
\end{equation}
Here, we have used Eq. \ref{eq:pkswitch}. $\Lambda_{n l}^{L M}$ can be determined analytically and the values for $n\leq 1$ are
\begin{eqnarray}
\Lambda_{0 l}^{0 0}&=&4\pi Y_{0 0}^{*}(\hat p_\Lambda), \nonumber \\
\Lambda_{1 l}^{0 0}&=&4\pi Y_{0 0}^{*}(\hat p_\Lambda) \left[l+\frac{3}{2}-\frac{1}{\beta^2}(k^2+c^2 p_\Lambda^2)\right], \nonumber \\
\Lambda_{1 l}^{1 M}&=&\frac{8\pi}{3}\frac{c p_\Lambda k}{\beta^2} Y_{1 M}^{*}(\hat p_\Lambda).
\end{eqnarray}
Substituting Eqs. \ref{eq:qcex} and \ref{eq:nlex} into Eq. \ref{eq:jint2}, we get
\begin{eqnarray}
{\cal J}_{\Gamma;n_\lambda l_\lambda s_q:\lambda \mu_\lambda}^{n_{\lambda'} l_{\lambda'} s_{q'}:\lambda' \mu_{\lambda'}}(\kappa_\lambda^2)&=&N_{n_{\lambda'} l_{\lambda'}}^{*}(\beta')N_{n_{\lambda} l_{\lambda}}(\beta) \nonumber \\ &&  \times \sum_{\ell m_\ell}\sum_{L M_L}\sum_{L_{a'} M_{a'}}\sum_{L_{a} M_{a}}\int d^3 k e^{-\kappa_\lambda^2 k^2} \zeta_{\Gamma;\ell m_\ell L M_L}^{s_{q'}s_q}(k,\vec p_\Lambda) \nonumber \\ && \times \Lambda_{n_{\lambda'} l_{\lambda'}}^{L_{a'} M_{a'}}(k,c'\vec p_\Lambda)\Lambda_{n_{\lambda} l_{\lambda}}^{L_{a} M_{a}}(k,c\vec p_\Lambda) \nonumber \\ && \times {\cal Y}_{\lambda' \mu_{\lambda'}}^{*}(\vec k) Y_{L_{a'} M_{a'}}(\hat k) Y_{L M_L}(\hat k) {\cal Y}_{\ell m_\ell}(\vec k) Y_{L_a M_a}(\hat k) {\cal Y}_{\lambda \mu_{\lambda}}(\vec k).\nonumber \\
\end{eqnarray}
Recall that, ${\cal Y}_{l m}=k^l Y_{l m}$. Making this substitution leads to
\begin{eqnarray}
{\cal J}_{\Gamma;n_\lambda l_\lambda s_q:\lambda \mu_\lambda}^{n_{\lambda'} l_{\lambda'} s_{q'}:\lambda' \mu_{\lambda'}}(\kappa_\lambda^2)&=&N_{n_{\lambda'} l_{\lambda'}}^{*}(\beta')N_{n_{\lambda} l_{\lambda}}(\beta) \nonumber \\ && \times  \sum_{\ell m_\ell}\sum_{L M_L}\sum_{L_{a'} M_{a'}}\sum_{L_{a} M_{a}}{\cal D}_{\Gamma;\ell m_\ell L M_L n_{\lambda'} l_{\lambda'} n_{\lambda} l_{\lambda}}^{s_{q'}s_q L_{a'} M_{a'} L_{a} M_{a}}(\lambda'+\ell+\lambda+2;\kappa_\lambda^2) \nonumber \\ &&  \times \langle Y_{\lambda'\mu_{\lambda'}}\mid Y_{L_{a'} M_{a'}} Y_{L M_L} Y_{\ell m_\ell} Y_{L_a M_a}\mid Y_{\lambda\mu_\lambda}\rangle,
\end{eqnarray}
where
\beqy
{\cal D}_{\Gamma;\ell m_\ell L M_L n_{\lambda'} l_{\lambda'} n_{\lambda} l_{\lambda}}^{s_{q'}s_q L_{a'} M_{a'} L_{a} M_{a}}(h;\kappa_\lambda^2)&=&\int_0^\infty d k e^{-\kappa_\lambda^2 k^2} k^h \zeta_{\Gamma;\ell m_\ell L M_L}^{s_{q'}s_q}(k,\vec p_\Lambda) \nn\\&&\times \Lambda_{n_{\lambda'} l_{\lambda'}}^{L_{a'} M_{a'}}(k,c'\vec p_\Lambda)\Lambda_{n_{\lambda} l_{\lambda}}^{L_{a} M_{a}}(k,c\vec p_\Lambda),
\eeqy
is the momentum integral, which is done numerically. The bra-ket term is the angular integral over the spherical harmonics which can be handled analytically. We note here that the integral over the spherical harmonics provides a cut-off for the infinite sum over $L$ in Eq. \ref{eq:xi}; this sum can be truncated at $L=\lambda+L_a+\ell+L_{a'}+\lambda'$.

\section{Analytic Form Factors\label{sec:aff}}

Here we present the tensor form factors obtained from single-component quark model wave functions in the oscillator basis. The form factors for the vector and axial vector currents have the same form as those published in \cite{pervin}. These form factors were extracted using a nonrelativistic reduction of the quark current. The resulting form factors have the form of a polynomial in the daughter baryon momentum $\pls$ times a Gaussian. The polynomials in the form factors presented here are truncated at zeroth order in $\pls$.

\subsection{${1/2}^{+}$}

\begin{eqnarray*}
H_{1}&=&I_{H}\bigg[ \frac{\alpha _{\lambda }^2 \alpha _{\lambda '}^2}{12 m_b m_s \alpha
   _{\lambda \lambda '}^2}+1 \bigg]  \\
H_{2}&=&I_{H}\bigg[ \frac{m_q^2 \alpha _{\lambda }^2 \alpha _{\lambda '}^2}{m_b m_s \alpha
   _{\lambda \lambda '}^4}-\frac{\alpha _{\lambda }^2 \alpha _{\lambda
   '}^2}{12 m_b m_s \alpha _{\lambda \lambda '}^2}+\frac{m_q \alpha
   _{\lambda '}^2}{m_s \alpha _{\lambda \lambda '}^2} \bigg]  \\
H_{3}&=&I_{H}\bigg[ -\frac{m_q^2 \alpha _{\lambda }^2 \alpha _{\lambda '}^2}{m_b m_s \alpha
   _{\lambda \lambda '}^4}-\frac{m_q \alpha _{\lambda }^2}{m_b \alpha
   _{\lambda \lambda '}^2} \bigg]  \\
H_{4}&=&I_{H}\bigg[ -\frac{m_q^2 \alpha _{\lambda }^2 \alpha _{\lambda '}^2}{m_b m_s \alpha
   _{\lambda \lambda '}^4} \bigg]
\end{eqnarray*}

$$
I_{H}=\bigg(\frac{\alpha_{\lambda}\alpha_{\lambda'}}{\alpha_{\lambda\lambda'}^2}
\bigg)^{3/2}\exp\bigg(-\frac{3m_q^2}{2\widetilde{m}_{\Lambda}^2}
\frac{p_{\Lambda}^2}{\alpha_{\lambda\lambda'}^2}\bigg)
$$

\subsection{${1/2_{1}}^{+}$}

\begin{eqnarray*}
H_{1}&=&I_{H}\bigg[ \frac{\alpha _{\lambda }^2 \bigg(7 \alpha _{\lambda }^2-3 \alpha _{\lambda
   '}^2\bigg) \alpha _{\lambda '}^2}{72 m_b m_s \alpha _{\lambda \lambda
   '}^4}+\frac{\alpha _{\lambda }^2-\alpha _{\lambda '}^2}{2 \alpha
   _{\lambda \lambda '}^2} \bigg]  \\
H_{2}&=&I_{H}\bigg[ \frac{7 m_q^2 \alpha _{\lambda }^2 \bigg(\alpha _{\lambda }^2-\alpha
   _{\lambda '}^2\bigg) \alpha _{\lambda '}^2}{6 m_b m_s \alpha _{\lambda
   \lambda '}^6}-\frac{\alpha _{\lambda }^2 \bigg(7 \alpha _{\lambda }^2-3
   \alpha _{\lambda '}^2\bigg) \alpha _{\lambda '}^2}{72 m_b m_s \alpha
   _{\lambda \lambda '}^4}+\frac{m_q \bigg(7 \alpha _{\lambda }^2-3 \alpha
   _{\lambda '}^2\bigg) \alpha _{\lambda '}^2}{6 m_s \alpha _{\lambda
   \lambda '}^4} \bigg]  \\
H_{3}&=&I_{H}\bigg[ -\frac{7 m_q^2 \alpha _{\lambda }^2 \alpha _{\lambda '}^2 \bigg(\alpha
   _{\lambda }^2-\alpha _{\lambda '}^2\bigg)}{6 m_b m_s \alpha _{\lambda
   \lambda '}^6}-\frac{m_q \alpha _{\lambda }^2 \bigg(3 \alpha _{\lambda
   }^2-7 \alpha _{\lambda '}^2\bigg)}{6 m_b \alpha _{\lambda \lambda
   '}^4} \bigg]  \\
H_{4}&=&I_{H}\bigg[ -\frac{7 m_q^2 \alpha _{\lambda }^2 \alpha _{\lambda '}^2 \bigg(\alpha
   _{\lambda }^2-\alpha _{\lambda '}^2\bigg)}{6 m_b m_s \alpha _{\lambda
   \lambda '}^6} \bigg]
\end{eqnarray*}

$$
I_{H}=\sqrt{\frac{3}{2}}\bigg(\frac{\alpha_{\lambda}\alpha_{\lambda'}}{\alpha_{\lambda\lambda'}^2}
\bigg)^{3/2}\exp\bigg(-\frac{3m_q^2}{2\widetilde{m}_{\Lambda}^2}
\frac{p_{\Lambda}^2}{\alpha_{\lambda\lambda'}^2}\bigg)
$$

\subsection{${1/2}^{-}$}

\begin{eqnarray*}
H_{1}&=&I_{H}\bigg[ -\frac{m_q \alpha _{\lambda } \bigg(3 \alpha _{\lambda }^2-2 \alpha _{\lambda
   '}^2\bigg)}{6 m_b m_s \alpha _{\lambda \lambda '}^2}+\frac{\alpha _{\lambda }}{2
   m_b}-\frac{2 m_q}{\alpha _{\lambda }} \bigg]  \\
H_{2}&=&I_{H}\bigg[ \frac{\alpha _{\lambda }}{3 m_b}-\frac{2 m_q}{\alpha _{\lambda }}+\frac{\alpha _{\lambda
   }}{2 m_s} \bigg]  \\
H_{3}&=&I_{H}\bigg[ \frac{\alpha _{\lambda }}{3 m_b}-\frac{m_q \alpha _{\lambda } \bigg(3 \alpha _{\lambda
   }^2-2 \alpha _{\lambda '}^2\bigg)}{6 m_b m_s \alpha _{\lambda \lambda '}^2} \bigg]  \\
H_{4}&=&I_{H}\bigg[ \frac{m_q \alpha _{\lambda } \bigg(3 \alpha _{\lambda }^2-2 \alpha _{\lambda '}^2\bigg)}{6
   m_b m_s \alpha _{\lambda \lambda '}^2}-\frac{2 m_q^2 \alpha _{\lambda }}{m_b \alpha
   _{\lambda \lambda '}^2}-\frac{\alpha _{\lambda }}{3 m_b} \bigg]
\end{eqnarray*}

$$
I_{H}=\bigg(\frac{\alpha_{\lambda}\alpha_{\lambda'}}{\alpha_{\lambda\lambda'}^2}
\bigg)^{5/2}\exp\bigg(-\frac{3m_q^2}{2\widetilde{m}_{\Lambda}^2}
\frac{p_{\Lambda}^2}{\alpha_{\lambda\lambda'}^2}\bigg)
$$

\subsection{${3/2}^{-}$}

\begin{eqnarray*}
H_{1}&=&I_{H}\bigg[ -\frac{7 m_q \alpha _{\lambda } \alpha _{\lambda '}^2}{4 m_b m_s \alpha
   _{\lambda \lambda '}^2}-\frac{3 \alpha _{\lambda }}{2 m_b}-\frac{3
   m_q}{\alpha _{\lambda }} \bigg]  \\
H_{2}&=&I_{H}\bigg[ -\frac{3 m_q^3 \alpha _{\lambda } \alpha _{\lambda '}^2}{m_b m_s \alpha
   _{\lambda \lambda '}^4}+\frac{9 m_q \alpha _{\lambda } \alpha _{\lambda
   '}^2}{4 m_b m_s \alpha _{\lambda \lambda '}^2}+\frac{2 \alpha _{\lambda
   }}{m_b}-\frac{3 m_q^2 \alpha _{\lambda '}^2}{m_s \alpha _{\lambda }
   \alpha _{\lambda \lambda '}^2} \bigg]  \\
H_{3}&=&I_{H}\bigg[ \frac{3 m_q^3 \alpha _{\lambda } \alpha _{\lambda '}^2}{m_b m_s \alpha
   _{\lambda \lambda '}^4}+\frac{m_q \alpha _{\lambda } \alpha _{\lambda
   '}^2}{2 m_b m_s \alpha _{\lambda \lambda '}^2}+\frac{3 m_q^2 \alpha
   _{\lambda }}{m_b \alpha _{\lambda \lambda '}^2}+\frac{\alpha _{\lambda
   }}{2 m_b} \bigg]  \\
H_{4}&=&I_{H}\bigg[ \frac{3 m_q^3 \alpha _{\lambda } \alpha _{\lambda '}^2}{m_b m_s \alpha
   _{\lambda \lambda '}^4}+\frac{m_q \alpha _{\lambda } \alpha _{\lambda
   '}^2}{2 m_b m_s \alpha _{\lambda \lambda '}^2} \bigg]  \\
H_{5}&=&I_{H}\bigg[-\frac{m_q \alpha _{\lambda } \alpha _{\lambda '}^2}{m_b m_s \alpha
   _{\lambda \lambda '}^2}-\frac{\alpha _{\lambda }}{m_b}  \bigg]  \\
H_{6}&=&I_{H}\bigg[ \frac{m_q \alpha _{\lambda } \alpha _{\lambda '}^2}{m_b m_s \alpha
   _{\lambda \lambda '}^2} \bigg]
\end{eqnarray*}

$$
I_{H}=-\frac{1}{\sqrt{3}}\bigg(\frac{\alpha_{\lambda}\alpha_{\lambda'}}{\alpha_{\lambda\lambda'}^2}
\bigg)^{5/2}\exp\bigg(-\frac{3m_q^2}{2\widetilde{m}_{\Lambda}^2}
\frac{p_{\Lambda}^2}{\alpha_{\lambda\lambda'}^2}\bigg)
$$

\subsection{${3/2}^{+}$}

\begin{eqnarray*}
H_{1}&=&I_{H}\bigg[ \frac{m_q^2 \bigg(5 \alpha _{\lambda }^2-2 \alpha _{\lambda '}^2\bigg)}{2
   m_b m_s \alpha _{\lambda \lambda '}^2}-\frac{3 m_q}{2 m_b}+\frac{6
   m_q^2}{\alpha _{\lambda }^2} \bigg]  \\
H_{2}&=&I_{H}\bigg[ -\frac{2 m_q}{m_b}+\frac{5 \alpha _{\lambda }^2}{12 m_b m_s}-\frac{5
   m_q}{2 m_s}+\frac{6 m_q^2}{\alpha _{\lambda }^2} \bigg]  \\
H_{3}&=&I_{H}\bigg[\frac{m_q^2 \bigg(5 \alpha _{\lambda }^2-2 \alpha _{\lambda '}^2\bigg)}{2
   m_b m_s \alpha _{\lambda \lambda '}^2}-\frac{m_q}{m_b}+\frac{5 \alpha
   _{\lambda }^2}{12 m_b m_s}  \bigg]  \\
H_{4}&=&I_{H}\bigg[ -\frac{m_q^2 \bigg(5 \alpha _{\lambda }^2-2 \alpha _{\lambda
   '}^2\bigg)}{2 m_b m_s \alpha _{\lambda \lambda '}^2}+\frac{6
   m_q^3}{m_b \alpha _{\lambda \lambda '}^2}+\frac{2 m_q}{m_b}-\frac{5
   \alpha _{\lambda }^2}{12 m_b m_s} \bigg]  \\
H_{5}&=&I_{H}\bigg[-\frac{5 \alpha _{\lambda }^2}{6 m_b m_s}  \bigg]  \\
H_{6}&=&I_{H}\bigg[ \frac{2 m_q}{m_b}-\frac{5 \alpha _{\lambda }^2}{6 m_b m_s} \bigg]
\end{eqnarray*}

$$
I_{H}=\frac{1}{\sqrt{5}}\bigg(\frac{\alpha_{\lambda}\alpha_{\lambda'}}{\alpha_{\lambda\lambda'}^2}
\bigg)^{7/2}\exp\bigg(-\frac{3m_q^2}{2\widetilde{m}_{\Lambda}^2}
\frac{p_{\Lambda}^2}{\alpha_{\lambda\lambda'}^2}\bigg)
$$

\subsection{${5/2}^{+}$}

\begin{eqnarray*}
H_{1}&=&I_{H}\bigg[\frac{13 m_q^2 \alpha _{\lambda '}^2}{4 m_b m_s \alpha _{\lambda \lambda
   '}^2}+\frac{3 m_q}{m_b}+\frac{3 m_q^2}{\alpha _{\lambda }^2}  \bigg]  \\
H_{2}&=&I_{H}\bigg[ \frac{3 m_q^4 \alpha _{\lambda '}^2}{m_b m_s \alpha _{\lambda \lambda
   '}^4}-\frac{17 m_q^2 \alpha _{\lambda '}^2}{4 m_b m_s \alpha _{\lambda
   \lambda '}^2}-\frac{4 m_q}{m_b}+\frac{3 m_q^3 \alpha _{\lambda
   '}^2}{m_s \alpha _{\lambda }^2 \alpha _{\lambda \lambda '}^2} \bigg]  \\
H_{3}&=&I_{H}\bigg[ -\frac{3 m_q^4 \alpha _{\lambda '}^2}{m_b m_s \alpha _{\lambda \lambda
   '}^4}-\frac{m_q^2 \alpha _{\lambda '}^2}{m_b m_s \alpha _{\lambda
   \lambda '}^2}-\frac{3 m_q^3}{m_b \alpha _{\lambda \lambda
   '}^2}-\frac{m_q}{m_b} \bigg]  \\
H_{4}&=&I_{H}\bigg[ -\frac{3 m_q^4 \alpha _{\lambda '}^2}{m_b m_s \alpha _{\lambda \lambda
   '}^4}-\frac{m_q^2 \alpha _{\lambda '}^2}{m_b m_s \alpha _{\lambda
   \lambda '}^2} \bigg]  \\
H_{5}&=&I_{H}\bigg[\frac{2 m_q^2 \alpha _{\lambda '}^2}{m_b m_s \alpha _{\lambda \lambda
   '}^2}+\frac{2 m_q}{m_b}  \bigg]  \\
H_{6}&=&I_{H}\bigg[ -\frac{2 m_q^2 \alpha _{\lambda '}^2}{m_b m_s \alpha _{\lambda \lambda
   '}^2} \bigg]
\end{eqnarray*}

$$
I_{H}=\frac{1}{\sqrt{2}}\bigg(\frac{\alpha_{\lambda}\alpha_{\lambda'}}{\alpha_{\lambda\lambda'}^2}
\bigg)^{7/2}\exp\bigg(-\frac{3m_q^2}{2\widetilde{m}_{\Lambda}^2}
\frac{p_{\Lambda}^2}{\alpha_{\lambda\lambda'}^2}\bigg)
$$

\section{Hadronic Tensors\label{sec:hadten}}

In this appendix we present the hadronic tensors we use in this work. Recall that the squared amplitude for dileptonic decays with unpolarized baryons is given by
\beq
|\overline{\cal M}|^2= \frac{G_F^2\alpha_{em}^2}{2^4\pi^2}\left|V_{tb}V_{ts}^{*}\right|^2\left(H_a^{\mu\nu}L_{a\mu\nu}+H_b^{\mu\nu}L_{b\mu\nu}+H_c^{\mu\nu}L_{c\mu\nu}+H_d^{\mu\nu}L_{d\mu\nu}\right),
\eeq
where $L_{f\mu\nu}$ are the leptonic tensors. The most general Lorentz structure for each hadronic tensor $H_f^{\mu\nu}$ is
\begin{equation}
H_f^{\mu\nu}=-\alpha^f g^{\mu\nu}+\beta^f_{++} Q^\mu Q^\nu+\beta^f_{+-} Q^\mu q^\nu+\beta^f_{-+} q^\mu Q^\nu+\beta^f_{--} q^\mu q^\nu+i\gamma^f\varepsilon^{\mu\nu\alpha\beta}Q_\alpha q_\beta,
\end{equation}
where $Q=\plb+\pls$ is the total baryon 4-momentum and $q=\plb-\pls$ is the 4-momentum transfer. The coefficients $\alpha$, $\beta_{\pm\pm}$, and $\gamma$ are written in terms of products of form factors. A generic coefficient $\lambda$ in the tensor $H_a^{\mu\nu}$ takes the form
\begin{eqnarray}
\lambda^a&=&\sum_{j k}\left[\eta_{j k}^{VV(\lambda)}A_j^{*} A_k+\eta_{j k}^{AA(\lambda)}B_j^{*} B_k+\eta_{j k}^{VA(\lambda)}A_j^{*} B_k+\eta_{j k}^{AV(\lambda)}B_j^{*} A_k\right],
\end{eqnarray}
and similarly for, $H_b^{\mu\nu}$, we have
\begin{eqnarray}
\lambda^b&=&\sum_{j k}\left[\eta_{j k}^{VV(\lambda)}D_j^{*} D_k+\eta_{j k}^{AA(\lambda)}E_j^{*} E_k+\eta_{j k}^{VA(\lambda)}D_j^{*} E_k+\eta_{j k}^{AV(\lambda)}E_j^{*} D_k\right].
\end{eqnarray}
For $H_c^{\mu\nu}$ and $H_d^{\mu\nu}$, the coefficients have the form
\begin{eqnarray}
\lambda^c&=&\sum_{j k}\left[\eta_{j k}^{VV(\lambda)}A_j^{*} D_k+\eta_{j k}^{AA(\lambda)}B_j^{*} E_k+\eta_{j k}^{VA(\lambda)}A_j^{*} E_k+\eta_{j k}^{AV(\lambda)}B_j^{*} D_k\right], \nonumber \\
\lambda^d&=&\sum_{j k}\left[\eta_{j k}^{VV(\lambda)}D_j^{*} A_k+\eta_{j k}^{AA(\lambda)}E_j^{*} B_k+\eta_{j k}^{VA(\lambda)}D_j^{*} B_k+\eta_{j k}^{AV(\lambda)}E_j^{*} A_k\right],
\end{eqnarray}
respectively. For states with natural parity spinors,
\beqy
A_i&=&-\frac{2m_b}{q^2}C_7F^{T}_i+C_9F_i,\nn\\
B_i&=&-\frac{2m_b}{q^2}C_7G^{T}_i-C_9G_i,\nn\\
D_i&=&C_{10}F_i, \,\,\,\,\ E_i=-C_{10}G_i,
\eeqy
and for states with unnatural parity,
\beqy
A_i&=&-\frac{2m_b}{q^2}C_7G^{T}_i-C_9G_i,\nn\\
B_i&=&-\frac{2m_b}{q^2}C_7F^{T}_i+C_9F_i,\nn\\
D_i&=&-C_{10}G_i, \,\,\,\,\ E_i=C_{10}F_i.
\eeqy
The nonzero values of $\eta_{j k}$ for the various final states are given below.

\subsection{$J={1/2}$}

\beqynn
\eta^{VV(\alpha)}_{1,1}&=&2 m_{\Lambda_{b}}^2 \left(r-2 \sqrt{r}-\sh+1\right), \,\,\,\,
\eta^{AA(\alpha)}_{1,1}=2 m_{\Lambda_{b}}^2 \left(r+2 \sqrt{r}-\sh+1\right).
\eeqynn

\beqynn
\eta^{VV(\beta_{++})}_{1,1}&=&2, \,\,\,\,
\eta^{VV(\beta_{++})}_{2,2}=\frac{1}{2} \left(r+2 \sqrt{r}-\sh+1\right), \,\,\,\,
\eta^{VV(\beta_{++})}_{3,3}=\frac{r+2 \sqrt{r}-\sh+1}{2 r}, \\
\eta^{AA(\beta_{++})}_{1,1}&=&2, \,\,\,\,
\eta^{AA(\beta_{++})}_{2,2}=\frac{1}{2} \left(r-2 \sqrt{r}-\sh+1\right), \,\,\,\,
\eta^{AA(\beta_{++})}_{3,3}=\frac{r-2 \sqrt{r}-\sh+1}{2 r}.
\eeqynn

\beqynn
\eta^{VV(\beta_{++})}_{1,2}&=&\eta^{VV(\beta_{++})}_{2,1}=\sqrt{r}+1, \,\,\,\,
\eta^{VV(\beta_{++})}_{1,3}=\eta^{VV(\beta_{++})}_{3,1}=\frac{1}{\sqrt{r}}+1, \\
\eta^{VV(\beta_{++})}_{2,3}&=&\eta^{VV(\beta_{++})}_{3,2}=\frac{r+2 \sqrt{r}-\sh+1}{2 \sqrt{r}}, \\
\eta^{AA(\beta_{++})}_{1,2}&=&\eta^{AA(\beta_{++})}_{2,1}=\sqrt{r}-1, \,\,\,\,
\eta^{AA(\beta_{++})}_{1,3}=\eta^{AA(\beta_{++})}_{3,1}=1-\frac{1}{\sqrt{r}}, \\
\eta^{AA(\beta_{++})}_{2,3}&=&\eta^{AA(\beta_{++})}_{3,2}=\frac{r-2 \sqrt{r}-\sh+1}{2 \sqrt{r}}.
\eeqynn

\beqynn
\eta^{VV(\beta_{+-})}_{2,2}&=&\detvvmp_{2,2}=\frac{1}{2} \left(r+2 \sqrt{r}-\sh+1\right), \,\,\,\,
\eta^{VV(\beta_{+-})}_{3,3}=\detvvmp_{3,3}=-\frac{r+2 \sqrt{r}-\sh+1}{2 r}, \\
\eta^{AA(\beta_{+-})}_{2,2}&=&\detaamp_{2,2}=\frac{1}{2} \left(r-2 \sqrt{r}-\sh+1\right), \,\,\,\,
\eta^{AA(\beta_{+-})}_{3,3}=\detaamp_{3,3}=\frac{-r+2 \sqrt{r}+\sh-1}{2 r}.
\eeqynn

\beqynn
\eta^{VV(\beta_{+-})}_{1,2}&=&\detvvmp_{2,1}=\sqrt{r}+1, \,\,\,\,\,\,\,\,\,\,
\eta^{VV(\beta_{+-})}_{2,1}=\detvvmp_{1,2}=\sqrt{r}-1, \\
\eta^{VV(\beta_{+-})}_{1,3}&=&\detvvmp_{3,1}=-\frac{1}{\sqrt{r}}-1, \,\,\,\,
\eta^{VV(\beta_{+-})}_{3,1}=\detvvmp_{1,3}=-\frac{1}{\sqrt{r}}+1, \\
\eta^{VV(\beta_{+-})}_{2,3}&=&\detvvmp_{3,2}=\frac{-r-2 \sqrt{r}+\sh-1}{2 \sqrt{r}}, \,\,\,\,
\eta^{VV(\beta_{+-})}_{3,2}=\detvvmp_{2,3}=\frac{r+2 \sqrt{r}-\sh+1}{2 \sqrt{r}}, \\
\eta^{AA(\beta_{+-})}_{1,2}&=&\detaamp_{2,1}=\sqrt{r}-1, \,\,\,\,
\eta^{AA(\beta_{+-})}_{2,1}=\detaamp_{1,2}=\sqrt{r}+1, \\
\eta^{AA(\beta_{+-})}_{1,3}&=&\detaamp_{3,1}=\frac{1}{\sqrt{r}}-1, \,\,\,\,
\eta^{AA(\beta_{+-})}_{3,1}=\detaamp_{1,3}=\frac{1}{\sqrt{r}}+1, \\
\eta^{AA(\beta_{+-})}_{2,3}&=&\detaamp_{3,2}=\frac{-r+2 \sqrt{r}+\sh-1}{2 \sqrt{r}}, \,\,\,\,
\eta^{AA(\beta_{+-})}_{3,2}=\detaamp_{2,3}=\frac{r-2 \sqrt{r}-\sh+1}{2 \sqrt{r}}.
\eeqynn

\beqynn
\eta^{VV(\beta_{--})}_{1,1}&=&-2, \,\,\,\,
\eta^{VV(\beta_{--})}_{2,2}=\frac{1}{2} \left(r+2 \sqrt{r}-\sh+1\right), \,\,\,\,
\eta^{VV(\beta_{--})}_{3,3}=\frac{r+2 \sqrt{r}-\sh+1}{2 r}, \\
\eta^{AA(\beta_{--})}_{1,1}&=&-2, \,\,\,\,
\eta^{AA(\beta_{--})}_{2,2}=\frac{1}{2} \left(r-2 \sqrt{r}-\sh+1\right), \,\,\,\,
\eta^{AA(\beta_{--})}_{3,3}=\frac{r-2 \sqrt{r}-\sh+1}{2 r}.
\eeqynn

\beqynn
\eta^{VV(\beta_{--})}_{1,2}&=&\eta^{VV(\beta_{--})}_{2,1}=\sqrt{r}-1, \,\,\,\,
\eta^{VV(\beta_{--})}_{1,3}=\eta^{VV(\beta_{--})}_{3,1}=\frac{1}{\sqrt{r}}-1, \\
\eta^{VV(\beta_{--})}_{2,3}&=&\eta^{VV(\beta_{--})}_{3,2}=\frac{-r-2 \sqrt{r}+\sh-1}{2 \sqrt{r}}, \\
\eta^{AA(\beta_{--})}_{1,2}&=&\eta^{AA(\beta_{--})}_{2,1}=\sqrt{r}+1, \,\,\,\,
\eta^{AA(\beta_{--})}_{1,3}=\eta^{AA(\beta_{--})}_{3,1}=-\frac{1}{\sqrt{r}}-1, \\
\eta^{AA(\beta_{--})}_{2,3}&=&\eta^{AA(\beta_{--})}_{3,2}=\frac{-r+2 \sqrt{r}+\sh-1}{2 \sqrt{r}}.
\eeqynn

\beqynn
\eta^{VA(\gamma)}_{1,1}=\eta^{AV(\gamma)}_{1,1}=-2.
\eeqynn

\subsection{$J={3/2}$}

\beqynn
\detvva_{1,1}\=\frac{\mlb^2}{3 r} \left(r-2 \sqrt{r}-\sh+1\right)^2 \left(r+2 \sqrt{r}-\sh+1\right),\,\,\,\,\,
\detvva_{4,4}=\frac{4}{3} \mlb^2 \left(r+2 \sqrt{r}-\sh+1\right),\\
\detaaa_{1,1}\=\frac{\mlb^2}{3 r} \left(r-2 \sqrt{r}-\sh+1\right) \left(r+2 \sqrt{r}-\sh+1\right)^2,\,\,\,\,\,
\detaaa_{4,4}=\frac{4}{3} \mlb^2 \left(r-2 \sqrt{r}-\sh+1\right),\\
\detvva_{1,4}\=\detvva_{4,1}=-\frac{\mlb^2}{3 \sqrt{r}} \left(r^2-2 r (\sh+1)+(\sh-1)^2\right),\\
\detaaa_{1,4}\=\detaaa_{4,1}=-\frac{\mlb^2}{3 \sqrt{r}} \left(r^2-2 r (\sh+1)+(\sh-1)^2\right).
\eeqynn

\beqynn
\detvvpp_{1,1}\=\frac{1}{3 r}\left(r^2-2 r (\sh+1)+(\sh-1)^2\right),\\
\detvvpp_{2,2}\=\frac{1}{12 r}\left(r-2 \sqrt{r}-\sh+1\right) \left(r+2 \sqrt{r}-\sh+1\right)^2,\\
\detvvpp_{3,3}\=\frac{1}{12 r^2}\left(r-2 \sqrt{r}-\sh+1\right) \left(r+2 \sqrt{r}-\sh+1\right)^2,\\
\detvvpp_{4,4}\=\frac{1}{3 r}\left(r+2 \sqrt{r}-\sh+1\right),\\
\detaapp_{1,1}\=\frac{1}{3 r}\left(r^2-2 r (\sh+1)+(\sh-1)^2\right),\\
\detaapp_{2,2}\=\frac{1}{12 r}\left(r-2 \sqrt{r}-\sh+1\right)^2 \left(r+2 \sqrt{r}-\sh+1\right),\\
\detaapp_{3,3}\=\frac{1}{12 r^2}\left(r-2 \sqrt{r}-\sh+1\right)^2 \left(r+2 \sqrt{r}-\sh+1\right),\\
\detaapp_{4,4}\=\frac{1}{3 r}\left(r-2 \sqrt{r}-\sh+1\right).
\eeqynn

\beqynn
\detvvpp_{1,2}\=\detvvpp_{2,1}=\frac{1}{6 r}\left(\sqrt{r}+1\right) \left(r^2-2 r (\sh+1)+(\sh-1)^2\right),\\
\detvvpp_{1,3}\=\detvvpp_{3,1}=\frac{1}{6 r^{3/2}}\left(\sqrt{r}+1\right) \left(r^2-2 r (\sh+1)+(\sh-1)^2\right), \\
\detvvpp_{1,4}\=\detvvpp_{4,1}=-\frac{r^{3/2}+r-\sqrt{r}+\sh-1}{3 r}, \\
\detvvpp_{2,3}\=\detvvpp_{3,2}=\frac{1}{12 r^{3/2}}\left(r-2 \sqrt{r}-\sh+1\right) \left(r+2 \sqrt{r}-\sh+1\right)^2,\\
\detvvpp_{2,4}\=\detvvpp_{4,2}=-\frac{1}{6 r}\left(2 r^{3/2}+r^2+2 \sqrt{r} (\sh-1)-(\sh-1)^2\right),\\
\detvvpp_{3,4}\=\detvvpp_{4,3}=\frac{1}{6 r^{3/2}}\left(-2 r^{3/2}-r^2-2 \sqrt{r} (\sh-1)+(\sh-1)^2\right).
\eeqynn

\beqynn
\detaapp_{1,2}\=\detaapp_{2,1}=\frac{1}{6 r}\left(\sqrt{r}-1\right) \left(r^2-2 r (\sh+1)+(\sh-1)^2\right),\\
\detaapp_{1,3}\=\detaapp_{3,1}=\frac{1}{6 r^{3/2}}\left(\sqrt{r}-1\right) \left(r^2-2 r (\sh+1)+(\sh-1)^2\right), \\
\detaapp_{1,4}\=\detaapp_{4,1}=\frac{-r^{3/2}+r+\sqrt{r}+\sh-1}{3 r}, \\
\detaapp_{2,3}\=\detaapp_{3,2}=\frac{1}{12 r^{3/2}}\left(r-2 \sqrt{r}-\sh+1\right)^2 \left(r+2 \sqrt{r}-\sh+1\right),\\
\detaapp_{2,4}\=\detaapp_{4,2}=\frac{1}{6 r}\left(2 r^{3/2}-r^2+2 \sqrt{r} (\sh-1)+(\sh-1)^2\right),\\
\detaapp_{3,4}\=\detaapp_{4,3}=\frac{1}{6 r^{3/2}}\left(2 r^{3/2}-r^2+2 \sqrt{r} (\sh-1)+(\sh-1)^2\right).
\eeqynn

\beqynn
\detvvpm_{2,2}\=\detvvmp_{2,2}=\frac{1}{12 r}\left(r-2 \sqrt{r}-\sh+1\right) \left(r+2 \sqrt{r}-\sh+1\right)^2,\\
\detvvpm_{3,3}\=\detvvmp_{3,3}=-\frac{1}{12 r^2}\left(r-2 \sqrt{r}-\sh+1\right) \left(r+2 \sqrt{r}-\sh+1\right)^2,\\
\detvvpm_{4,4}\=\detvvmp_{4,4}=-\frac{1}{3 r}\left(r+2 \sqrt{r}-\sh+1\right),\\
\detaapm_{2,2}\=\detaamp_{2,2}=\frac{1}{12 r}\left(r-2 \sqrt{r}-\sh+1\right)^2 \left(r+2 \sqrt{r}-\sh+1\right),\\
\detaapm_{3,3}\=\detaamp_{3,3}=-\frac{1}{12 r^2}\left(r-2 \sqrt{r}-\sh+1\right)^2 \left(r+2 \sqrt{r}-\sh+1\right),\\
\detaapm_{4,4}\=\detaamp_{4,4}=\frac{1}{3 r}\left(-r+2 \sqrt{r}+\sh-1\right).
\eeqynn

\beqynn
\detvvpm_{1,2}\=\detvvmp_{2,1}=\frac{1}{6 r}\left(\sqrt{r}+1\right) \left(r^2-2 r (\sh+1)+(\sh-1)^2\right),\\
\detvvpm_{1,3}\=\detvvmp_{3,1}=-\frac{1}{6 r^{3/2}}\left(\sqrt{r}+1\right) \left(r^2-2 r (\sh+1)+(\sh-1)^2\right),\\
\detvvpm_{1,4}\=\detvvmp_{4,1}=-\frac{1}{3 r}\left(\sqrt{r}+1\right) \left(2 r+\sqrt{r}-\sh+1\right).
\eeqynn

\beqynn
\detvvpm_{2,1}\=\detvvmp_{1,2}=\frac{1}{6 r}\left(\sqrt{r}-1\right) \left(r^2-2 r (\sh+1)+(\sh-1)^2\right),\\
\detvvpm_{2,3}\=\detvvmp_{3,2}=-\frac{1}{12 r^{3/2}}\left(r-2 \sqrt{r}-\sh+1\right) \left(r+2 \sqrt{r}-\sh+1\right)^2,\\
\detvvpm_{2,4}\=\detvvmp_{4,2}=-\frac{1}{6 r}\left(6 r^{3/2}+3 r^2-4 r (\sh-1)-2 \sqrt{r} (\sh-1)+(\sh-1)^2\right).
\eeqynn

\beqynn
\detvvpm_{3,1}\=\detvvmp_{1,3}=\frac{1}{6 r^{3/2}}\left(\sqrt{r}-1\right) \left(r^2-2 r (\sh+1)+(\sh-1)^2\right),\\
\detvvpm_{3,2}\=\detvvmp_{2,3}=\frac{1}{12 r^{3/2}}\left(r-2 \sqrt{r}-\sh+1\right) \left(r+2 \sqrt{r}-\sh+1\right)^2,\\
\detvvpm_{3,4}\=\detvvmp_{4,3}=\frac{1}{6 r^{3/2}}\left(-6 r^{3/2}-3 r^2+4 r (\sh-1)+2 \sqrt{r} (\sh-1)-(\sh-1)^2\right).
\eeqynn

\beqynn
\detvvpm_{4,1}\=\detvvmp_{1,4}=\frac{1}{3 r}\left(-\sqrt{r} \sh+r+\sh-1\right),\\
\detvvpm_{4,2}\=\detvvmp_{2,4}=-\frac{1}{6 r}\left(2 r^{3/2}+r^2+2 \sqrt{r} (\sh-1)-(\sh-1)^2\right),\\
\detvvpm_{4,3}\=\detvvmp_{3,4}=\frac{1}{6 r^{3/2}}\left(2 r^{3/2}+r^2+2 \sqrt{r} (\sh-1)-(\sh-1)^2\right).
\eeqynn

\beqynn
\detaapm_{1,2}\=\detaamp_{2,1}=\frac{1}{6 r}\left(\sqrt{r}-1\right) \left(r^2-2 r (\sh+1)+(\sh-1)^2\right),\\
\detaapm_{1,3}\=\detaamp_{3,1}=-\frac{1}{6 r^{3/2}}\left(\sqrt{r}-1\right) \left(r^2-2 r (\sh+1)+(\sh-1)^2\right),\\
\detaapm_{1,4}\=\detaamp_{4,1}=-\frac{1}{3 r}\left(\sqrt{r}-1\right) \left(2 r-\sqrt{r}-\sh+1\right).
\eeqynn

\beqynn
\detaapm_{2,1}\=\detaamp_{1,2}=\frac{1}{6 r}\left(\sqrt{r}+1\right) \left(r^2-2 r (\sh+1)+(\sh-1)^2\right),\\
\detaapm_{2,3}\=\detaamp_{3,2}=-\frac{1}{12 r^{3/2}}\left(r-2 \sqrt{r}-\sh+1\right)^2 \left(r+2 \sqrt{r}-\sh+1\right),\\
\detaapm_{2,4}\=\detaamp_{4,2}=-\frac{1}{6 r}\left(-6 r^{3/2}+3 r^2-4 r (\sh-1)+2 \sqrt{r} (\sh-1)+(\sh-1)^2\right).
\eeqynn

\beqynn
\detaapm_{3,1}\=\detaamp_{1,3}=\frac{1}{6 r^{3/2}}\left(\sqrt{r}+1\right) \left(r^2-2 r (\sh+1)+(\sh-1)^2\right),\\
\detaapm_{3,2}\=\detaamp_{2,3}=\frac{1}{12 r^{3/2}}\left(r-2 \sqrt{r}-\sh+1\right)^2 \left(r+2 \sqrt{r}-\sh+1\right),\\
\detaapm_{3,4}\=\detaamp_{4,3}=\frac{1}{6 r^{3/2}}\left(6 r^{3/2}-3 r^2+4 r (\sh-1)-2 \sqrt{r} (\sh-1)-(\sh-1)^2\right).
\eeqynn

\beqynn
\detaapm_{4,1}\=\detaamp_{1,4}=-\frac{1}{3 r}\left(\sqrt{r}+1\right) \left(\sqrt{r}+\sh-1\right),\\
\detaapm_{4,2}\=\detaamp_{2,4}=\frac{1}{6 r}\left(2 r^{3/2}-r^2+2 \sqrt{r} (\sh-1)+(\sh-1)^2\right),\\
\detaapm_{4,3}\=\detaamp_{3,4}=\frac{1}{6 r^{3/2}}\left(-2 r^{3/2}+r^2-2 \sqrt{r} (\sh-1)-(\sh-1)^2\right).
\eeqynn

\beqynn
\detvvmm_{1,1}\=-\frac{1}{3 r}\left(r^2-2 r (\sh+1)+(\sh-1)^2\right),\\
\detvvmm_{2,2}\=\frac{1}{12 r}\left(r-2 \sqrt{r}-\sh+1\right) \left(r+2 \sqrt{r}-\sh+1\right)^2,\\
\detvvmm_{3,3}\=\frac{1}{12 r^2}\left(r-2 \sqrt{r}-\sh+1\right) \left(r+2 \sqrt{r}-\sh+1\right)^2,\\
\detvvmm_{4,4}\=\frac{1}{3 r}\left(r+2 \sqrt{r}-\sh+1\right),\\
\detaamm_{1,1}\=-\frac{1}{3 r}\left(r^2-2 r (\sh+1)+(\sh-1)^2\right),\\
\detaamm_{2,2}\=\frac{1}{12 r}\left(r-2 \sqrt{r}-\sh+1\right)^2 \left(r+2 \sqrt{r}-\sh+1\right),\\
\detaamm_{3,3}\=\frac{1}{12 r^2}\left(r-2 \sqrt{r}-\sh+1\right)^2 \left(r+2 \sqrt{r}-\sh+1\right),\\
\detaamm_{4,4}\=\frac{1}{3 r}\left(r-2 \sqrt{r}-\sh+1\right).
\eeqynn

\beqynn
\detvvmm_{1,2}\=\detvvmm_{2,1}=\frac{1}{6 r}\left(\sqrt{r}-1\right) \left(r^2-2 r (\sh+1)+(\sh-1)^2\right),\\
\detvvmm_{1,3}\=\detvvmm_{3,1}=-\frac{1}{6 r^{3/2}}\left(\sqrt{r}-1\right) \left(r^2-2 r (\sh+1)+(\sh-1)^2\right), \\
\detvvmm_{1,4}\=\detvvmm_{4,1}=\frac{-r^{3/2}+3 r+\sqrt{r}-\sh+1}{3 r}, \\
\detvvmm_{2,3}\=\detvvmm_{3,2}=-\frac{1}{12 r^{3/2}}\left(r-2 \sqrt{r}-\sh+1\right) \left(r+2 \sqrt{r}-\sh+1\right)^2,\\
\detvvmm_{2,4}\=\detvvmm_{4,2}=-\frac{1}{6 r}\left(6 r^{3/2}+3 r^2-4 r (\sh-1)-2 \sqrt{r} (\sh-1)+(\sh-1)^2\right),\\
\detvvmm_{3,4}\=\detvvmm_{4,3}=\frac{1}{6 r^{3/2}}\left(6 r^{3/2}+3 r^2-4 r (\sh-1)-2 \sqrt{r} (\sh-1)+(\sh-1)^2\right).
\eeqynn

\beqynn
\detaamm_{1,2}\=\detaamm_{2,1}=\frac{1}{6 r}\left(\sqrt{r}+1\right) \left(r^2-2 r (\sh+1)+(\sh-1)^2\right),\\
\detaamm_{1,3}\=\detaamm_{3,1}=-\frac{1}{6 r^{3/2}}\left(\sqrt{r}+1\right) \left(r^2-2 r (\sh+1)+(\sh-1)^2\right), \\
\detaamm_{1,4}\=\detaamm_{4,1}=\frac{-r^{3/2}-3 r+\sqrt{r}+\sh-1}{3 r}, \\
\detaamm_{2,3}\=\detaamm_{3,2}=-\frac{1}{12 r^{3/2}}\left(r-2 \sqrt{r}-\sh+1\right)^2 \left(r+2 \sqrt{r}-\sh+1\right),\\
\detaamm_{2,4}\=\detaamm_{4,2}=-\frac{1}{6 r}\left(-6 r^{3/2}+3 r^2-4 r (\sh-1)+2 \sqrt{r} (\sh-1)+(\sh-1)^2\right),\\
\detaamm_{3,4}\=\detaamm_{4,3}=\frac{1}{6 r^{3/2}}\left(-6 r^{3/2}+3 r^2-4 r (\sh-1)+2 \sqrt{r} (\sh-1)+(\sh-1)^2\right).
\eeqynn

\beqynn
\detvag_{1,1}\=\detavg_{1,1}=-\frac{r^2-2 r (\sh+1)+(\sh-1)^2}{3 r},\,\,\,\,\,
\detvag_{4,4}=\detavg_{4,4}=\frac{2}{3},\\
\detvag_{1,4}\=\detavg_{4,1}=\frac{r-2 \sqrt{r}-\sh+1}{3 \sqrt{r}},\,\,\,\,\,
\detvag_{4,1}=\detavg_{1,4}=\frac{r+2 \sqrt{r}-\sh+1}{3 \sqrt{r}}.
\eeqynn

\subsection{$J={5/2}$}

\beqynn
\detvva_{1,1}\=\frac{\mlb^2}{20 r^2} \left(r-2 \sqrt{r}-\sh+1\right)^3 \left(r+2 \sqrt{r}-\sh+1\right)^2,\\
\detvva_{4,4}\=\frac{3 \mlb^2}{20 r} \left(r-2 \sqrt{r}-\sh+1\right) \left(r+2 \sqrt{r}-\sh+1\right)^2,\\
\detaaa_{1,1}\=\frac{\mlb^2}{20 r^2} \left(r-2 \sqrt{r}-\sh+1\right)^2 \left(r+2 \sqrt{r}-\sh+1\right)^3,\\
\detaaa_{4,4}\=\frac{3 \mlb^2}{20 r} \left(r-2 \sqrt{r}-\sh+1\right)^2 \left(r+2 \sqrt{r}-\sh+1\right).
\eeqynn

\beqynn
\detvva_{1,4}\=\detvva_{4,1}=-\frac{\mlb^2}{20 r^{3/2}} \left(r^2-2 r (\sh+1)+(\sh-1)^2\right)^2,\\
\detaaa_{1,4}\=\detaaa_{4,1}=-\frac{\mlb^2}{20 r^{3/2}} \left(r^2-2 r (\sh+1)+(\sh-1)^2\right)^2.
\eeqynn

\beqynn
\detvvpp_{1,1}\=\frac{1}{20 r^2}\left(r^2-2 r (\sh+1)+(\sh-1)^2\right)^2,\\
\detvvpp_{2,2}\=\frac{1}{80 r^2}\left(r-2 \sqrt{r}-\sh+1\right)^2 \left(r+2 \sqrt{r}-\sh+1\right)^3,\\
\detvvpp_{3,3}\=\frac{1}{80 r^3}\left(r-2 \sqrt{r}-\sh+1\right)^2 \left(r+2 \sqrt{r}-\sh+1\right)^3,\\
\detvvpp_{4,4}\=\frac{1}{20 r^2}\bigg(2 r^{5/2}-2 r^{3/2} (\sh+2)+r^3-r^2 (2 \sh+1)+r \left(2 \sh^2-\sh-1\right)+\\&&2 \sqrt{r} (\sh-1)^2-(\sh-1)^3\bigg).
\eeqynn

\beqynn
\detaapp_{1,1}\=\frac{1}{20 r^2}\left(r^2-2 r (\sh+1)+(\sh-1)^2\right)^2,\\
\detaapp_{2,2}\=\frac{1}{80 r^2}\left(r-2 \sqrt{r}-\sh+1\right)^3 \left(r+2 \sqrt{r}-\sh+1\right)^2,\\
\detaapp_{3,3}\=\frac{1}{80 r^3}\left(r-2 \sqrt{r}-\sh+1\right)^3 \left(r+2 \sqrt{r}-\sh+1\right)^2,\\
\detaapp_{4,4}\=\frac{1}{20 r^2}\bigg(-2 r^{5/2}+2 r^{3/2} (\sh+2)+r^3-r^2 (2 \sh+1)+r \left(2 \sh^2-\sh-1\right)-\\&&2 \sqrt{r} (\sh-1)^2-(\sh-1)^3\bigg).
\eeqynn

\beqynn
\detvvpp_{1,2}\=\detvvpp_{2,1}=\frac{1}{40 r^2}\left(\sqrt{r}+1\right) \left(r^2-2 r (\sh+1)+(\sh-1)^2\right)^2,\\
\detvvpp_{1,3}\=\detvvpp_{3,1}=\frac{1}{40 r^{5/2}}\left(\sqrt{r}+1\right) \left(r^2-2 r (\sh+1)+(\sh-1)^2\right)^2, \\
\detvvpp_{1,4}\=\detvvpp_{4,1}=-\frac{1}{20 r^2}\bigg(r^{7/2}-r^{5/2} (2 \sh+3)+r^{3/2} \left(\sh^2+3\right)+r^3-r^2 (\sh+3)-\\&&r \left(\sh^2+2 \sh-3\right)-\sqrt{r} (\sh-1)^2+(\sh-1)^3\bigg)^2, \\
\detvvpp_{2,3}\=\detvvpp_{3,2}=\frac{1}{80 r^{5/2}}\left(r-2 \sqrt{r}-\sh+1\right)^2 \left(r+2 \sqrt{r}-\sh+1\right)^3,\\
\detvvpp_{2,4}\=\detvvpp_{4,2}=-\frac{1}{40 r^2}\left(-2 r^{3/2}+r^2-2 \sqrt{r} (\sh-1)-(\sh-1)^2\right) \left(r+2 \sqrt{r}-\sh+1\right)^2,\\
\detvvpp_{3,4}\=\detvvpp_{4,3}=-\frac{1}{40 r^{5/2}}\left(-2 r^{3/2}+r^2-2 \sqrt{r} (\sh-1)-(\sh-1)^2\right) \left(r+2 \sqrt{r}-\sh+1\right)^2.
\eeqynn

\beqynn
\detaapp_{1,2}\=\detaapp_{2,1}=\frac{1}{40 r^2}\left(\sqrt{r}-1\right) \left(r^2-2 r (\sh+1)+(\sh-1)^2\right)^2,\\
\detaapp_{1,3}\=\detaapp_{3,1}=\frac{1}{40 r^{5/2}}\left(\sqrt{r}-1\right) \left(r^2-2 r (\sh+1)+(\sh-1)^2\right)^2, \\
\detaapp_{1,4}\=\detaapp_{4,1}=\frac{1}{20 r^2}\bigg(-r^{7/2}+r^{5/2} (2 \sh+3)-r^{3/2} \left(\sh^2+3\right)+r^3-r^2 (\sh+3)-\\&&r \left(\sh^2+2 \sh-3\right)+\sqrt{r} (\sh-1)^2+(\sh-1)^3\bigg), \\
\detaapp_{2,3}\=\detaapp_{3,2}=\frac{1}{80 r^{5/2}}\left(r-2 \sqrt{r}-\sh+1\right)^3 \left(r+2 \sqrt{r}-\sh+1\right)^2,\\
\detaapp_{2,4}\=\detaapp_{4,2}=-\frac{1}{40 r^2}\left(2 r^{3/2}+r^2+2 \sqrt{r} (\sh-1)-(\sh-1)^2\right) \left(r-2 \sqrt{r}-\sh+1\right)^2,\\
\detaapp_{3,4}\=\detaapp_{4,3}=-\frac{1}{40 r^{5/2}}\left(2 r^{3/2}+r^2+2 \sqrt{r} (\sh-1)-(\sh-1)^2\right) \left(r-2 \sqrt{r}-\sh+1\right)^2.
\eeqynn

\beqynn
\detvvpm_{2,2}\=\detvvmp_{2,2}=\frac{1}{80 r^2}\left(r-2 \sqrt{r}-\sh+1\right)^2 \left(r+2 \sqrt{r}-\sh+1\right)^3,\\
\detvvpm_{3,3}\=\detvvmp_{3,3}=-\frac{1}{80 r^3}\left(r-2 \sqrt{r}-\sh+1\right)^2 \left(r+2 \sqrt{r}-\sh+1\right)^3,\\
\detvvpm_{4,4}\=\detvvmp_{4,4}=\frac{1}{20 r^2}\bigg(r^{3/2} (4 \sh+2)+r^2 (2 \sh+1)-2 \sqrt{r} (\sh-1)^2-\\&& 3 r \sh (\sh-1)+(\sh-1)\bigg)^3,\\
\detaapm_{2,2}\=\detaamp_{2,2}=\frac{1}{80 r^2}\left(r-2 \sqrt{r}-\sh+1\right)^3 \left(r+2 \sqrt{r}-\sh+1\right)^2,\\
\detaapm_{3,3}\=\detaamp_{3,3}=-\frac{1}{80 r^3}\left(r-2 \sqrt{r}-\sh+1\right)^3 \left(r+2 \sqrt{r}-\sh+1\right)^2,\\
\detaapm_{4,4}\=\detaamp_{4,4}=\frac{1}{20 r^2}\bigg(-2 r^{3/2} (2 \sh+1)+r^2 (2 \sh+1)+2 \sqrt{r} (\sh-1)^2-\\&& 3 r \sh (\sh-1)+(\sh-1)^3\bigg).
\eeqynn

\beqynn
\detvvpm_{1,2}\=\detvvmp_{2,1}=\frac{1}{40 r^2}\left(\sqrt{r}+1\right) \left(r^2-2 r (\sh+1)+(\sh-1)^2\right)^2,\\
\detvvpm_{1,3}\=\detvvmp_{3,1}=-\frac{1}{40 r^{5/2}}\left(\sqrt{r}+1\right) \left(r^2-2 r (\sh+1)+(\sh-1)^2\right)^2,\\
\detvvpm_{1,4}\=\detvvmp_{4,1}=-\frac{1}{20 r^2}\left(\sqrt{r}+1\right) \bigg(r^{5/2}-2 r^{3/2} (\sh+1)+2 r^3-r^2 (5 \sh+3)+\\&& 4 r (\sh-1) \sh+\sqrt{r} (\sh-1)^2-(\sh-1)^3\bigg).
\eeqynn

\beqynn
\detvvpm_{2,1}\=\detvvmp_{1,2}=\frac{1}{40 r^2}\left(\sqrt{r}-1\right) \left(r^2-2 r (\sh+1)+(\sh-1)^2\right)^2,\\
\detvvpm_{2,3}\=\detvvmp_{3,2}=-\frac{1}{80 r^{5/2}}\left(r-2 \sqrt{r}-\sh+1\right)^2 \left(r+2 \sqrt{r}-\sh+1\right)^3,\\
\detvvpm_{2,4}\=\detvvmp_{4,2}=-\frac{1}{40 r^2}\bigg(-6 r^{3/2}+3 r^2-4 r (\sh-1)+2 \sqrt{r} (\sh-1)+\\&&(\sh-1)^2\bigg) \left(r+2 \sqrt{r}-\sh+1\right)^2.
\eeqynn

\beqynn
\detvvpm_{3,1}\=\detvvmp_{1,3}=\frac{1}{40 r^{5/2}}\left(\sqrt{r}-1\right) \left(r^2-2 r (\sh+1)+(\sh-1)^2\right)^2,\\
\detvvpm_{3,2}\=\detvvmp_{2,3}=\frac{1}{80 r^{5/2}}\left(r-2 \sqrt{r}-\sh+1\right)^2 \left(r+2 \sqrt{r}-\sh+1\right)^3,\\
\detvvpm_{3,4}\=\detvvmp_{4,3}=-\frac{1}{40 r^{5/2}}\bigg(-6 r^{3/2}+3 r^2-4 r (\sh-1)+2 \sqrt{r} (\sh-1)+\\&&(\sh-1)^2\bigg) \left(r+2 \sqrt{r}-\sh+1\right)^2.
\eeqynn

\beqynn
\detvvpm_{4,1}\=\detvvmp_{1,4}=\frac{1}{20 r^2}\left(\sqrt{r}-1\right) \bigg(r^{5/2}-2 r^{3/2} (\sh+1)-r^2 (\sh-1)+2 r \left(\sh^2-1\right)+\\&&\sqrt{r} (\sh-1)^2-(\sh-1)^3\bigg),\\
\detvvpm_{4,2}\=\detvvmp_{2,4}=-\frac{1}{40 r^2}\left(-2 r^{3/2}+r^2-2 \sqrt{r} (\sh-1)-(\sh-1)^2\right) \left(r+2 \sqrt{r}-\sh+1\right)^2,\\
\detvvpm_{4,3}\=\detvvmp_{3,4}=\frac{1}{40 r^{5/2}}\left(-2 r^{3/2}+r^2-2 \sqrt{r} (\sh-1)-(\sh-1)^2\right) \left(r+2 \sqrt{r}-\sh+1\right)^2.
\eeqynn

\beqynn
\detaapm_{1,2}\=\detaamp_{2,1}=\frac{1}{40 r^2}\left(\sqrt{r}-1\right) \left(r^2-2 r (\sh+1)+(\sh-1)^2\right)^2,\\
\detaapm_{1,3}\=\detaamp_{3,1}=-\frac{1}{40 r^{5/2}}\left(\sqrt{r}-1\right) \left(r^2-2 r (\sh+1)+(\sh-1)^2\right)^2,\\
\detaapm_{1,4}\=\detaamp_{4,1}=-\frac{1}{20 r^2}\left(\sqrt{r}-1\right) \bigg(-r^{5/2}+2 r^{3/2} (\sh+1)+2 r^3-r^2 (5 \sh+3)+\\&&4 r (\sh-1) \sh-\sqrt{r} (\sh-1)^2-(\sh-1)^3\bigg).
\eeqynn

\beqynn
\detaapm_{2,1}\=\detaamp_{1,2}=\frac{1}{40 r^2}\left(\sqrt{r}+1\right) \left(r^2-2 r (\sh+1)+(\sh-1)^2\right)^2,\\
\detaapm_{2,3}\=\detaamp_{3,2}=-\frac{1}{80 r^{5/2}}\left(r-2 \sqrt{r}-\sh+1\right)^3 \left(r+2 \sqrt{r}-\sh+1\right)^2,\\
\detaapm_{2,4}\=\detaamp_{4,2}=-\frac{1}{40 r^2}\bigg(6 r^{3/2}+3 r^2-4 r (\sh-1)-2 \sqrt{r} (\sh-1)+\\&&(\sh-1)^2\bigg) \left(r-2 \sqrt{r}-\sh+1\right)^2.
\eeqynn

\beqynn
\detaapm_{3,1}\=\detaamp_{1,3}=\frac{1}{40 r^{5/2}}\left(\sqrt{r}+1\right) \left(r^2-2 r (\sh+1)+(\sh-1)^2\right)^2,\\
\detaapm_{3,2}\=\detaamp_{2,3}=\frac{1}{80 r^{5/2}}\left(r-2 \sqrt{r}-\sh+1\right)^3 \left(r+2 \sqrt{r}-\sh+1\right)^2,\\
\detaapm_{3,4}\=\detaamp_{4,3}=-\frac{1}{40 r^{5/2}}\bigg(6 r^{3/2}+3 r^2-4 r (\sh-1)-2 \sqrt{r} (\sh-1)+\\&&(\sh-1)^2\bigg) \left(r-2 \sqrt{r}-\sh+1\right)^2.
\eeqynn

\beqynn
\detaapm_{4,1}\=\detaamp_{1,4}=-\frac{1}{20 r^2}\left(\sqrt{r}+1\right) \bigg(r^{5/2}-2 r^{3/2} (\sh+1)+r^2 (\sh-1)-2 r \left(\sh^2-1\right)+\\&&\sqrt{r} (\sh-1)^2+(\sh-1)^3\bigg),\\
\detaapm_{4,2}\=\detaamp_{2,4}=-\frac{1}{40 r^2}\left(2 r^{3/2}+r^2+2 \sqrt{r} (\sh-1)-(\sh-1)^2\right) \left(r-2 \sqrt{r}-\sh+1\right)^2,\\
\detaapm_{4,3}\=\detaamp_{3,4}=\frac{1}{40 r^{5/2}}\left(2 r^{3/2}+r^2+2 \sqrt{r} (\sh-1)-(\sh-1)^2\right) \left(r-2 \sqrt{r}-\sh+1\right)^2.
\eeqynn

\beqynn
\detvvmm_{1,1}\=-\frac{1}{20 r^2}\left(r^2-2 r (\sh+1)+(\sh-1)^2\right)^2,\\
\detvvmm_{2,2}\=\frac{1}{80 r^2}\left(r-2 \sqrt{r}-\sh+1\right)^2 \left(r+2 \sqrt{r}-\sh+1\right)^3,\\
\detvvmm_{3,3}\=\frac{1}{80 r^3}\left(r-2 \sqrt{r}-\sh+1\right)^2 \left(r+2 \sqrt{r}-\sh+1\right)^3,\\
\detvvmm_{4,4}\=\frac{1}{20 r^2}\bigg(6 r^{5/2}-6 r^{3/2} \sh+3 r^3+r^2 (3-6 \sh)+r \left(4 \sh^2-5 \sh+1\right)+\\&&2 \sqrt{r} (\sh-1)^2-(\sh-1)^3\bigg).
\eeqynn

\beqynn
\detaamm_{1,1}\=-\frac{1}{20 r^2}\left(r^2-2 r (\sh+1)+(\sh-1)^2\right)^2,\\
\detaamm_{2,2}\=\frac{1}{80 r^2}\left(r-2 \sqrt{r}-\sh+1\right)^3 \left(r+2 \sqrt{r}-\sh+1\right)^2,\\
\detaamm_{3,3}\=\frac{1}{80 r^3}\left(r-2 \sqrt{r}-\sh+1\right)^3 \left(r+2 \sqrt{r}-\sh+1\right)^2,\\
\detaamm_{4,4}\=\frac{1}{20 r^2}\bigg(-6 r^{5/2}+6 r^{3/2} \sh+3 r^3+r^2 (3-6 \sh)+r \left(4 \sh^2-5 \sh+1\right)-\\&&2 \sqrt{r} (\sh-1)^2-(\sh-1)^3\bigg).
\eeqynn

\beqynn
\detvvmm_{1,2}\=\detvvmm_{2,1}=\frac{1}{40 r^2}\left(\sqrt{r}-1\right) \left(r^2-2 r (\sh+1)+(\sh-1)^2\right)^2,\\
\detvvmm_{1,3}\=\detvvmm_{3,1}=-\frac{1}{40 r^{5/2}}\left(\sqrt{r}-1\right) \left(r^2-2 r (\sh+1)+(\sh-1)^2\right)^2, \\
\detvvmm_{1,4}\=\detvvmm_{4,1}=-\frac{1}{20 r^2}\bigg(r^{7/2}-r^{5/2} (2 \sh+3)+r^{3/2} \left(\sh^2+3\right)-3 r^3+r^2 (7 \sh+5)+\\&&r \left(-5 \sh^2+6 \sh-1\right)-\sqrt{r} (\sh-1)^2+(\sh-1)^3\bigg), \\
\detvvmm_{2,3}\=\detvvmm_{3,2}=-\frac{1}{80 r^{5/2}}\left(r-2 \sqrt{r}-\sh+1\right)^2 \left(r+2 \sqrt{r}-\sh+1\right)^3,\\
\detvvmm_{2,4}\=\detvvmm_{4,2}=-\frac{1}{40 r^2}\bigg(-6 r^{3/2}+3 r^2-4 r (\sh-1)+2 \sqrt{r} (\sh-1)+\\&&(\sh-1)^2\bigg) \left(r+2 \sqrt{r}-\sh+1\right)^2,\\
\detvvmm_{3,4}\=\detvvmm_{4,3}=\frac{1}{40 r^{5/2}}\bigg(-6 r^{3/2}+3 r^2-4 r (\sh-1)+2 \sqrt{r} (\sh-1)+\\&&(\sh-1)^2\bigg) \left(r+2 \sqrt{r}-\sh+1\right)^2.
\eeqynn

\beqynn
\detaamm_{1,2}\=\detaamm_{2,1}=\frac{1}{40 r^2}\left(\sqrt{r}+1\right) \left(r^2-2 r (\sh+1)+(\sh-1)^2\right)^2,\\
\detaamm_{1,3}\=\detaamm_{3,1}=-\frac{1}{40 r^{5/2}}\left(\sqrt{r}+1\right) \left(r^2-2 r (\sh+1)+(\sh-1)^2\right)^2, \\
\detaamm_{1,4}\=\detaamm_{4,1}=\frac{1}{20 r^2}\bigg(-r^{7/2}+r^{5/2} (2 \sh+3)-r^{3/2} \left(\sh^2+3\right)-3 r^3+r^2 (7 \sh+5)+\\&&r \left(-5 \sh^2+6 \sh-1\right)+\sqrt{r} (\sh-1)^2+(\sh-1)^3\bigg), \\
\detaamm_{2,3}\=\detaamm_{3,2}=-\frac{1}{80 r^{5/2}}\left(r-2 \sqrt{r}-\sh+1\right)^3 \left(r+2 \sqrt{r}-\sh+1\right)^2,\\
\detaamm_{2,4}\=\detaamm_{4,2}=-\frac{1}{40 r^2}\bigg(6 r^{3/2}+3 r^2-4 r (\sh-1)-2 \sqrt{r} (\sh-1)+\\&&(\sh-1)^2\bigg) \left(r-2 \sqrt{r}-\sh+1\right)^2,\\
\detaamm_{3,4}\=\detaamm_{4,3}=\frac{1}{40 r^{5/2}}\bigg(6 r^{3/2}+3 r^2-4 r (\sh-1)-2 \sqrt{r} (\sh-1)+\\&&(\sh-1)^2\bigg) \left(r-2 \sqrt{r}-\sh+1\right)^2.
\eeqynn

\beqynn
\detvag_{1,1}\=\detavg_{1,1}=-\frac{\left(r^2-2 r (\sh+1)+(\sh-1)^2\right)^2}{20 r^2},\\
\detvag_{4,4}\=\detavg_{4,4}=\frac{r^2-2 r (\sh+1)+(\sh-1)^2}{20 r},\\
\detvag_{1,4}\=\detavg_{4,1}=-\frac{r^2-2 r (\sh+1)+(\sh-1)^2}{10 r},\\
\detvag_{4,1}\=\detavg_{1,4}=\frac{r^2-2 r (\sh+1)+(\sh-1)^2}{10 r}.
\eeqynn

\end{document}